# When Darwin meets Lorenz: Evolving new chaotic attractors through genetic programming

Indranil Pan[a,*] and Saptarshi Das[b]

a) *Department of Earth Science and Engineering, Imperial College London, Exhibition Road, London SW7 2AZ, United Kingdom (Email: i.pan11@imperial.ac.uk )*

b) *School of Electronics and Computer Science, University of Southampton, Southampton SO17 1BJ, United Kingdom (Email: s.das@soton.ac.uk )*

**Abstract:**

In this paper, we propose a novel methodology for automatically finding new chaotic attractors through a computational intelligence technique known as multi-gene genetic programming (MGGP). We apply this technique to the case of the Lorenz attractor and evolve several new chaotic attractors based on the basic Lorenz template. The MGGP algorithm automatically finds new nonlinear expressions for the different state variables starting from the original Lorenz system. The Lyapunov exponents of each of the attractors are calculated numerically based on the time series of the state variables using time delay embedding techniques. The MGGP algorithm tries to search the functional space of the attractors by aiming to maximise the largest Lyapunov exponent (LLE) of the evolved attractors. To demonstrate the potential of the proposed methodology, we report over one hundred new chaotic attractor structures along with their parameters, which are evolved from just the Lorenz system alone.

## 1. Introduction:

An important research theme in non-linear dynamics is to identify sets of differential equations along with their parameters which give rise to chaos. Starting from the advent of Lorenz attractors in three dimensional nonlinear differential equations [1][2], its several other family of attractors have been developed like Rossler, Rucklidge, Chen, Lu, Liu, Sprott, Genesio-Tesi, Shimizu-Morioka etc. [3]. Extension of the basic attractors to four or even higher dimensional systems has resulted in a similar family of hyper-chaotic systems [4]. In addition, several other structures like multi-scroll [5] and multi-wing [6] versions of the Lorenz family of attractors have also been developed by increasing the number of equilibrium points. Development of new chaotic attractors has huge application especially in data encryption, secure communication [7] etc. and in understanding the dynamics of many real world systems whose governing equations match with the template of these chaotic systems [8]. There has been several research reports on the application of master-slave chaos synchronization in secure communication [9], where the fresh set of chaotic attractors can play a big role due to their rich phase space dynamics. Here we explore the potential of automatic generation of chaotic attractors which is developed from the basic three dimensional structure of the Lorenz system.

We use the Lyapunov exponent – the most popular signature of chaos, to judge whether a computer generated arbitrary nonlinear structure of third order differential equation along with some chosen parameters exhibits chaotic motion in the phase space. Since there could be chaotic behaviour or complex limit cycles or even stable/unstable motions in the phase space, for unknown mathematical expressions of similar third order dynamical system, the computationally tractable way for the investigation of the chaos seems to be the characterisation of the observed time series. The Taken's theorem says that the original attractor could be reconstructed from the observation of just one state variable using the time delay embedding method [10]. In the time delay reconstruction method, the dynamics of the chaotic attractors is approximated in the phase space by plotting the observed time series and its delayed versions along orthogonal axes and finding the delay that has got maximum span in the phase space [10], [11].

We predominantly report similar chaotic attractors evolved over the basic Lorenz system of equations by changing two and three state equations together using the GP. We use standard nonlinear terms in the Lorenz family like the cross product and square terms in combination with sinusoidal terms, giving rise to multiple equilibrium points and hence very complex dynamical behaviour in the phase space e.g. with infinite number of equilibrium points [12][13].

Previous attempts have looked at chaotic dynamical systems modelling through the use of genetic programming [14], [15] using nonlinear autoregressive moving average with exogenous inputs or NARMAX models. The papers try to reproduce the dynamics of the Chua circuit through the nonlinear autoregressive models with different lags and orders. They also use multi-objective genetic algorithms to obtain a set of parsimonious models which can reproduce the original behaviour of the known chaotic attractor. The disadvantage of the paper is that the obtained expressions have multiple time delay terms in them, which are more complex than the ones obtained intuitively (i.e. which have three coupled first order nonlinear differential terms). Also, due to the method employed, the obtained chaotic dynamics are very similar to the original attractor and their phase portraits do not differ drastically, i.e. new types of chaotic attractors with totally different phase space dynamics are difficult to obtain using this method.

There have been attempts of evolving new single state discrete time chaotic dynamical systems using the concept of GP and study of their bifurcation diagrams [16]. Similar work has been done in [17] using evolutionary algorithms and in [18] using analytical programming techniques. The technique has also been extended to higher number of states and used for evolutionary reconstruction of continuous time chaotic systems [19]. The present paper uses a MGGP paradigm to evolve multiple expressions for the state variables simultaneously. The multi-gene approach helps in obtaining new sets of dynamical equations which can show completely different phase space dynamics. This fact is exploited in the GP algorithm to find new sets of chaotic dynamical systems by maximising the LLE.

This is the first paper of its kind that reports not only a single or a handful of attractors by varying the basic Lorenz system of equations like Rössler, Chen, Lu, Liu etc., but over hundreds of new interesting differential equation structures along with their parameters and Lyapunov exponents. The generic method proposed in this paper can be viewed as the first step towards explaining very complex chaotic motions hidden in various physical systems, which could be obtained by mixing the Lorenz equation with simple transcendental like sinusoids [20], which has got infinite number of equilibria.

## 2. Genetic programming to evolve new chaotic attractors
### 2.1. Basics of Genetic programming

Genetic programming is an intelligent algorithm which is capable of automatically evolving computer programs to perform a given task [21], [22]. The GP algorithm has been applied to a variety of practical applications in human competitive engineering design [23]. GP has been used for symbolic regression to fit analytical nonlinear expressions to do prediction for any given experimental input-output data set [24][25]. Essentially it can evolve the structure and parameters of a nonlinear expression by minimising the mean squared error (MSE) between the predicted and the observed values. This expression is analytic in nature and is therefore amenable to mathematical analysis. The present work exploits this paradigm to evolve analytical expressions for chaotic attractors, by trying to maximise the LLE of each of the nonlinear dynamical systems.

The GP algorithm is based on the Darwinian principle of evolution and survival of the fittest. The recently introduced multi-gene GP or MGGP [24] is used in the present study. Each state equation which the GP searches is represented by one gene. If the GP simultaneously searches for two state equations, then they are represented by two separate genes and together they constitute one individual. These genes can be represented in the form of a tree structure as shown in Figure 1. In the beginning, the individuals are randomly initialized within the feasible space. Then they undergo reproduction, crossover and mutation to evolve fitter individuals in the subsequent generations. Crossover refers to the interchange of genetic material among the solutions. Mutation on the other hand refers to a random change within a gene itself. The crossover and mutation operations are stochastic ones and their probability of occurrence is pre-specified by the user. The tree structure representation as shown in Figure 1 is useful for doing the cross over and mutation operations algorithmically. Generally the maximum number of levels in a tree is confined to a specific small number to decrease the bloat in the solutions and also to reduce the run time of the algorithms. Here bloat refers to the case where the GP goes on evolving very complicated nonlinear expressions, without significant increase in the performance metric (i.e. the objective function) [23].

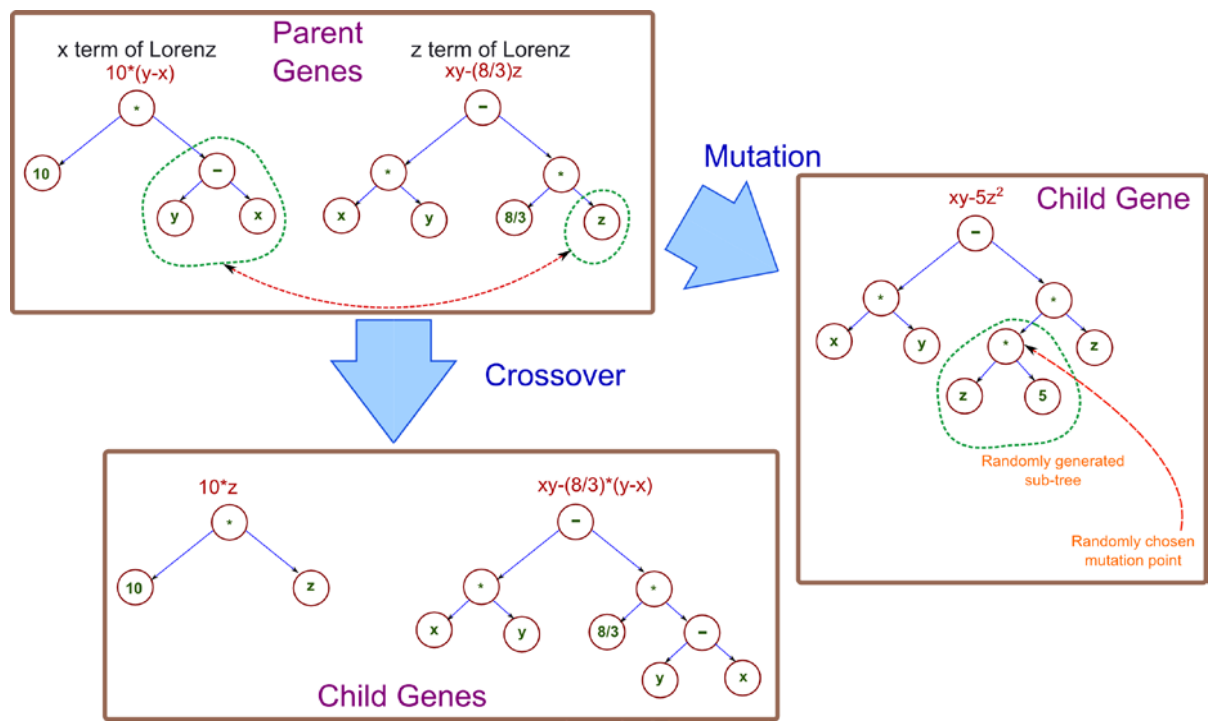

Figure 1: Schematic of the automatic evolution of new chaotic attractors using GP, based on the Lorenz system

The MGGP algorithm, as introduced in [24], uses a tree structure to represent each sub-gene and a weighted combination of these sub-genes are used to represent one individual gene. In this paper, we use the sub-genes to represent each of the state equations of the chaotic attractor, but we do not use the weights. Therefore the whole dynamical system is represented by one gene which has multiple sub-genes for each of the different state variables. So for example, if the objective is to evolve only two of the three state equations of the dynamical system, then two sub-genes would represent the overall system (since the other state equation is constant).

### 2.2. Objective function

For any experimentally observed time series, the Lyapunov exponent of the reconstructed attractor having value greater than one is considered as a genuine signature of chaos [26]. In the present paper, we exploit this fact to identify whether a given set of equations are chaotic or not. There are many different methods of calculating the LLE or the Largest Lyapunov Exponent from the analytical set of differential equations. However, most of them require calculation of the Jacobian, which is difficult to do automatically through a computer program as they might involve complicated nonlinear expressions and unnecessarily increase the computation time within each run of the GP. Therefore to circumvent this problem we resorted to a numerical method of evaluating the LLE [27]. We simulated each of the GP evolved expression with an ordinary differential equation solver using a fixed time step of 0.01 seconds for a total time of 25 seconds. Then from the short time series obtained from the $x$-state variable [28], we used the time delay embedding method to reconstruct the attractor dynamics as done in Rosenstein et al. [27] and computed the LLE numerically. The GP algorithm aims to maximise the LLE in order to obtain systems which exhibit strong chaotic behaviour.

It is to be noted that since the LLE is calculated from only one of the state variables by reconstructing the original attractor dynamics, it might not always give the true Lyapunov exponent of the system. However, we found that this is able to easily distinguish between chaotic and non-chaotic sequences and is therefore suitable as a performance metric to evaluate the chaotic dynamics of a given system. One of the problems is that the GP algorithm frequently evolves expressions whose solutions blow up. These expressions are not useful and a high value of fitness function is assigned to these solutions to discourage the algorithm from exploring these parts of the search space.

### 2.3. Different cases considered in this study

Depending on which state equations are modified in the original Lorenz system (1), four different cases are identified as shown below in equations (2)-(5) by modifying two/three states together in the template of the celebrated Lorenz system [1].

$$\dot{x} = 10(y-x), \dot{y} = 28x - xz - y, \dot{z} = xy - (8/3)z \tag{1}$$

i) *Case A: Generalized type y-z Lorenz family*

$$\dot{x} = 10(y-x), \dot{y} = f(x,y,z), \dot{z} = h(x,y,z) \tag{2}$$

ii) *Case B: Generalized type x-y Lorenz family*

$$\dot{x} = g(x,y,z), \dot{y} = f(x,y,z), \dot{z} = xy - (8/3)z \tag{3}$$

iii) *Case C: Generalized type x-z Lorenz family*

$$\dot{x} = g(x,y,z), \dot{y} = 28x - xz - y, \dot{z} = h(x,y,z) \qquad (4)$$

iv) *Case D: Generalized x-y-z family (no structure)*

$$\dot{x} = g(x,y,z), \dot{y} = f(x,y,z), \dot{z} = h(x,y,z) \qquad (5)$$

In case A (Lorenz *y-z* family), the original expression of the Lorenz system is retained for the *x* state variable, while the expressions for the *y* and *z* states are evolved by the GP algorithm in each iteration. For cases B (Lorenz *x-y* family) and C (Lorenz *x-z* family) the algorithm evolves the equations for the states *x-y* and *x-z* respectively. Case D (Lorenz *x-y-z* family) is a generalized template which can evolve any chaotic attractor and the algorithm evolves all the three state equations involving the rate of *x*, *y* and *z*. Therefore this template has the potential to evolve even the other well-known chaotic attractors like Rössler, Chen, Lu, Rabinovich-Fabrikant systems etc.

## 3. Simulation results

The GP algorithm is run for the different cases A to D for 100 times each and the unique structures with positive LLEs are identified and saved for post-processing. The parameter settings of the GP algorithm are shown in Table 1. One of the important parameters is the ratio of the mathematical operators to that of the numeric values in each gene which is taken as 0.5. A higher value would imply more mathematical operators which would make the expressions more complicated. A lower value would have more number of constants in the expression, which would make the expressions simpler but would reduce the capability of the expression to exhibit rich time domain and phase space dynamics.

Table 1: Parameter settings for the MGGP algorithm

| GP Algorithm parameters | Parameter settings |
|---|---|
| Population size | 25 |
| Number of generation | 50 |
| Selection method | Plain lexicographic tournament selection |
| Tournament size | 3 |
| Termination criteria | Maximum number of iterations |
| Maximum depth of tree | 5 |
| Maximum tree depth for mutation | 2 |
| Ratio of mathematical operators to numeric values | 0.5 |
| Mathematical operators | $\{+, -, \times, \div, \sin\}$ |

In general the objective is to obtain all dynamical systems which have a positive Lyapunov exponent and not the one with the maximum Lyapunov exponent alone. Therefore we added a history variable in the GP algorithm which keeps track of all the nonlinear expressions which had a Lyapunov exponent greater than zero. It is essential to preserve the chaotic attractors in the intermediate generations as subsequent generations might destroy the attractor structure and evolve more fitter structures (with higher LLE) through the crossover and mutation operations. The simulations were run on a 64 bit Windows desktop with 16 GB memory and an Intel I7, 3.4 GHz processor. 100 runs of each generalized family of attractors (i.e. each of the cases A, B, C, D) took approximately two days each.

After the simulation runs, we screened the attractors which have a positive LLE as given by the time delay embedding technique. However the LLE obtained from this method is not reliable since it is computed by the phase space reconstruction using time delay embedding of just one of the state variables using a very short time series. Therefore we symbolically computed the Jacobian of each of the automatically evolved differential equations and calculated all the Lyapunov exponents and hence the LLE numerically, using the standard method as reported in [29]. We have also validated our algorithm with the Lyapunov exponents of standard attractors given in [29]. The algorithm to find the LLE of the new structures comprises of symbolic differentiation of the state equations to obtain the Jacobian, as an intermediate step. All our simulation results are reported with an initial condition of $\{x_0 = 2, y_0 = 3, z_0 = 1\}$ using a fixed step-size of 0.5 sec in the ode45 routine of Matlab (which uses Dormand-Price algorithm). For this numerical integration a total time span of 10000 second is simulated.

Table 2: List of generalized Lorenz *x-z* family of attractors

| Name | $\dot{x}$ expression | $\dot{z}$ expression | Coefficient values | LLE |
|---|---|---|---|---|
| LorXZ1 | $ay - bx + \sin(z) - c$ | $xy - dz$ | a=10.0, b=10.0, c=5.9666, d=2.6667 | 0.8824 |
| LorXZ2 | $ay - bx + \sin(\sin(y)) - c$ | $xy - dz$ | a=10.0, b=10.0, c=2.6667, d=2.6667 | 0.9040 |
| LorXZ3 | $ay - bx + \sin(z) - c$ | $x^d - ez$ | a=10.0, b=10.0, c=5.9666, d=2, e=2.6667 | 0.9809 |
| LorXZ4 | $ay - bx - cz + \sin(z) - d$ | $xy - ez$ | a=10.0, b=10.0, c=1.0, d=2.6667, e=2.6667 | 0.8952 |
| LorXZ5 | $x - ay + bz + \sin(\sin(z) - cy + d)$ | $\sin(x)$ | a=1.0, b=7.6825, c=1.0, d=3.9819 | 0.0606 |
| LorXZ6 | $ay - bx$ | $x(y+c) - dz$ | a=10.0, b=10.0, c=0.24176, d=2.6667 | 0.8985 |
| LorXZ7 | $ay - bx$ | $x(x+c) - dz$ | a=10.0, b=10.0, c=0.24176, d=2.6667 | 0.9626 |
| LorXZ8 | $ay - bx$ | $xy - cz - dy + x(\sin(z) - e)$ | a=10.0, b=10.0, c=2.6667, d=1.0, e=2.3465 | 0.7488 |
| LorXZ9 | $a\sin(z)$ | $b\sin(y)$ | a=10.0, b=100.0 | 0.0036 |
| LorXZ10 | $ay - bx + \sin(\sin(x)) + \sin(x - cy)$ | $xy - dz$ | a=10.0, b=10.0, c=1.0, d=2.6667 | 0.9031 |
| LorXZ11 | $ay - bx + \sin(\sin(x))$ | $xy - cz$ | a=10.0, b=10.0, c=2.6667 | 0.9053 |
| LorXZ12 | $ay - bx + \sin(x - cy)$ | $xy - dz$ | a=10.0, b=10.0, c=1.0, d=2.6667 | 0.8985 |
| LorXZ13 | $ay - bx + \sin(x - c)$ | $x^d - ez$ | a=10.0, b=10.0, c=4.5177, d=2, e=2.6667 | 0.9863 |
| LorXZ14 | $ay - bx - c\sin(y - dz)$ | $xy - ez$ | a=11.0, b=10.0, c=1.0, d=1.0, e=2.6667 | 0.9674 |
| LorXZ15 | $ax(z - b)$ | $c\sin(dy) - ex$ | a=10.0, b=1.0, c=10.0, d=6.6163, e=10.0 | 0.0722 |
| LorXZ16 | $ay - bx$ | $x(x + \sin(z)) - cz$ | a=10.0, b=10.0, c=2.6667 | 0.9325 |

Here we explore the four different cases – Lorenz XZ, XY and YZ and XYZ family of attractors evolved using GP. These are reported in Table 2-Table 5 respectively. In the first three cases, one expression is kept common as in the Lorenz system and in the fourth case, all the state equations are evolved using GP. Symbolically computed LLE and the co-efficient values (exhibiting chaos) are also reported along with their nomenclature in the following tables which describes their dynamical behaviour in the 3D phase space. It is to be noted that we have reported only those chaotic systems which has at least one sine term in any of its state equations because it opens up the possibility of exhibiting rich phase space dynamics [20].

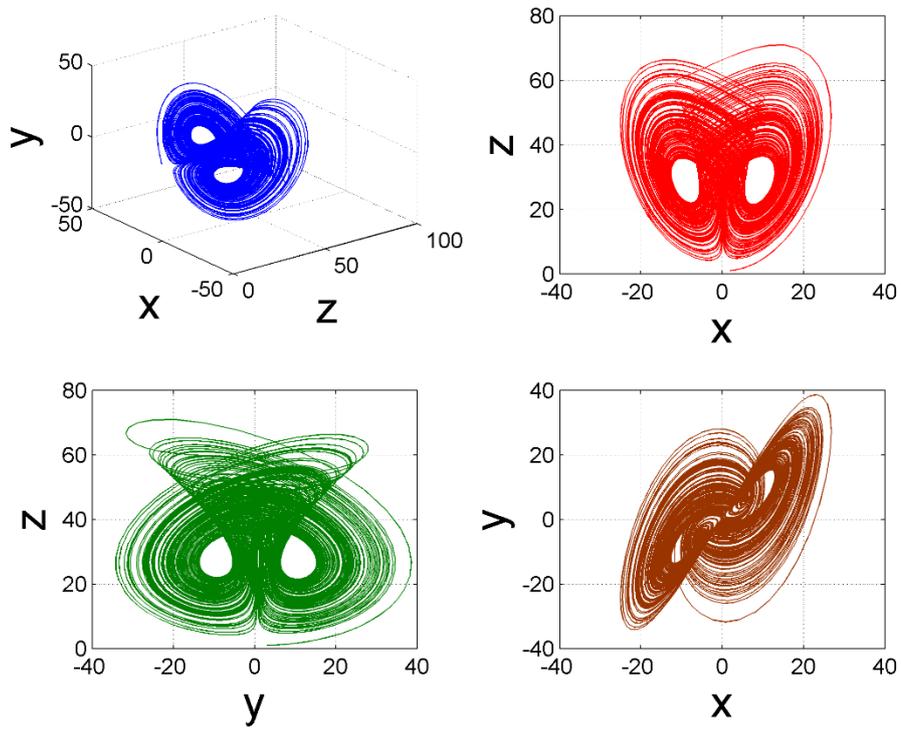

Figure 2: Phase space dynamics of LorXZ3 (*trunk*)

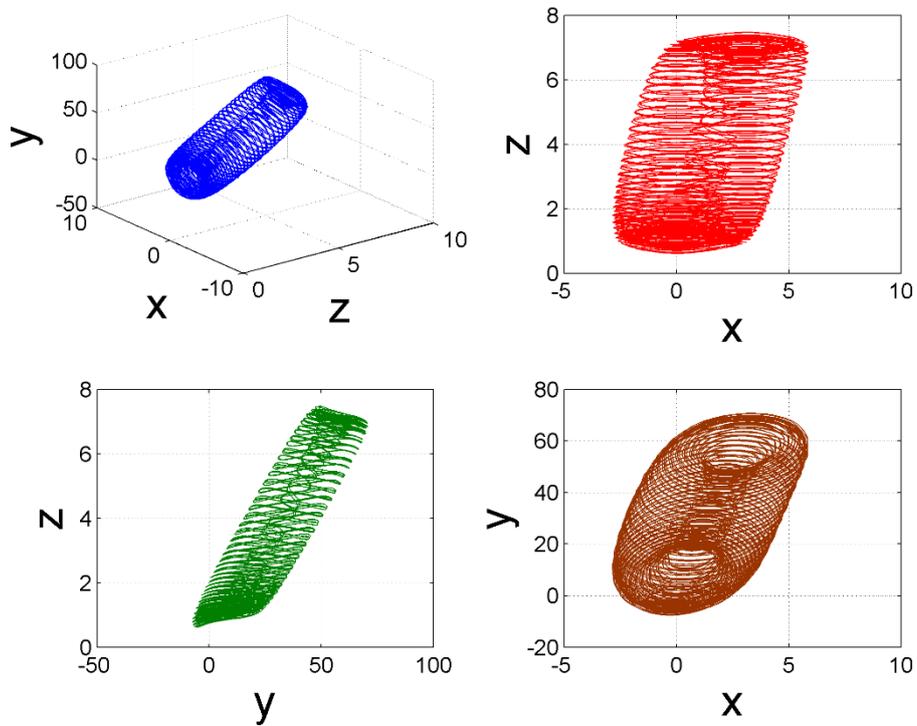

Figure 3: Phase space dynamics of LorXZ5 (*toroidal spring*)

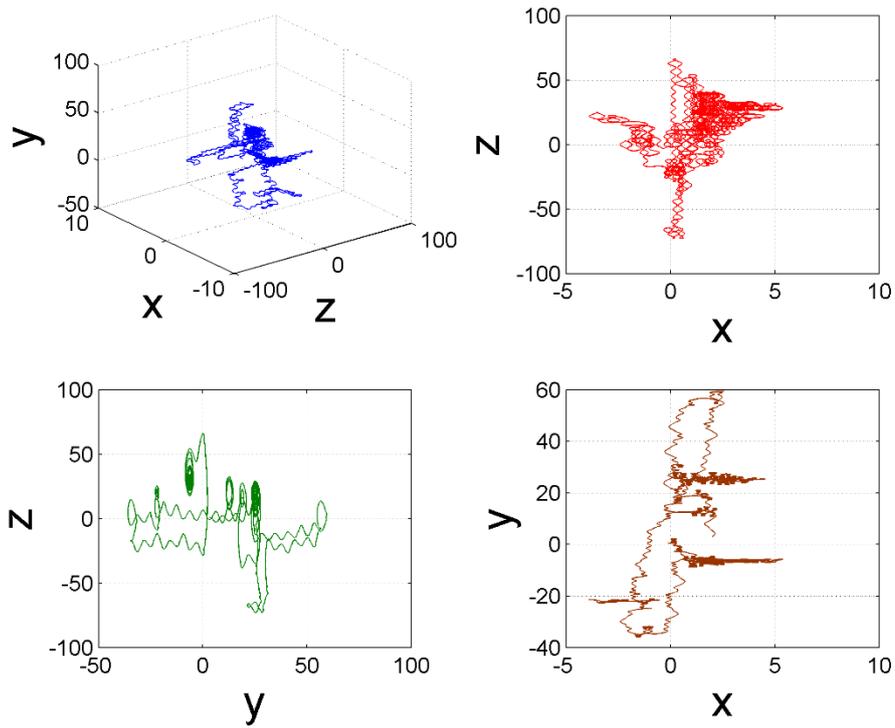

Figure 4: Phase space dynamics of LorXZ9 (*tangled string*)

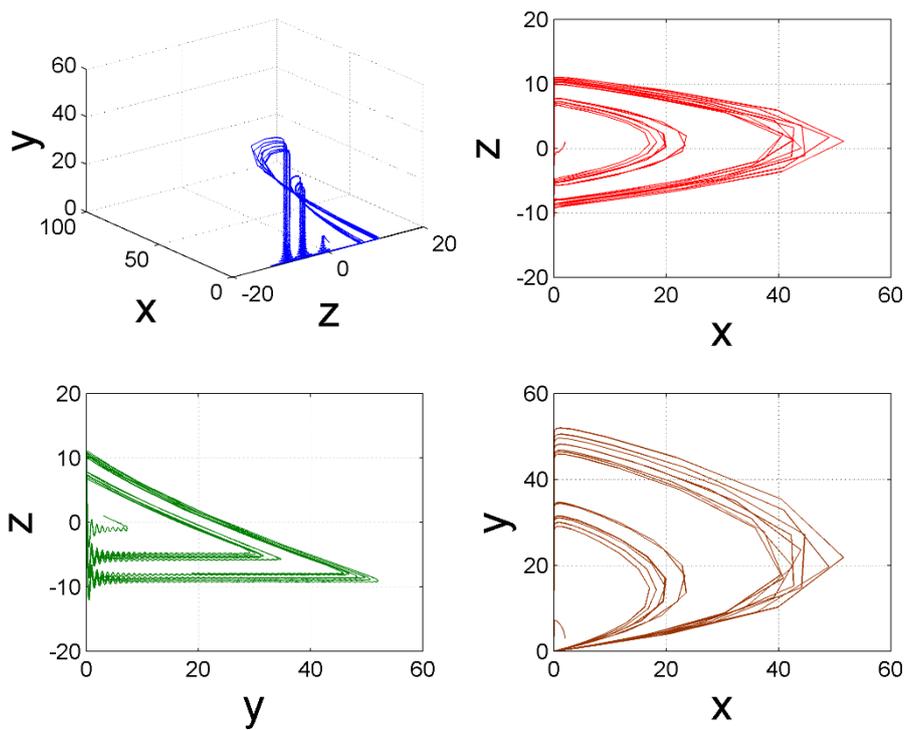

Figure 5: Phase space dynamics of LorXZ15

Due to the large number of chaotic attractors obtained through GP, most of the phase portraits of the new systems are shown in the supplementary material. We here show the 2D and 3D phase space diagrams of only those systems which has some interesting patterns in contrast with the standard Lorenz system. For example in Figure 2 for the LorXZ3 system, especially in the *y-z* plane, there is a prominently large '*trunk*' between the two wings and its LLE is 0.98 which is close to that of the classical Lorenz system. Whereas the LorXZ5 system in Figure 3 shows very complex dynamics like a '*toroidal spring*' and its LLE has deviated significantly from the original Lorenz system. LorXZ9 system in Figure 4 shows trapping of the phase space dynamics around several equilibrium points which takes a shape of a '*tangled string*'. System LorXZ15 shows high oscillations near low *y* values and takes multiple triangular excursions in the phase space combining both fast and slow dynamics together. Both the LorXZ5 and LorXZ15 have a small LLE of 0.06 and 0.07 respectively, whereas the LorXZ9 shows zero LLE up to two decimal place which refers to a weak chaos [30]. Phase portraits of the rest of the chaotic attractors in Table 2 are shown in the supplementary material and have the LLEs ranging between 0.74-0.98 which is pretty close to that of the original Lorenz system.

In Table 3, similar results are reported for the Lorenz *x-y* family of attractors. In Figure 6, the phase space diagram of LorXY10 is shown which has a relatively smaller LLE of 0.5803. This system shows '*uneven wings*' and the trajectories are more concentrated around one wing as compared to the other. A similar LLE of 0.58 and the resulting phase space dynamics can also be found in LorXY15. It is also interesting to note that in this family some of them have LLE greater than one e.g. LorXY4, LorXY5, LorXY21, LorXY22, LorXY23, LorXY25, LorXY28, LorXY29, LorXY31, LorXY33, LorXY34, indicating strong chaotic dynamics [30]. For example the LorXY21 system in Figure 7 shows a dense phase space diagram (especially in *y-z* plane) which is also confirmed by its high LLE of 1.1416. In Table 3, LorXY22 has the highest LLE of 1.2 among all the attractors in the LorXY family. Rest of the attractors' LLEs in the LorXY family vary between 0.87-1.06.

Table 3: List of generalized Lorenz *x-y* family of attractors

| Name | $\dot{x}$ expression | $\dot{y}$ expression | Coefficient values | LLE |
|---|---|---|---|---|
| LorXY1 | $ay - bx$ | $cx + \sin(dz) - exz$ | a=10.0, b=10.0, c=28.0, d=5.0302, e=1.0 | 0.9894 |
| LorXY2 | $ay - bx$ | $cx + \sin(dx) - exz$ | a=10.0, b=10.0, c=28.0, d=5.0302, e=1.0 | 0.9783 |
| LorXY3 | $ay - bx$ | $cx - d\sin(\sin(x+e)) - fxz$ | a=10.0, b=10.0, c=28.0, d=1.0, e=6.2452, f=1.0 | 0.9978 |
| LorXY4 | $ay - bx$ | $cx + y - d\sin(x) - exz - f$ | a=10.0, b=10.0, c=28.0, d=1.0, e=1.0, f=16.727 | 1.0598 |
| LorXY5 | $ay - bx$ | $cx + dy + e\sin(y) - fxz$ | a=10.0, b=10.0, c=28.0, d=0.044006, e=0.044006, f=1.0 | 1.0037 |
| LorXY6 | $ay - bx$ | $cx + \sin(y-d)(\sin(x)+e) - fxz$ | a=10.0, b=10.0, c=28.0, d=9.09, e=0.82679, f=1.0 | 0.9924 |
| LorXY7 | $ay - bx$ | $cx - dy - e\sin(\sin(y)) - fxz$ | a=10.0, b=10.0, c=28.0, d=1.0, e=1.0, f=1.0 | 0.8791 |
| LorXY8 | $ay - bx$ | $cx + \sin(x) - dxz$ | a=10.0, b=10.0, c=27.0, d=1.0 | 0.9765 |
| LorXY9 | $ay - bx$ | $cx + \sin(y+d) - exz$ | a=10.0, b=10.0, c=28.0, d=14.276, e=1.0 | 0.9858 |
| LorXY10 | $ay - bx - c$ | $dx + \sin(y+e) - fxz$ | a=10.0, b=10.0, c=62.569, d=28.0, e=14.276, f=1.0 | 0.5803 |
| LorXY11 | $ay - bx$ | $cx - d\sin(x+y-e) - fxz$ | a=10.0, b=10.0, c=28.0, d=1.0, e=6.2569, f=1.0 | 0.9919 |
| LorXY12 | $ay - bx$ | $cx + \sin(dx - ey) - fxz$ | a=10.0, b=10.0, c=28.0, d=10.0, e=10.0, f=1.0 | 0.9765 |
| LorXY13 | $ay - bx$ | $cx - d\sin(x - ey - f) - gxz$ | a=10.0, b=10.0, c=28.0, d=1.0, e=1.0, f=6.386, g=1.0 | 0.9965 |
| LorXY14 | $ay - bx$ | $cx - d\sin(x - ey) - fxz$ | a=10.0, b=10.0, c=28.0, d=1.0, e=1.0, f=1.0 | 0.9978 |
| LorXY15 | $ay - bx - c$ | $dx + \sin(ex - fy) - gxz$ | a=10.0, b=10.0, c=62.569, d=28.0, e=2.0, f=1.0, g=1.0 | 0.5800 |
| LorXY16 | $ay - bx$ | $cx - d\sin(\sin(x)) - exz$ | a=10.0, b=10.0, c=28.0, d=1.0, e=1.0 | 0.9970 |
| LorXY17 | $ay - bx$ | $cx - d\sin(z) - exz + f$ | a=10.0, b=10.0, c=28.0, d=1.0, e=1.0, f=9.3771 | 0.9618 |
| LorXY18 | $ay - bx$ | $cx - d\sin(ey) - fxz$ | a=10.0, b=10.0, c=28.0, d=1.0, e=4.8349, f=1.0 | 0.9155 |
| LorXY19 | $ay - bx$ | $cx - d\sin(z^e) - fxz$ | a=10.0, b=10.0, c=28.0, d=1.0, e=2, f=1.0 | 0.9684 |
| LorXY20 | $ay - bx$ | $cx - d\sin(xy) - exz$ | a=10.0, b=10.0, c=28.0, d=1.0, e=1.0 | 0.9755 |
| LorXY21 | $ay - bx$ | $cx - d\sin(yz) - exz^f$ | a=10.0, b=10.0, c=28.0, d=1.0, e=1.0, f=2 | 1.1416 |
| LorXY22 | $ay - bx$ | $-cxz^d + ex + \sin(x - fy)$ | a=10.0, b=10.0, c=1.0, d=2, e=28.0, f=1.0 | 1.2034 |
| LorXY23 | $ay - bx$ | $cx + y + \sin(x) - dxz + e$ | a=10.0, b=10.0, c=27.0, d=1.0, e=7.9648 | 1.0629 |
| LorXY24 | $ay - bx$ | $cx - d\sin(\sin(x^e)) - fxz$ | a=10.0, b=10.0, c=28.0, d=1.0, e=2, f=1.0 | 0.9791 |
| LorXY25 | $ay - bx$ | $cx - dz - e\sin(z) - fxz^g + h$ | a=10.0, b=10.0, c=28.0, d=1.0, e=1.0, f=1.0, g=2, h=1.8865 | 1.0457 |
| LorXY26 | $ay - bx$ | $cx + \sin(x - dy) + \sin(z) - exz - f$ | a=10.0, b=10.0, c=28.0, d=1.0, e=1.0, f=13.173 | 0.9085 |
| LorXY27 | $ay - bx$ | $cx + \sin(x - dy) - e\sin(y) + \sin(z) - fxz$ | a=10.0, b=10.0, c=28.0, d=1.0, e=1.0, f=1.0 | 0.9853 |
| LorXY28 | $ay - bx$ | $cx - d\sin(y) + \sin(z) - exz - f$ | a=10.0, b=10.0, c=29.0, d=1.0, e=1.0, f=6.6666 | 1.0073 |
| LorXY29 | $ay - bx$ | $cx + y - d\sin(y) + \sin(z) - exz - f$ | a=10.0, b=10.0, c=28.0, d=1.0, e=1.0, f=6.6666 | 1.0635 |
| LorXY30 | $ay - bx$ | $cx + \sin(x - d) - e\sin(y) + \sin(z) - fxz$ | a=10.0, b=10.0, c=28.0, d=8.2736, e=1.0, f=1.0 | 0.9907 |
| LorXY31 | $ay - bx$ | $cx + y + \sin(x - dy) - e\sin(y) - fxz$ | a=10.0, b=10.0, c=28.0, d=1.0, e=1.0, f=1.0 | 1.0438 |
| LorXY32 | $ay - bx$ | $cx + \sin(x - dy) - e\sin(x) + \sin(z) - fxz$ | a=10.0, b=10.0, c=28.0, d=1.0, e=1.0, f=1.0 | 0.9879 |
| LorXY33 | $ay - bx$ | $cx + y + \sin(z) - dxz - e$ | a=10.0, b=10.0, c=27.0, d=1.0, e=6.6666 | 1.0592 |
| LorXY34 | $ay - bx$ | $cx - d\sin(x) - exz$ | a=10.0, b=10.0, c=30.0, d=1.0, e=1.0 | 1.0373 |

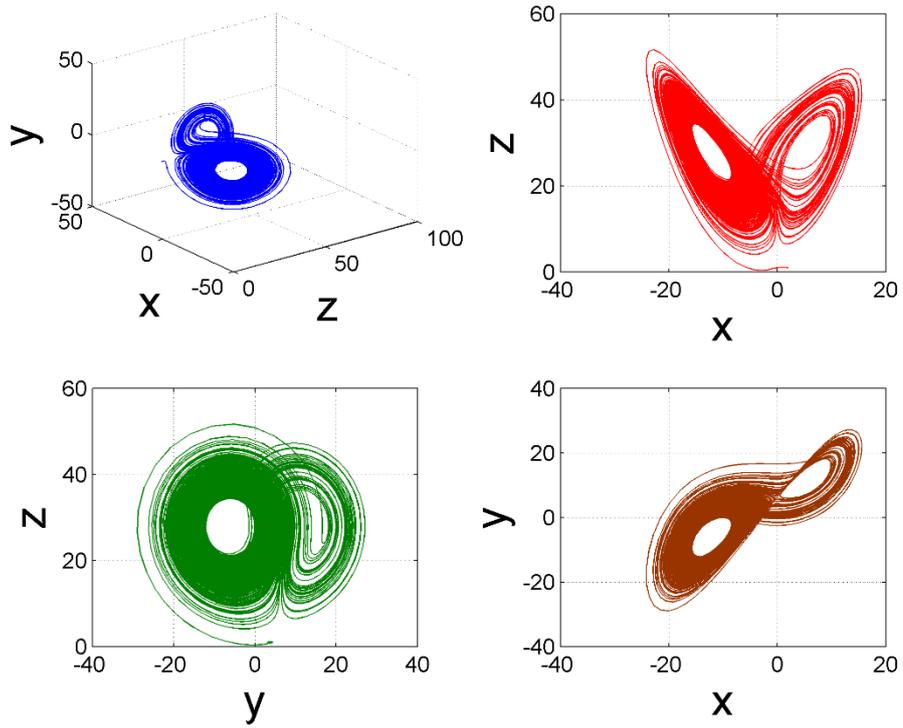

Figure 6: Phase space dynamics of LorXY10 (*uneven wings*)

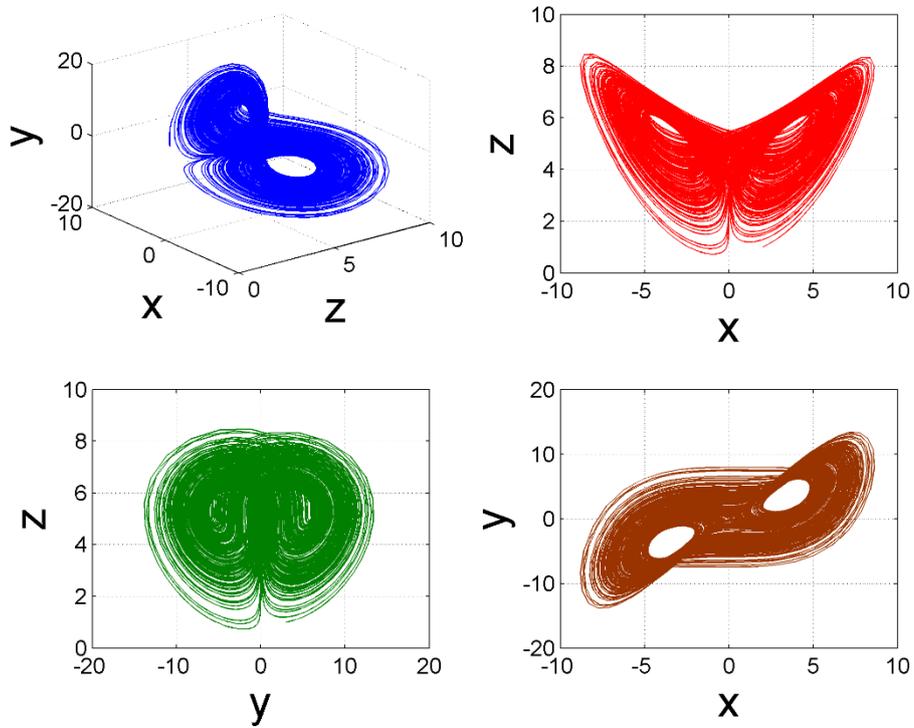

Figure 7: Phase space dynamics of LorXY21

Table 4: List of generalized Lorenz *y-z* family of attractors

| Name | $\dot{y}$ expression | $\dot{z}$ expression | Coefficient values | LLE |
|---|---|---|---|---|
| LorYZ1 | $ax + \sin(\sin(\sin(y))) - bxz$ | $xy - cz$ | a=28.0, b=1.0, c=2.6667 | 0.9942 |
| LorYZ2 | $ax + \sin(y) - bxz$ | $xy - cz$ | a=28.0, b=1.0, c=2.6667 | 0.9893 |
| LorYZ3 | $ax + \sin(\sin(z)) - bxz$ | $xy - cz$ | a=28.0, b=1.0, c=2.6667 | 0.9958 |
| LorYZ4 | $ax + \sin(\sin(x+b)) - cxz$ | $xy - dz$ | a=28.0, b=28.0, c=1.0, d=2.6667 | 0.9970 |
| LorYZ5 | $ax + \sin(\sin(\sin(y))) - bxz$ | $x^c - dz$ | a=28.0, b=1.0, c=2, d=2.6667 | 1.3102 |
| LorYZ6 | $ax - b\sin(x) - cxz$ | $xy - d$ | a=2.6667, b=1.0, c=1.0, d=74.746 | 1.6529 |
| LorYZ7 | $-ax(z + \sin(x) - b)$ | $x^c - dz$ | a=1.0, b=28.0, c=2, d=2.6667 | 1.1940 |
| LorYZ8 | $-ax(z - b)$ | $x(cx - d\sin(x)) - ez$ | a=1.0, b=28.0, c=2.6667, d=1.0, e=2.6667 | 1.3731 |
| LorYZ9 | $ax - b\sin(cz) - dxz$ | $xy - ez$ | a=28.0, b=1.0, c=2.6667, d=1.0, e=2.6667 | 0.9988 |
| LorYZ10 | $ax - b\sin(z(y+c)) - dxz$ | $xy - ez$ | a=28.0, b=1.0, c=5.1757, d=1.0, e=2.6667 | 0.8833 |
| LorYZ11 | $ax - b\sin(z^c) - dxz$ | $xy - ez$ | a=28.0, b=1.0, c=2, d=1.0, e=2.6667 | 0.9605 |
| LorYZ12 | $ax - b\sin(cx) - dxz$ | $xy - ez$ | a=28.0, b=1.0, c=2.6667, d=1.0, e=2.6667 | 0.9951 |
| LorYZ13 | $ax - b\sin(cx - d) - exz$ | $xy - fz$ | a=28.0, b=1.0, c=2.6667, d=10.882, e=1.0, f=2.6667 | 0.9987 |
| LorYZ14 | $ax - b\sin(cxz) - dxz$ | $xy - ez$ | a=28.0, b=1.0, c=2.6667, d=1.0, e=2.6667 | 0.8842 |
| LorYZ15 | $ax - b\sin(z) - cxz$ | $xy - d$ | a=28.0, b=1.0, c=1.0, d=74.667 | 1.6428 |
| LorYZ16 | $ax - b\sin(cx) - dxz$ | $xy - e$ | a=28.0, b=1.0, c=28.0, d=1.0, e=74.667 | 1.5952 |
| LorYZ17 | $ax + \sin(bx) - cxz$ | $xy - dz$ | a=28.0, b=4.7585, c=1.0, d=2.6667 | 0.9861 |
| LorYZ18 | $-ax(z + \sin(\sin(z)) - b)$ | $cx^d - ez$ | a=1.0, b=28.0, c=28.0, d=2, e=2.6667 | 1.1573 |
| LorYZ19 | $-a\sin(y+z)(z - b\sin(y) + c)$ | $-d\sin(x)$ | a=1.0, b=1.0, c=7.6203, d=34.054 | 0.4993 |
| LorYZ20 | $ax - b\sin(xy) - cxz$ | $xy - dz$ | a=28.0, b=1.0, c=1.0, d=2.6667 | 0.9760 |
| LorYZ21 | $ax - b\sin(y - cz + d) - exz$ | $xy - fz$ | a=28.0, b=1.0, c=1.0, d=2.6667, e=1.0, f=2.6667 | 0.9983 |
| LorYZ22 | $-axz^b + cx - d\sin(x)$ | $ex^f - gz$ | a=1.0, b=2, c=28.0, d=1.0, e=28.0, f=2, g=2.6667 | 2.4512 |
| LorYZ23 | $-axz^b + cx - d\sin(x)$ | $exy - fz$ | a=1.0, b=2, c=28.0, d=1.0, e=28.0, f=2.6667 | 1.0945 |
| LorYZ24 | $-axz^b + cx - d\sin(y)$ | $ex^f - gz$ | a=1.0, b=2, c=28.0, d=1.0, e=28.0, f=2, g=2.6667 | 2.3313 |
| LorYZ25 | $ax - b\sin(x) - cxz$ | $dx^e - fz$ | a=28.0, b=1.0, c=2.6667, d=28.0, e=2, f=2.6667 | 1.1666 |
| LorYZ26 | $\sin(y) - ay - bxz$ | $xy - c$ | a=1.0, b=1.0, c=74.667 | 1.4151 |
| LorYZ27 | $\sin(y) - ay - bxz$ | $y^c - d$ | a=1.0, b=1.0, c=2, d=74.667 | 1.0990 |
| LorYZ28 | $x\sin(y) - axz - by$ | $xy - c$ | a=1.0, b=1.0, c=74.667 | 1.4381 |
| LorYZ29 | $ax - b\sin(cx^d) - exz$ | $xy - fz$ | a=28.0, b=1.0, c=2.6667, d=2, e=1.0, f=2.6667 | 0.9423 |
| LorYZ30 | $ax - b\sin(cy) - dxz$ | $xy - ez$ | a=28.0, b=1.0, c=2.6667, d=1.0, e=2.6667 | 0.9579 |
| LorYZ31 | $\sin(z) - ay - bxz$ | $cxy - d$ | a=2.0, b=1.0, c=2.6667, d=74.667 | 1.1891 |
| LorYZ32 | $ax - b\sin(cz) - d\sin(x) - exz$ | $xy - fz$ | a=28.0, b=1.0, c=2.6667, d=4.6373, e=1.0, f=2.6667 | 0.9667 |
| LorYZ33 | $ax + \sin(y+b) - c\sin(dx) - exz$ | $xy - fz$ | a=28.0, b=8.0335, c=1.0, d=28.0, e=1.0, f=2.6667 | 0.9360 |
| LorYZ34 | $ax + \sin(y+b) - c\sin(dz) - exz$ | $x^f - gz$ | a=28.0, b=8.0335, c=1.0, d=28.0, e=1.0, f=2, g=2.6667 | 1.2302 |
| LorYZ35 | $ax + \sin(bz) - c\sin(dx) - exz$ | $xy - fz$ | a=28.0, b=2.6667, c=1.0, d=28.0, e=1.0, f=2.6667 | 0.9479 |
| LorYZ36 | $ax + \sin(y+b) - c\sin(y) - dxz$ | $xy - ez$ | a=28.0, b=8.0335, c=1.0, d=1.0, e=2.6667 | 0.9827 |
| LorYZ37 | $ax + \sin(y+b) - cxz + d$ | $xy - ez$ | a=28.0, b=8.0335, c=1.0, d=0.7661, e=2.6667 | 0.9869 |
| LorYZ38 | $ax + \sin(x+y) - bxz + c$ | $xy - dz$ | a=28.0, b=1.0, c=0.7661, d=2.6667 | 0.9972 |
| LorYZ39 | $ax + \sin(y+b) + \sin(cy) - dxz$ | $xy - ez$ | a=28.0, b=8.0335, c=2.9484, d=1.0, e=2.6667 | 0.9520 |
| LorYZ40 | $ax + \sin(y+b) - cxz - d$ | $xy - ez$ | a=28.0, b=8.0335, c=1.0, d=0.9016, e=2.6667 | 0.9875 |
| LorYZ41 | $ax - b\sin(y+z) - cxz$ | $xy - dz$ | a=28.0, b=1.0, c=1.0, d=2.6667 | 0.9955 |
| LorYZ42 | $ax - b\sin(y+z) - cxz$ | $y(dx + yz) - ez$ | a=28.0, b=1.0, c=1.0, d=28.0, e=2.6667 | 1.0601 |
| LorYZ43 | $ax - b\sin(\sin(x)) - cxz$ | $xy - dz$ | a=28.0, b=1.0, c=1.0, d=2.6667 | 0.9962 |
| LorYZ44 | $ax - b\sin(x + xy) - cxz$ | $xy - dz$ | a=28.0, b=1.0, c=1.0, d=2.6667 | 0.9671 |
| LorYZ45 | $ax - b\sin(x+y) - cxz$ | $y(dx + yz) - ez$ | a=28.0, b=1.0, c=1.0, d=28.0, e=2.6667 | 0.9556 |
| LorYZ46 | $ax - b\sin(x+c) - dxz$ | $exy - fz$ | a=28.0, b=1.0, c=28.0, d=1.0, e=56.0, f=2.6667 | 1.0033 |
| LorYZ47 | $ax - b\sin(xz\sin(z)) - cxz^d$ | $xy - ez$ | a=28.0, b=1.0, c=1.0, d=2, e=2.6667 | 1.1070 |
| LorYZ48 | $ax - b\sin(z^c\sin(z)) - dxz^e$ | $xy - fz$ | a=28.0, b=1.0, c=2, d=1.0, e=2, f=2.6667 | 1.0440 |
| LorYZ49 | $ax - b\sin(z) - cxz^d$ | $xy - ez$ | a=29.0, b=1.0, c=1.0, d=2, e=2.6667 | 1.1915 |
| LorYZ50 | $ax - b\sin(z) - cxz + d$ | $xy - ez$ | a=28.0, b=1.0, c=1.0, d=2.174, e=2.6667 | 0.9979 |
| LorYZ51 | $ax + \sin(xy) - bxz - c$ | $xy - dz$ | a=28.0, b=1.0, c=0.065552, d=2.6667 | 0.9698 |
| LorYZ52 | $\sin(x - ay) - by - cxz$ | $xy - d$ | a=1.0, b=1.0, c=1.0, d=74.667 | 1.2824 |
| LorYZ53 | $ax - b\sin(y) - cxz - d$ | $xy - ez$ | a=28.0, b=1.0, c=1.0, d=8.396, e=2.6667 | 0.9629 |
| LorYZ54 | $ax + \sin(x) - bxz$ | $xy - cz$ | a=27.0, b=1.0, c=2.6667 | 0.9779 |
| LorYZ55 | $ax - b\sin(y) - cxz$ | $xy - dz$ | a=28.0, b=1.0, c=1.0, d=1.8597 | 0.8615 |
| LorYZ56 | $-ax(z - b)$ | $x(x + \sin(y)) - cz$ | a=1.0, b=28.0, c=2.6667 | 1.2818 |
| LorYZ57 | $-ax(z - b)$ | $x(cx + \sin(y)) - dz$ | a=1.0, b=28.0, c=28.0, d=2.6667 | 1.3103 |

| LorYZ58 | $-ax(z-b)$ | $x(y+\sin(y))-cz$ | a=1.0, b=28.0, c=2.6667 | 0.9897 |

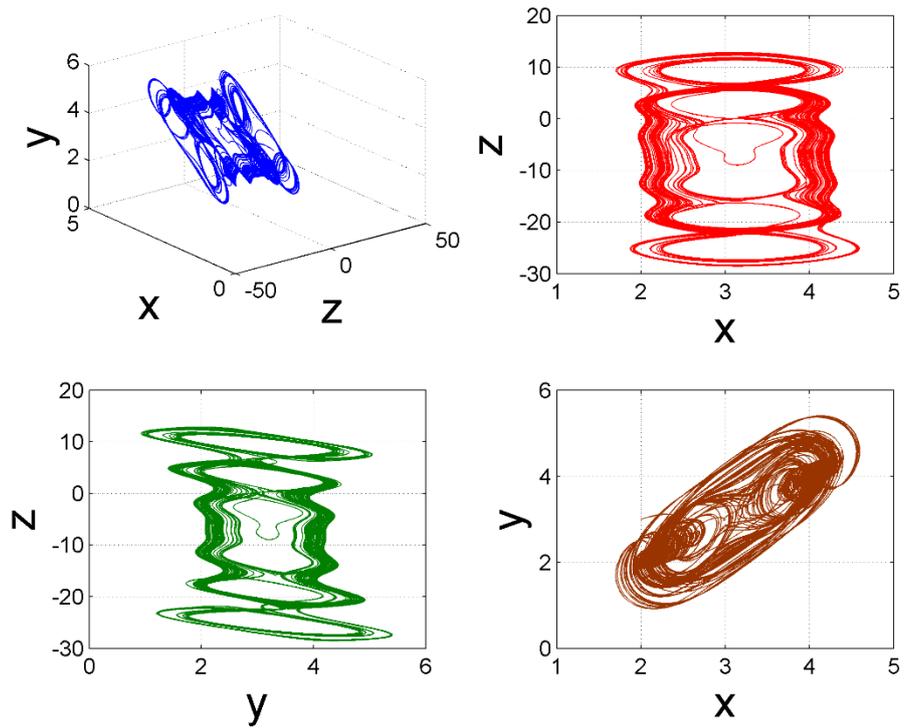

Figure 8: Phase space dynamics of LorYZ19 (*gaming console*)

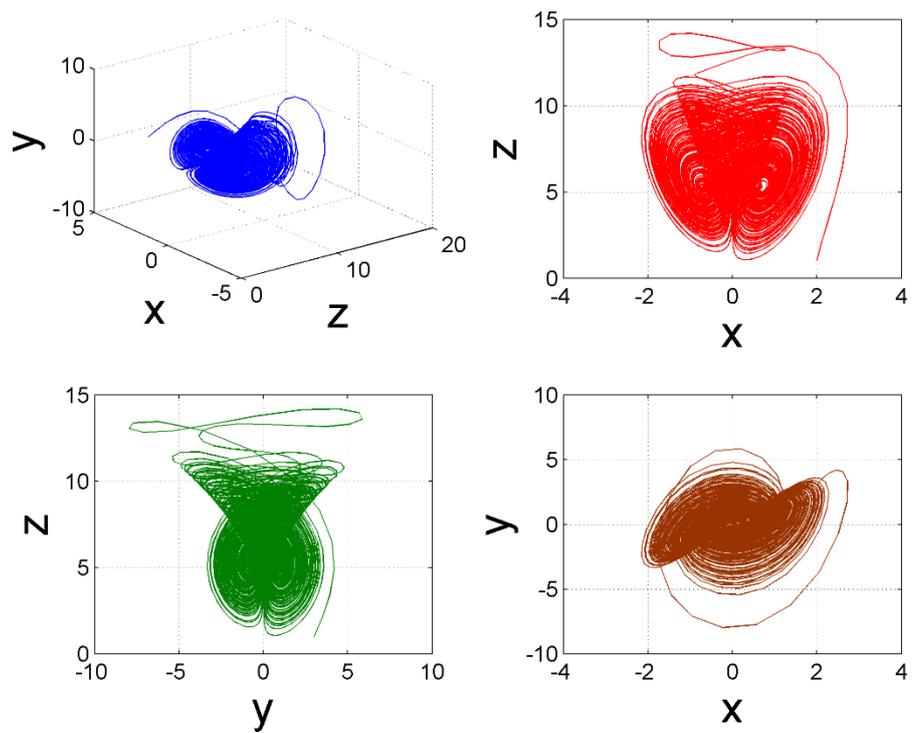

Figure 9: Phase space dynamics of LorYZ22 (*honey pot*)

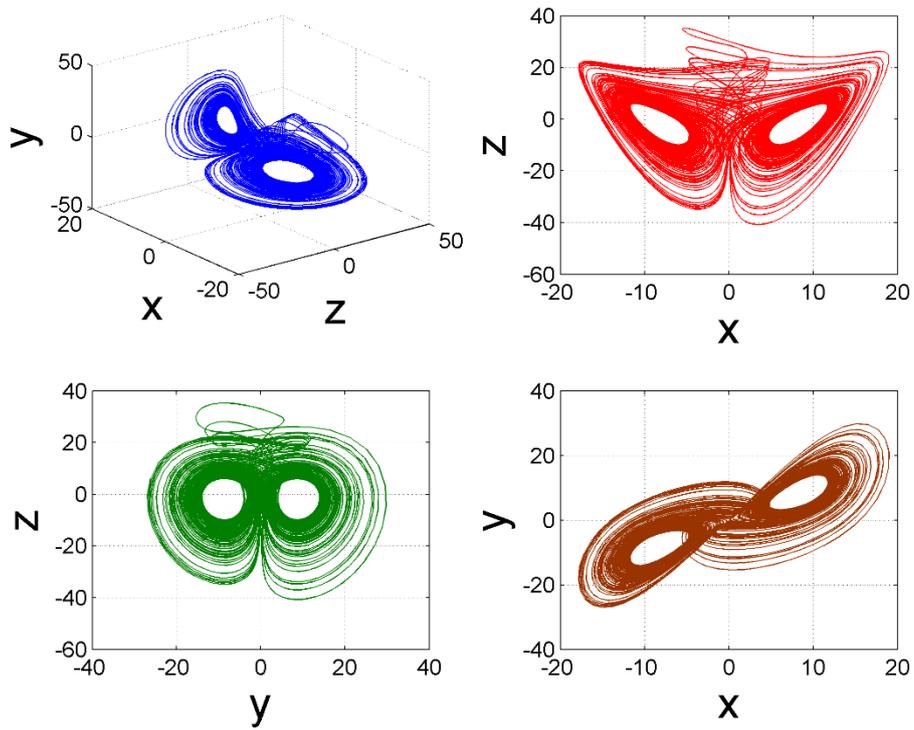

Figure 10: Phase space dynamics of LorYZ27 (*Batman's glasses*)

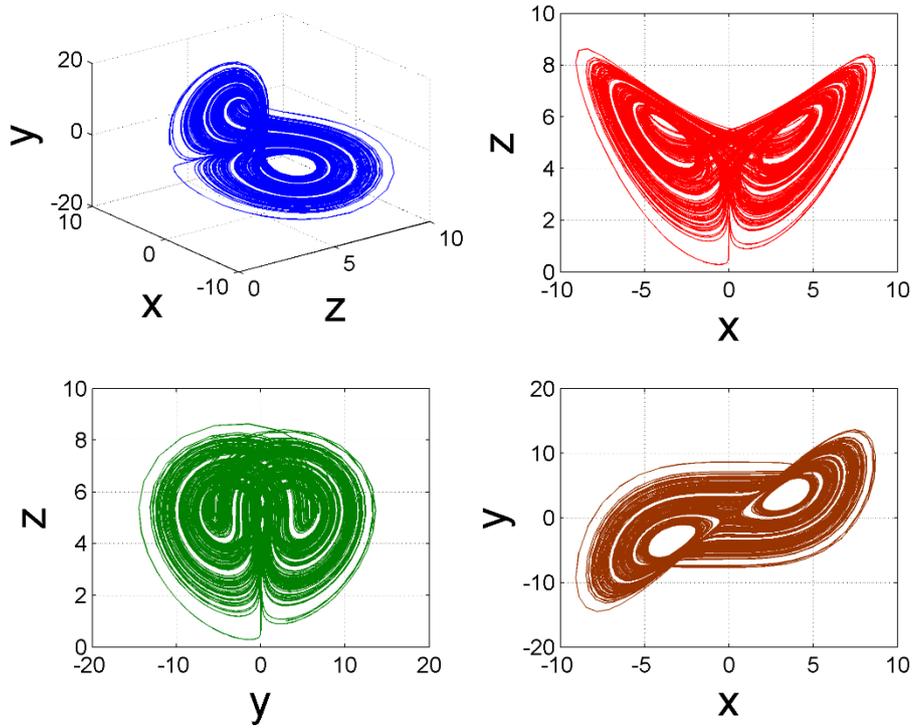

Figure 11: Phase space dynamics of LorYZ47 (*snake's hood*)

In the LorYZ family, the obtained LLE could be as low as 0.49 (LorYZ19) and as high as 2.45 (LorYZ22). In Figure 8 for the LorYZ19 system, the phase space dynamics takes a shape similar to a '*gaming console*'. It has an LLE of 0.49 which is a significant deviation from the Lorenz system. In Figure 9 for the LorYZ22, the LLE reaches 2.45 which is the maximum amongst all the attractors found through GP. Also, the phase space dynamics of the strong chaotic system takes a shape similar to a '*honey pot*' (in the *y-z* plane). Figure 10 for the LorYZ27 looks like '*Batman's glasses*' (in *x-z* plane) which has a LLE slightly higher than one. Also, a '*snake's hood*' like phase space diagram in the *y-z* plane has been observed in Figure 11, for the LorYZ47 system.

Table 5: List of generalized Lorenz *y-z* family of attractors

| Name | $\dot{x}$ expression | $\dot{y}$ expression | $\dot{z}$ expression | Coefficient values | LLE |
|---|---|---|---|---|---|
| LorXYZ1 | $ay - bx$ | $cx - d\sin(yz) - e\sin(y+z) - fxz$ | $xy - gz$ | a=10.0, b=10.0, c=28.0, d=1.0, e=1.0, f=1.0, g=2.6667 | 0.0177 |
| LorXYZ2 | $ay - bx$ | $\sin(y) - cy - dxz$ | $xy - e$ | a=10.0, b=10.0, c=1.0, d=1.0, e=74.667 | 1.4151 |
| LorXYZ3 | $ay - bx$ | $cx + \sin(z) - dxz + e$ | $xy - fz$ | a=10.0, b=10.0, c=29.0, d=1.0, e=10.434, f=2.6667 | 0.9821 |
| LorXYZ4 | $ay - bx$ | $cx + \sin(z) - dxz + e$ | $x^f - gz$ | a=10.0, b=10.0, c=29.0, d=1.0, e=10.434, f=2, g=2.6667 | 1.3045 |
| LorXYZ5 | $ay - bx$ | $cx + \sin(y) - dxz - e$ | $xy - fz$ | a=10.0, b=10.0, c=28.0, d=1.0, e=7.5891, f=2.6667 | 0.9625 |
| LorXYZ6 | $ay - bx$ | $cx + \sin(y) - dxz - e$ | $x^f - gz$ | a=10.0, b=10.0, c=28.0, d=1.0, e=7.5891, f=2, g=2.6667 | 0.1414 |
| LorXYZ7 | $ay - bx$ | $cx - d\sin(y+e) - fxz$ | $xy - gz$ | a=10.0, b=10.0, c=28.0, d=1.0, e=28.0, f=1.0, g=2.6667 | 0.9749 |
| LorXYZ8 | $ay - bx$ | $cx - d\sin(ey+f) - gxz$ | $xy - hz$ | a=10.0, b=10.0, c=28.0, d=1.0, e=10.0, f=28.0, g=1.0, h=2.6667 | 0.8807 |
| LorXYZ9 | $ay - bx$ | $cx - d\sin(y) - exz$ | $xy - fz$ | a=10.0, b=10.0, c=29.0, d=1.0, e=1.0, f=2.6667 | 1.0127 |
| LorXYZ10 | $ay - bx$ | $cx - d\sin(x) - exz$ | $fx - gz + xy$ | a=10.0, b=10.0, c=28.0, d=1.0, e=1.0, f=2.6667, g=2.6667 | 0.9462 |
| LorXYZ11 | $ay - bx$ | $cx - d\sin(z-e) - fxz$ | $x(x+g) - hz$ | a=10.0, b=10.0, c=28.0, d=1.0, e=2.6674, f=1.0, g=1.2624, h=2.6667 | 1.2253 |
| LorXYZ12 | $ay - bx$ | $cx - d\sin(x-e) - fxz$ | $x(x+g) - hz$ | a=10.0, b=10.0, c=28.0, d=1.0, e=2.6674, f=1.0, g=1.2624, h=2.6667 | 1.1576 |
| LorXYZ13 | $ay - bx$ | $cx - d\sin(x+y) - exz$ | $x(x+f) - gz$ | a=10.0, b=10.0, c=28.0, d=1.0, e=1.0, | 1.2002 |
| | | | | f=1.2624, g=2.6667 | |
| LorXYZ14 | $ay - bx + c$ | $dx - e\sin(x - f) - gxz$ | $hx^i - jz$ | a=10.0, b=10.0, c=26.674, d=28.0, e=1.0, f=2.6674, g=1.0, h=2.0, i=2, j=2.6667 | 1.2119 |
| LorXYZ15 | $z + \sin(y^a)(z + \sin(y))$ | $z + \sin(xy)(\sin(y) - b)$ | $x(\sin(y) - c)$ | a=2, b=0.82942, c=7.178 | 0.0005 |
| LorXYZ16 | $z + \sin(y^a)(z + \sin(y))$ | $z + \sin(yz)(\sin(y) - b)$ | $x(\sin(y) - c)$ | a=2, b=0.82942, c=7.178 | 0.0170 |
| LorXYZ17 | $ay - bx - c\sin(y)$ | $-dx(z - e)$ | $fxy - gz$ | a=10.0, b=10.0, c=10.0, d=1.0, e=27.0, f=5.6746, g=2.6667 | 0.0482 |
| LorXYZ18 | $ay - bx$ | $cx - d\sin(z - e) - fxz$ | $xy - gz$ | a=10.0, b=10.0, c=28.0, d=1.0, e=4.2886, f=1.0, g=2.6667 | 0.9970 |
| LorXYZ19 | $ay\sin(y + b)$ | $c - dx$ | $ey - fx$ | a=0.76197, b=9.9326, c=1.5855, d=8.466, e=0.99989, f=9.465 | 0.0026 |

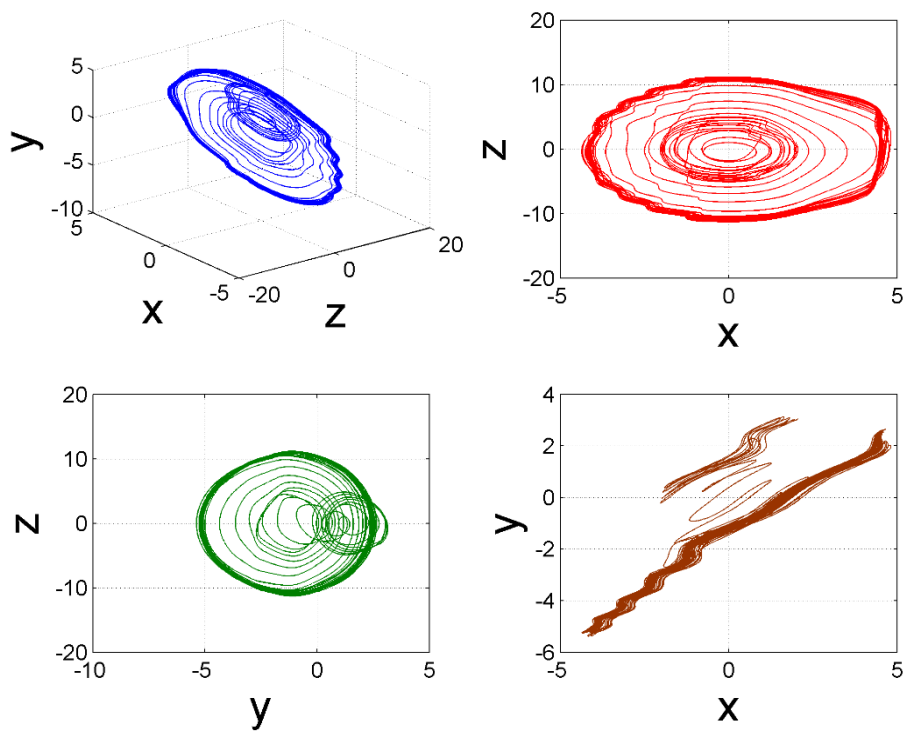

Figure 12: Phase space dynamics of LorXYZ15 (*lasso*).

For the MGGP based evolution of all the three state variables, the chaotic dynamical systems with sine terms are reported in Table 5. It is interesting to note that all of them have a simple straight line state trajectory in the *x-y* plane. For the LorXYZ family also, high LLE values like 1.3 (LorXYZ4) is obtained. Here we report some interesting patterns whose LLEs are relatively less than the Lorenz system. Drawing comparison of the phase space dynamics with physical objects, we can identify several interesting attractors like the '*lasso*' in LorXYZ15, '*gramophone record*' in LorXYZ16,

'*wrinkled butterfly*' in LorXYZ17, '*sieve*' in LorXYZ19 etc. LorXYZ19 has a low LLE of 0.0026 and the evidence of chaos is even weaker in the LorXYZ15 which needs to be considered with caution and further investigation with regards to its dynamical behaviour needs to be done.

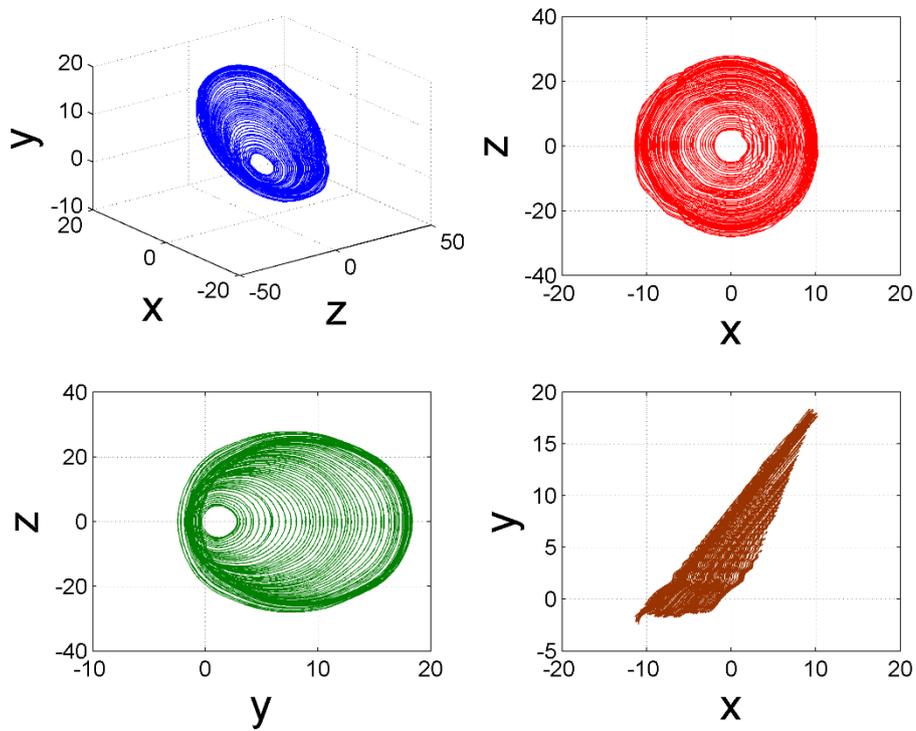

Figure 13: Phase space dynamics of LorXYZ16 (*gramophone record*)

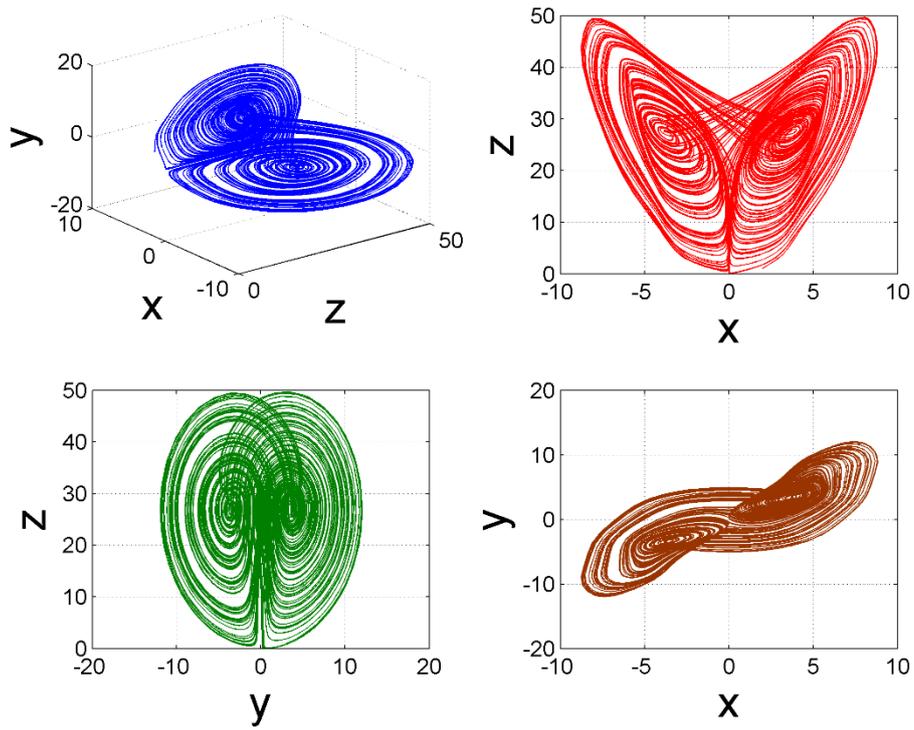

Figure 14: Phase space dynamics of LorXYZ17 (*wrinkled butterfly*)

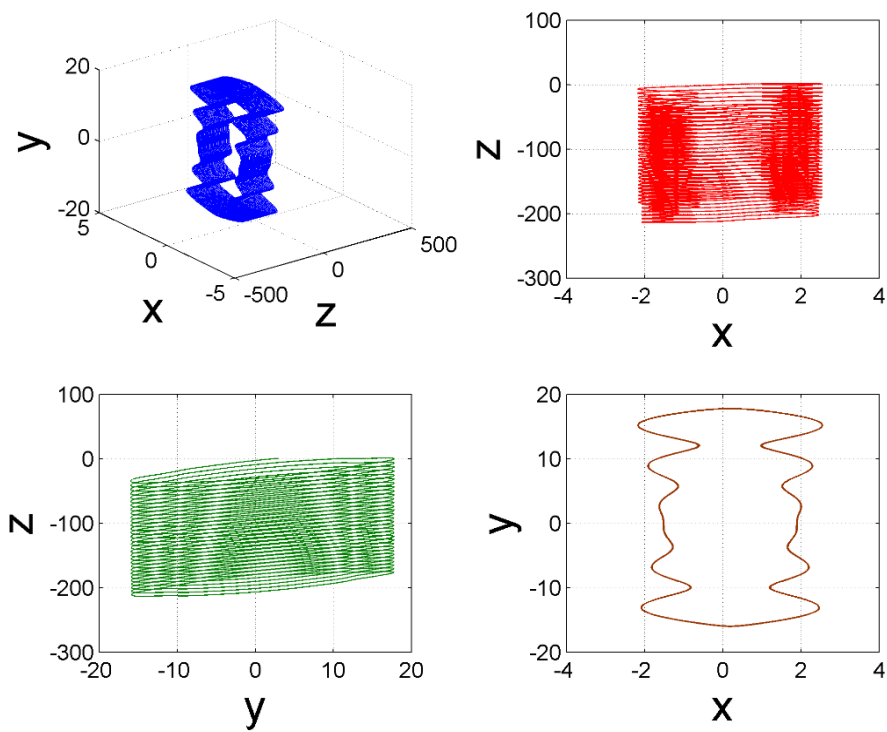

Figure 15: Phase space dynamics of LorXYZ19 (*sieve*)

## 4. Discussions

We ran the GP algorithm for the cases of individual state variables as well, i.e. where the expressions of two of the variables are fixed (e.g. $x$ and $y$ states are fixed) and the expression for the remaining state variable (e.g. $z$) is evolved through GP. However, we obtained very few new chaotic attractors for these cases. Most of the obtained expressions were slight variation from the nominal case of the Lorenz system and therefore we did not report them. One of the reasons for this can be that since the dynamical system is constrained by the other two expressions, the search space is relatively limited and there is not much potential for the GP algorithm to explore and evolve new chaotic structures.

For the case of two and three variables, the search space is very large and the GP was not able to find good candidate solutions which exhibit chaos. Therefore we gave the Lorenz attractor as an initial guess value to the GP algorithm. This ensured that the GP found a good starting solution and in subsequent generations it was able to evolve good chaotic attractors.

The two variable cases seems to be the most conducive for evolving chaotic attractors and we obtained many new chaotic structures for these cases. However for the two variable cases, there is a significant difference in the number of chaotic attractors produced. For example, the $x$-$y$ family has much more attractors than $y$-$z$ or $x$-$z$, even though all are run for 100 times each. This implies that some structures or expressions are more favourable in obtaining chaotic attractors than others. This is an interesting insight and can be leveraged to evolve more chaotic attractors for a particular family by putting them as an initial guess, in the first iteration of the GP algorithm.

Many other interesting limit cycles and oscillators were evolved during the runs which are interesting in their own right, but are not reported here as they do not exhibit chaos. It would be interesting to explore chaotic attractors other than the Lorenz system to get new and richer dynamics. In some cases, the numerically calculated Lyapunov exponent is positive but on screening them manually, by looking at the phase portraits, we found several of them were complicated limit cycles. Therefore, after the machine generated a list of possible new chaotic attractor candidates, a manual screening was necessary to verify whether the expressions were actually chaotic or not. We found that in most cases if the LLE (with the time delay embedding method) was less than 0.1, the evolved dynamical systems were not chaotic. This can be mostly attributed to numerical issues, since the analysed time series was of considerably short duration (25 seconds) and also we are essentially trying to infer the chaotic properties of a three dimensional system from the observation of just the first state variable. The time delay reconstruction method for estimating LLE does not give the actual Lyapunov exponent as the delay embedding is on one of the states and not on all of them. Therefore we used LLE = 0.1 with the delay embedding method within the GP cost function, as a threshold for a first hand screening of the expressions before looking at the phase portraits. We also report here, only those expressions which has at least one sine term in it. We observed that the other expressions which only involved the terms +, -, ×, ÷, without the sine term, were mostly variants of the Lorenz system with almost similar time domain dynamics. After screening with the evolved attractors with LLE > 0.1 in the time delay embedding technique, their true LLE was again computed directly from the structure of the differential equations using symbolic differentiation and are reported in Table 1-4. There are a few research results on the increase in the number of equilibrium points in the phase space due to the sine terms which is commonly known as the '*Labyrinth chaos*' [31], [32]. Similar complex dynamics have also been observed in the phase space due to increase in the number of equilibrium points beside the standard two wing Lorenz like attractor dynamics.

Since the reported chaotic attractors have been obtained by keeping a history list of all those which were evolved during one GP run, some of them have nearly similar structure with slight modification in some of the terms. This is due to the fact that once the GP algorithm finds a good solution with a high value of Lyapunov exponent, it gives more importance to this structure and tries to find solutions in the vicinity of the obtained one. Also, we used only a limited set of mathematical functions for evolving the chaotic attractors, i.e. only +, -, ×, ÷, sine. It is possible to use other mathematical functions like cosine, tangent, hyperbolic sine, hyperbolic tangent, signum function, exponential, logarithm etc. to obtain a diverse variety of chaotic attractors exhibiting rich dynamics. This might result in more complex expressions and can be pursued as scope of future work.

Since our proposed methodology mostly relies on a numerical technique, one of its drawbacks is that it reports only a limited set of parameters for a chaotic structure. To make each structure truly a generalised one, each of the individual parameter sets needs to be varied and detailed bifurcation analysis needs to be done to obtain the properties of each dynamical system. This would give detailed results like the range of parameters that results in chaotic behaviour for each system etc. Though this kind of analysis is important and interesting in its own right, it is a separate stream of work and is therefore not included in the present paper.

## 5. Conclusions

We proposed a GP based intelligent technique to evolve new chaotic dynamical systems based on the original Lorenz system of equations. We reported over one hundred new chaotic attractors to validate our proposed methodology. The GP algorithms employed a strategy of maximising the largest Lyapunov exponent to evolve these new attractors. The GP runs also simultaneously evolved new types of limit cycles and oscillators but they were not reported as they did not exhibit chaotic dynamics. The study also showed that some of the structures of the attractors (like the x-y family) were favourable in terms of producing a larger number of attractors. For most of the newly found chaotic attractors, we obtained phase space dynamics which are qualitatively similar in nature to that of the original Lorenz attractors. However some of them resulted in completely new kind of dynamics which could not be apprehended before. GP being a numerical algorithm reports only one set of parameters for which the system of equations exhibit chaos. If the numerical parameters for each family of attractors are varied, a different type of chaotic dynamics might be obtained or the chaos might disappear altogether. This needs further investigation.

We believe that such symbiotic confluence of machine intelligence and classical dynamical systems theory, would play a big role in investigating complexity in nature, in the near future. One stream of extension of this work can look at mathematical analysis of such structures to find out properties of the dynamical system, carry out bifurcation analysis etc. Another extension of this work can investigate more new chaotic attractors which are evolved from other established chaotic systems like the unified chaotic systems, chaotic neural networks etc.

## References

[1]  E. N. Lorenz, "Deterministic nonperiodic flow," *Journal of the atmospheric sciences*, vol. 20, no. 2, pp. 130–141, 1963.

[2]  I. Stewart, "Mathematics: The Lorenz attractor exists," *Nature*, vol. 406, no. 6799, pp. 948–949, 2000.

[3]  J. Lü, G. Chen, and D. Cheng, "A new chaotic system and beyond: the generalized Lorenz-like system," *International Journal of Bifurcation and Chaos*, vol. 14, no. 05, pp. 1507–1537, 2004.

[4]  S. Yu, J. Lu, X. Yu, and G. Chen, "Design and implementation of grid multiwing hyperchaotic Lorenz system family via switching control and constructing super-heteroclinic loops," *Circuits and Systems I: Regular Papers, IEEE Transactions on*, vol. 59, no. 5, pp. 1015–1028, 2012.

[5]  J. Lü and G. Chen, "Generating multiscroll chaotic attractors: theories, methods and applications," *International Journal of Bifurcation and Chaos*, vol. 16, no. 04, pp. 775–858, 2006.

[6]  S. Yu, W. K. Tang, J. Lü, and G. Chen, "Generating 2n-wing attractors from Lorenz-like systems," *International Journal of Circuit Theory and Applications*, vol. 38, no. 3, pp. 243–258, 2010.

[7]  A. Abel and W. Schwarz, "Chaos communications-principles, schemes, and system analysis," *Proceedings of the IEEE*, vol. 90, no. 5, pp. 691–710, 2002.

[8]  C. Letellier, *Chaos in nature*, vol. 81. World Scientific, 2013.

[9]  B. Jovic, *Synchronization Techniques for Chaotic Communication Systems*. Springer, 2011.

[10] M. Small, *Applied nonlinear time series analysis: applications in physics, physiology and finance*, vol. 52. World Scientific, 2005.

[11] H. Kantz and T. Schreiber, *Nonlinear time series analysis*, vol. 7. Cambridge university press, 2004.

[12] P. Zhou and F. Yang, "Hyperchaos, chaos, and horseshoe in a 4D nonlinear system with an infinite number of equilibrium points," *Nonlinear Dynamics*, vol. 76, no. 1, pp. 473–480, 2014.

[13] P. Zhou, K. Huang, and C. Yang, "A fractional-order chaotic system with an infinite number of equilibrium points," *Discrete Dynamics in Nature and Society*, vol. 2013, 2013.

[14] K. Rodriguez-Vázquez and P. J. Fleming, "Evolution of mathematical models of chaotic systems based on multiobjective genetic programming," *Knowledge and Information Systems*, vol. 8, no. 2, pp. 235–256, 2005.

[15] K. Rodriguez-Vazquez and P. J. Fleming, "Genetic programming for dynamic chaotic systems modelling," in *Evolutionary Computation, 1999. CEC 99. Proceedings of the 1999 Congress on*, vol. 1, 1999.

[16] I. Zelinka, S. Celikovsky, H. Richter, and G. Chen, *Evolutionary algorithms and chaotic systems*, vol. 267. Springer, 2010.

[17] I. Zelinka, G. Chen, and S. Celikovsky, "Chaos synthesis by means of evolutionary algorithms," *International Journal of Bifurcation and Chaos*, vol. 18, no. 04, pp. 911–942, 2008.

[18] I. Zelinka, D. Davendra, R. Senkerik, R. Jasek, and Z. Oplatkova, "Analytical programming-a novel approach for evolutionary synthesis of symbolic structures," *Evolutionary Algorithms. InTech*, 2011.

[19] I. Zelinka, M. Chadli, D. Davendra, R. Senkerik, and R. Jasek, "An investigation on evolutionary reconstruction of continuous chaotic systems," *Mathematical and Computer Modelling*, vol. 57, no. 1, pp. 2–15, 2013.

[20] W. K. Tang, G. Zhong, G. Chen, and K. Man, "Generation of n-scroll attractors via sine function," *Circuits and Systems I: Fundamental Theory and Applications, IEEE Transactions on*, vol. 48, no. 11, pp. 1369–1372, 2001.

[21] J. R. Koza, F. H. Bennett III, and O. Stiffelman, "Genetic programming as a Darwinian invention machine," in *Genetic Programming*, Springer, 1999, pp. 93–108.

[22] J. R. Koza, *Genetic programming III: Darwinian invention and problem solving*, vol. 3. Morgan Kaufmann, 1999.

[23] J. R. Koza, *Genetic programming: on the programming of computers by means of natural selection*, vol. 1. MIT press, 1992.

[24] D. P. Searson, D. E. Leahy, and M. J. Willis, "GPTIPS: an open source genetic programming toolbox for multigene symbolic regression," *International MultiConference of Engineers and Computer Scientists 2010*, 2010.

[25] I. Pan, D. S. Pandey, and S. Das, "Global solar irradiation prediction using a multi-gene genetic programming approach," *Journal of Renewable and Sustainable Energy*, vol. 5, no. 6, p. 063129, 2013.

[26]   A. Wolf, J. B. Swift, H. L. Swinney, and J. A. Vastano, "Determining Lyapunov exponents from a time series," *Physica D: Nonlinear Phenomena*, vol. 16, no. 3, pp. 285–317, 1985.

[27]   M. T. Rosenstein, J. J. Collins, and C. J. De Luca, "A practical method for calculating largest Lyapunov exponents from small data sets," *Physica D: Nonlinear Phenomena*, vol. 65, no. 1, pp. 117–134, 1993.

[28]   M. Banbrook, G. Ushaw, and S. McLaughlin, "How to extract Lyapunov exponents from short and noisy time series," *Signal Processing, IEEE Transactions on*, vol. 45, no. 5, pp. 1378–1382, 1997.

[29]   K. Ramasubramanian and M. Sriram, "A comparative study of computation of Lyapunov spectra with different algorithms," *Physica D: Nonlinear Phenomena*, vol. 139, no. 1, pp. 72–86, 2000.

[30]   S. Heiligenthal et al., "Strong and weak chaos in nonlinear networks with time-delayed couplings," *Physical review letters*, vol. 107, no. 23, p. 234102, 2011.

[31]   J. C. Sprott, *Elegant chaos: algebraically simple chaotic flows*. World Scientific, 2010.

[32]   J. C. Sprott and K. E. Chlouverakis, "Labyrinth chaos," *International Journal of Bifurcation and Chaos*, vol. 17, no. 06, pp. 2097–2108, 2007.

# Supplementary Material

### a) Phase portraits of generalized Lorenz x-z family of attractors

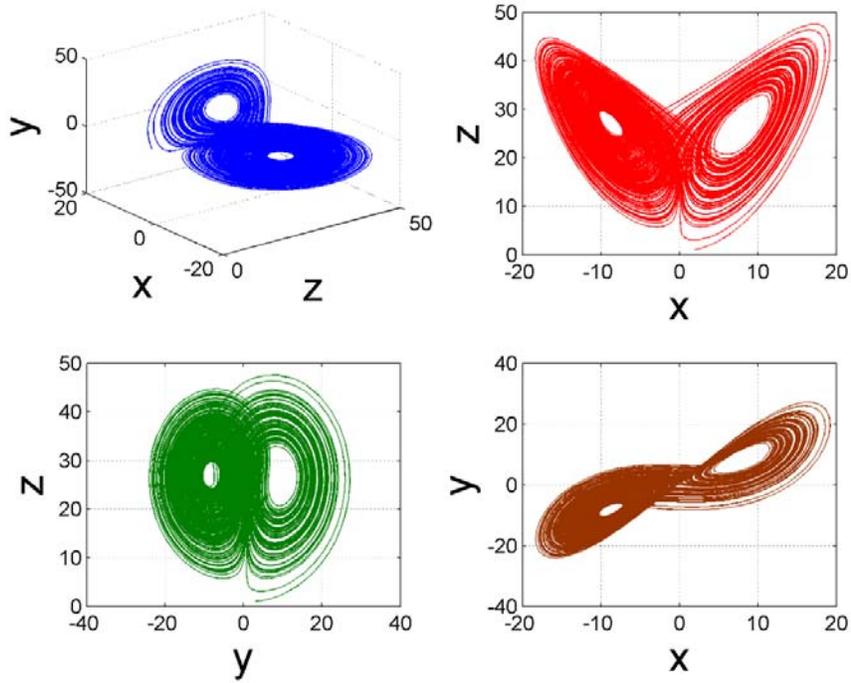

*Figure 1: Phase space dynamics of LorXZ1*

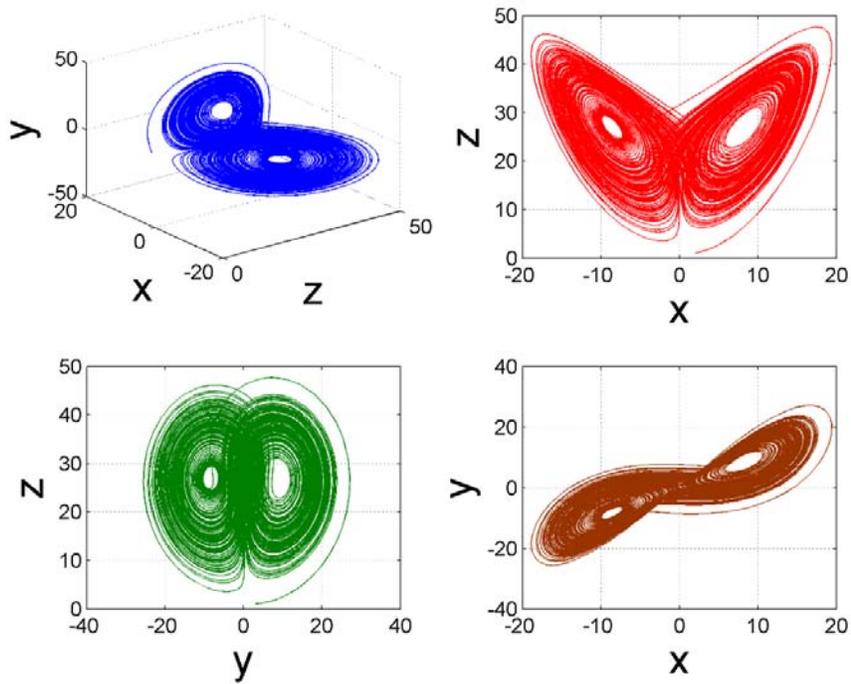

*Figure 2: Phase space dynamics of LorXZ2*

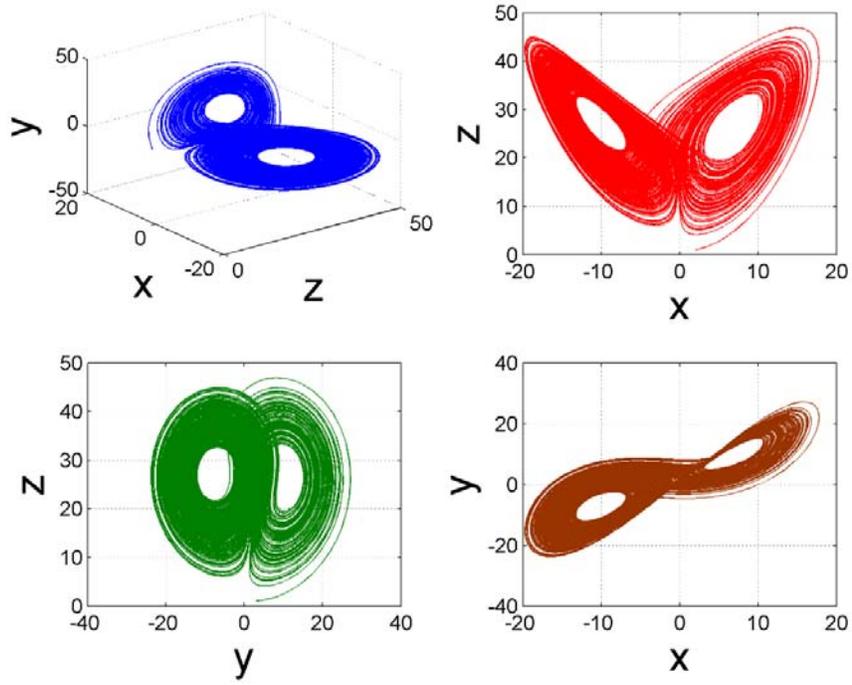

*Figure 3: Phase space dynamics of LorXZ4*

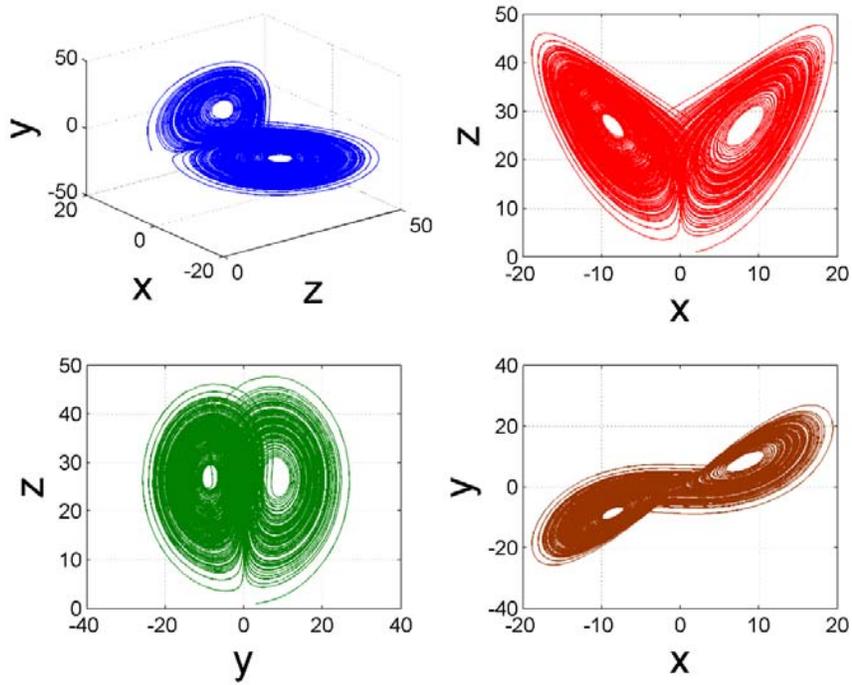

*Figure 4: Phase space dynamics of LorXZ6*

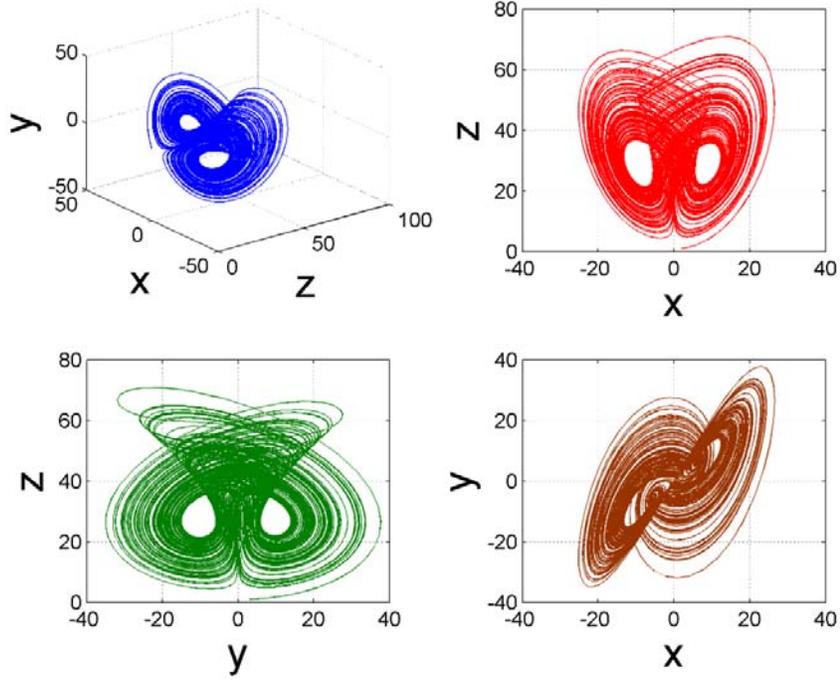

*Figure 5: Phase space dynamics of LorXZ7*

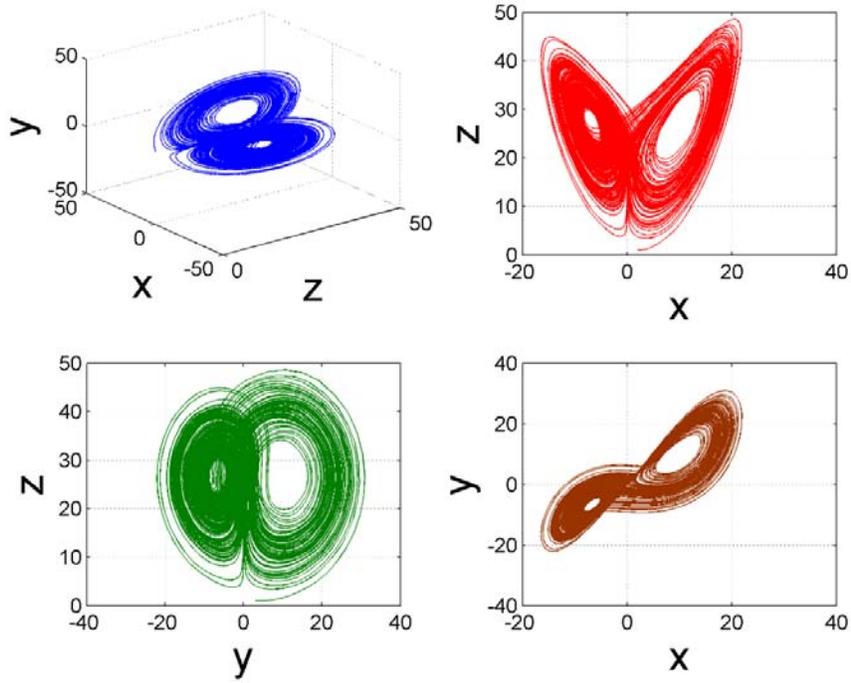

*Figure 6: Phase space dynamics of LorXZ8*

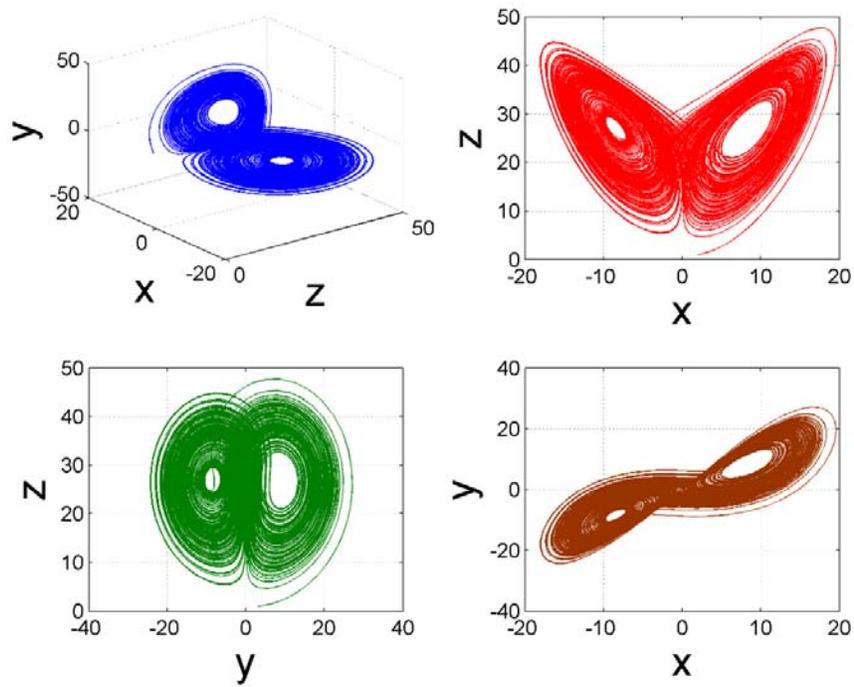

*Figure 7: Phase space dynamics of LorXZ10*

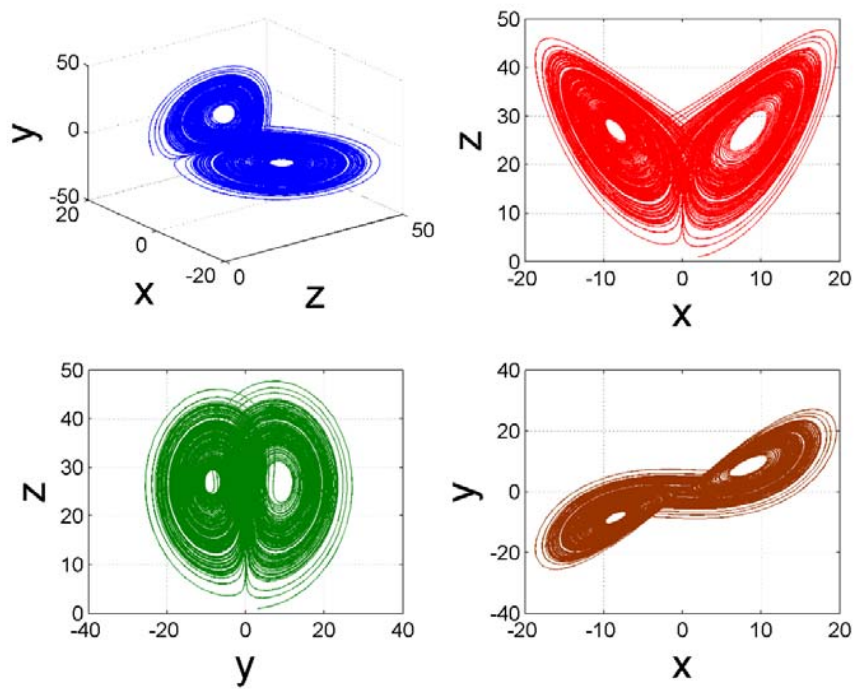

*Figure 8: Phase space dynamics of LorXZ11*

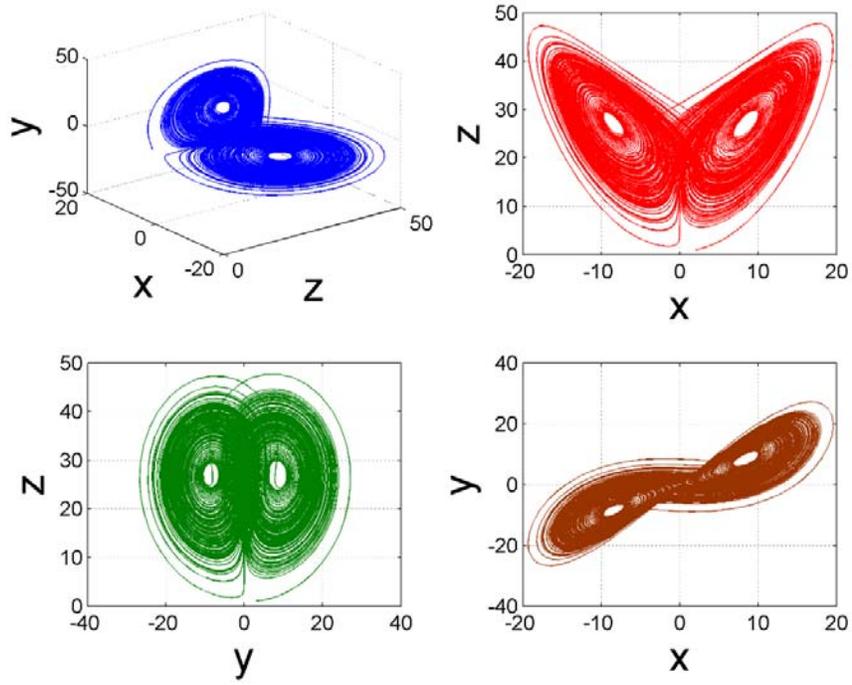

*Figure 9: Phase space dynamics of LorXZ12*

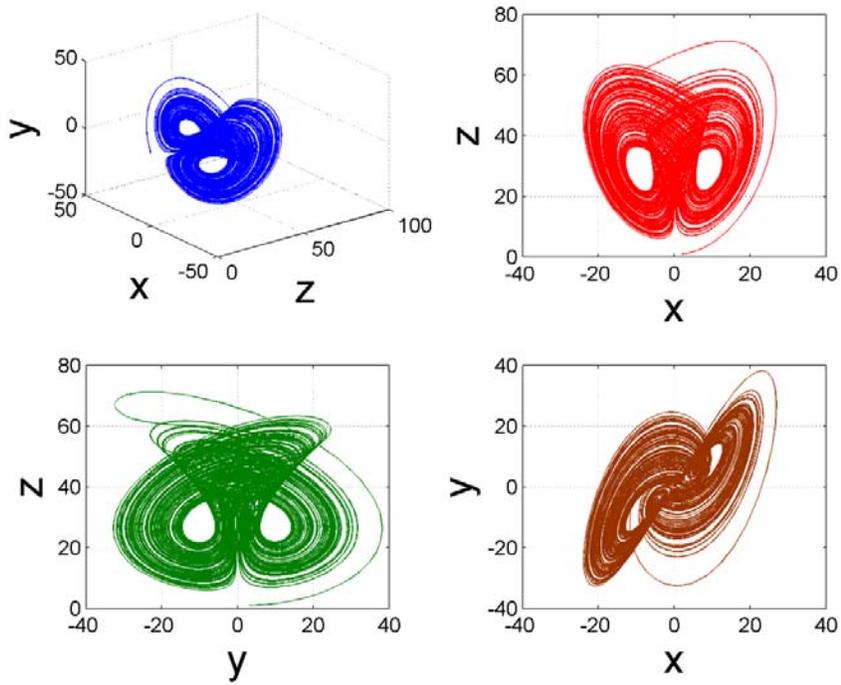

*Figure 10: Phase space dynamics of LorXZ13*

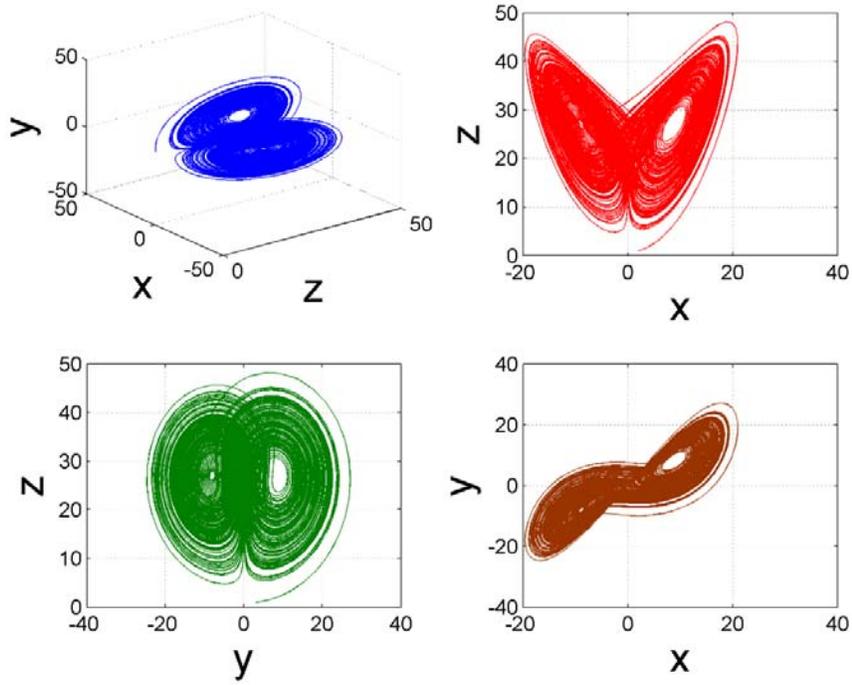

*Figure 11: Phase space dynamics of LorXZ14*

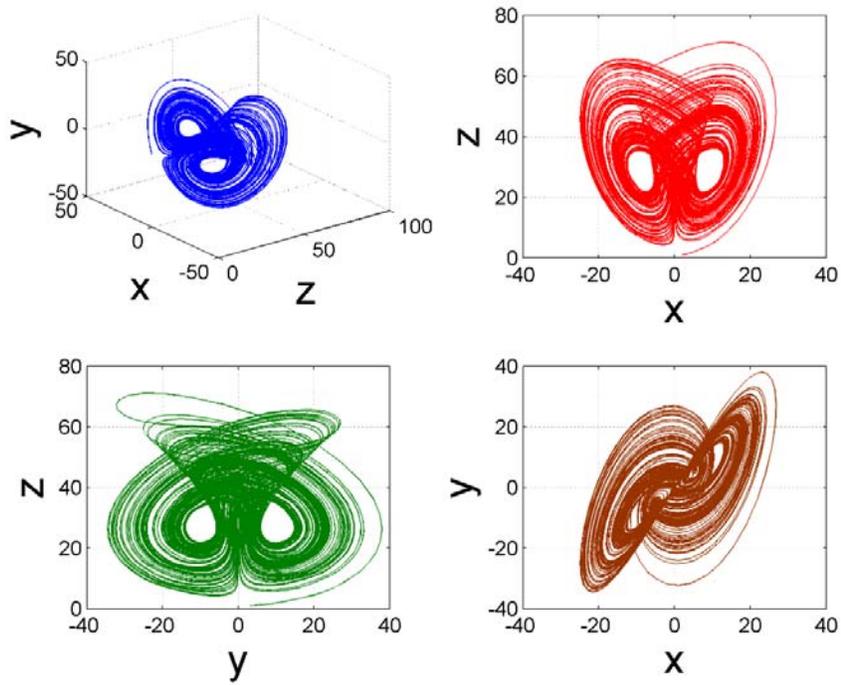

*Figure 12: Phase space dynamics of LorXZ16*

## b) *Phase portraits of generalized Lorenz x-y family of attractors*

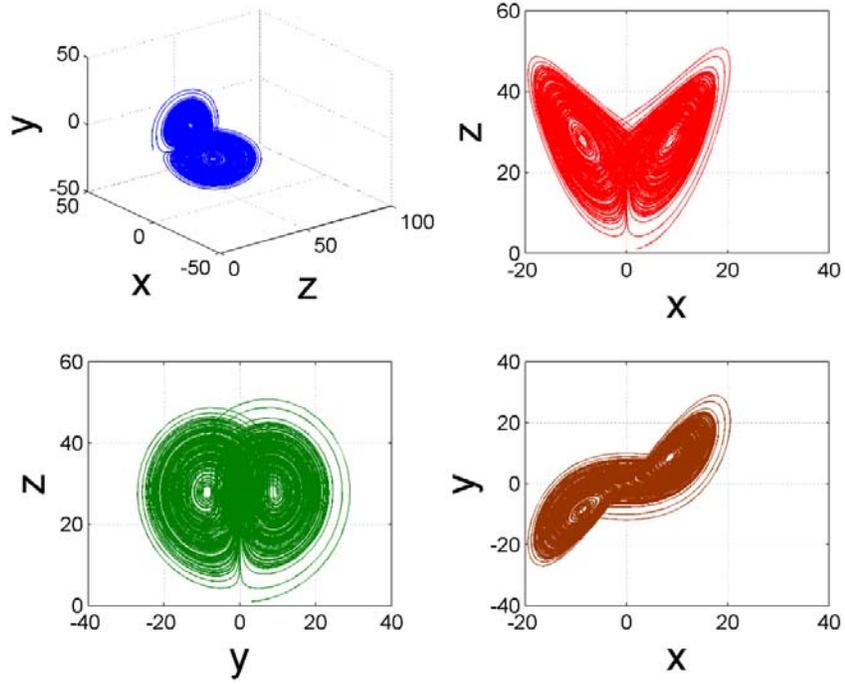

*Figure 13: Phase space dynamics of LorXY1*

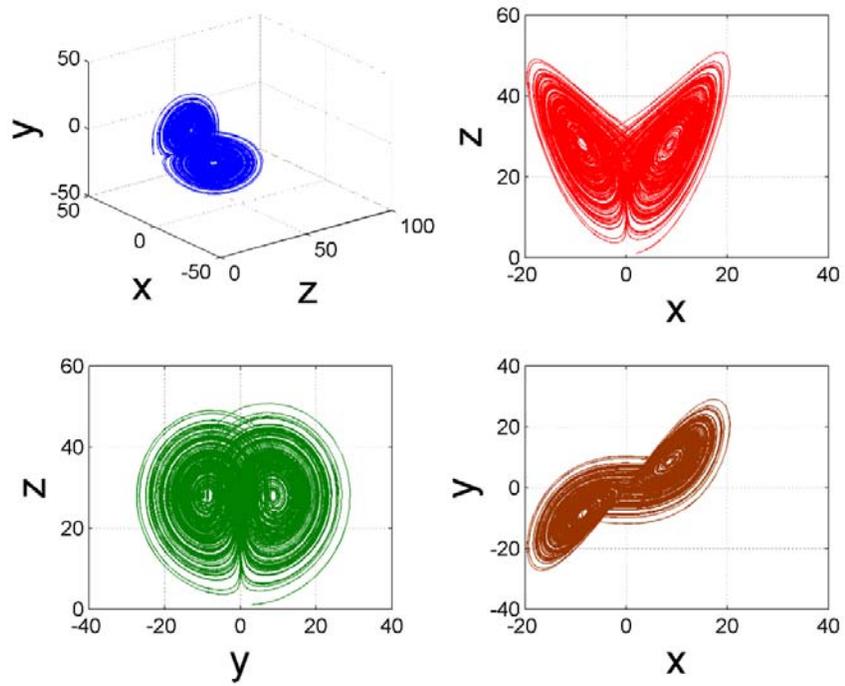

*Figure 14: Phase space dynamics of LorXY2*

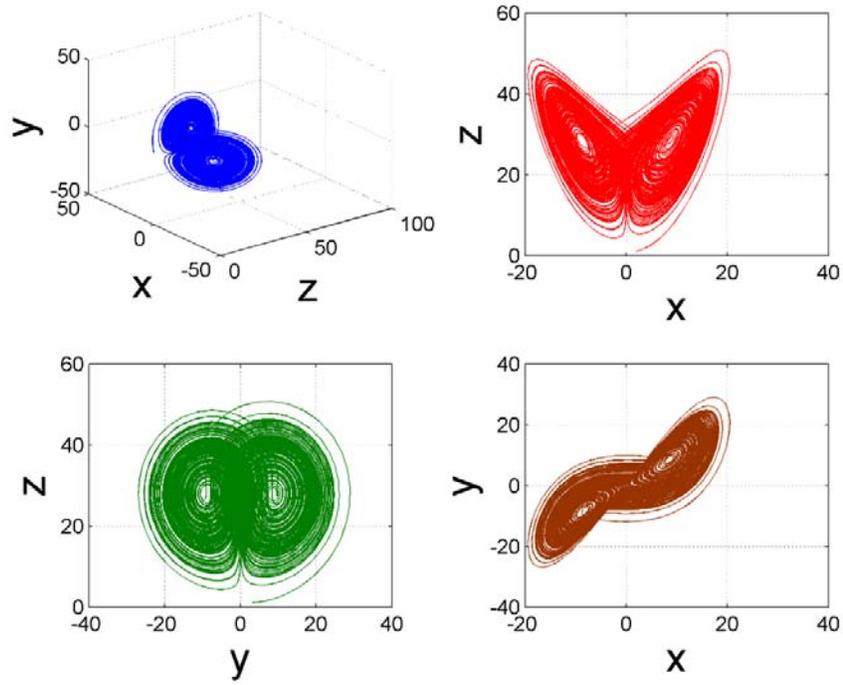

*Figure 15: Phase space dynamics of LorXY3*

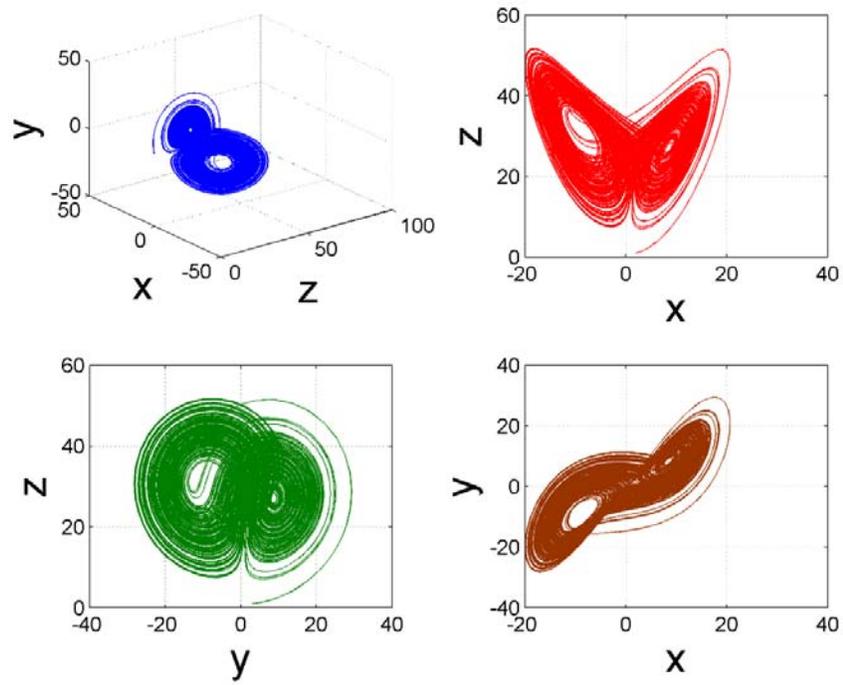

*Figure 16: Phase space dynamics of LorXY4*

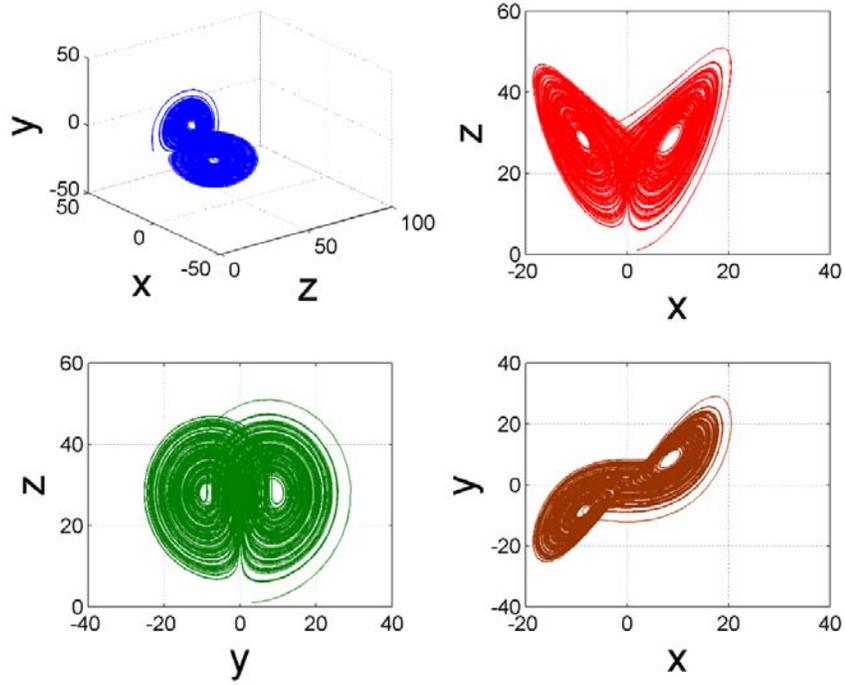

*Figure 17: Phase space dynamics of LorXY5*

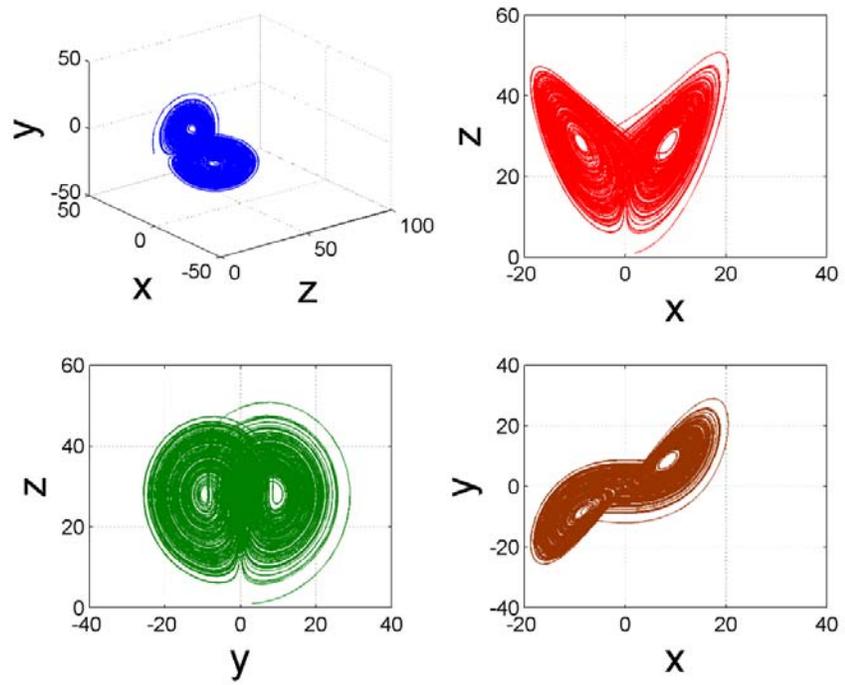

*Figure 18: Phase space dynamics of LorXY6*

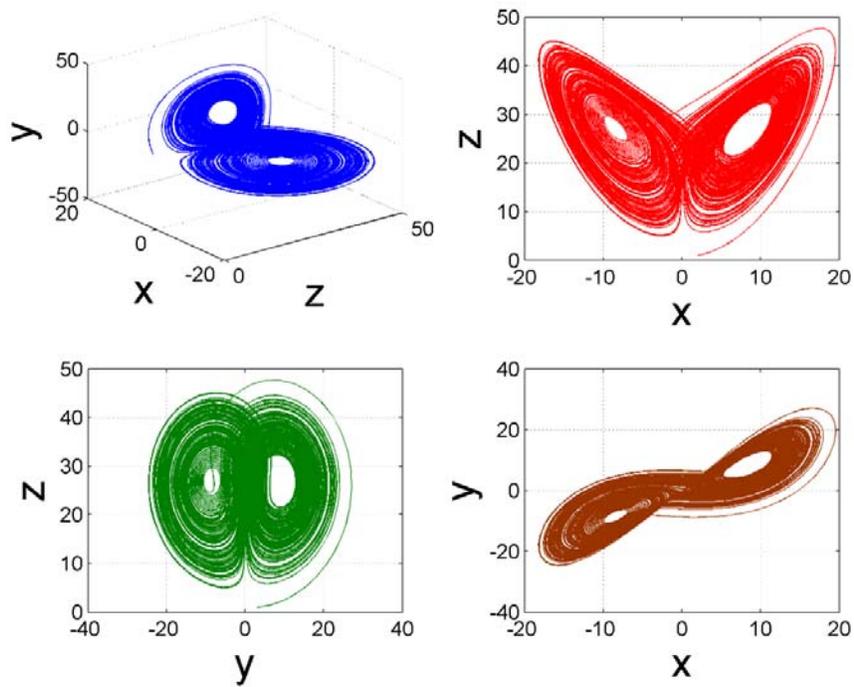

*Figure 19: Phase space dynamics of LorXY7*

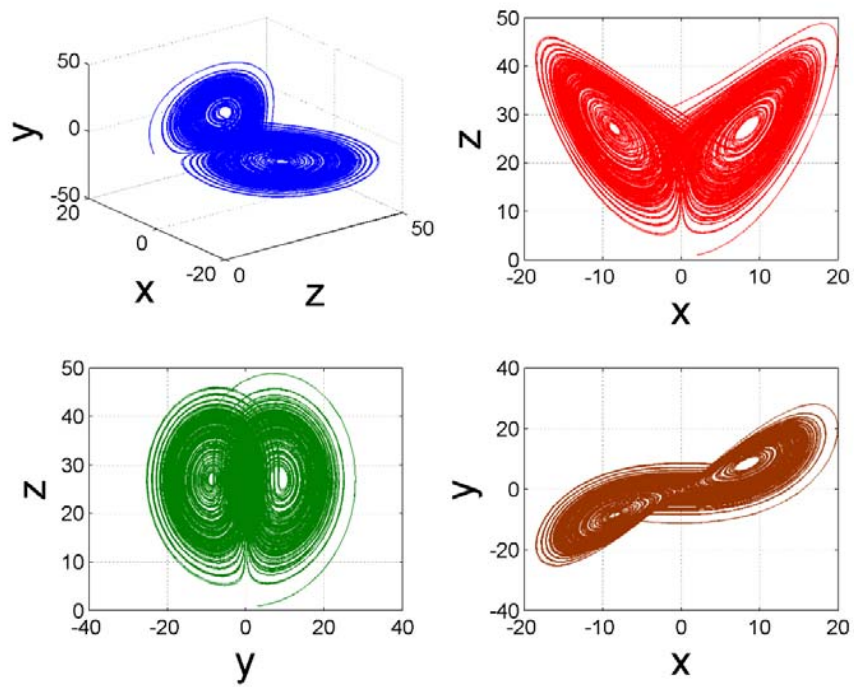

*Figure 20: Phase space dynamics of LorXY8*

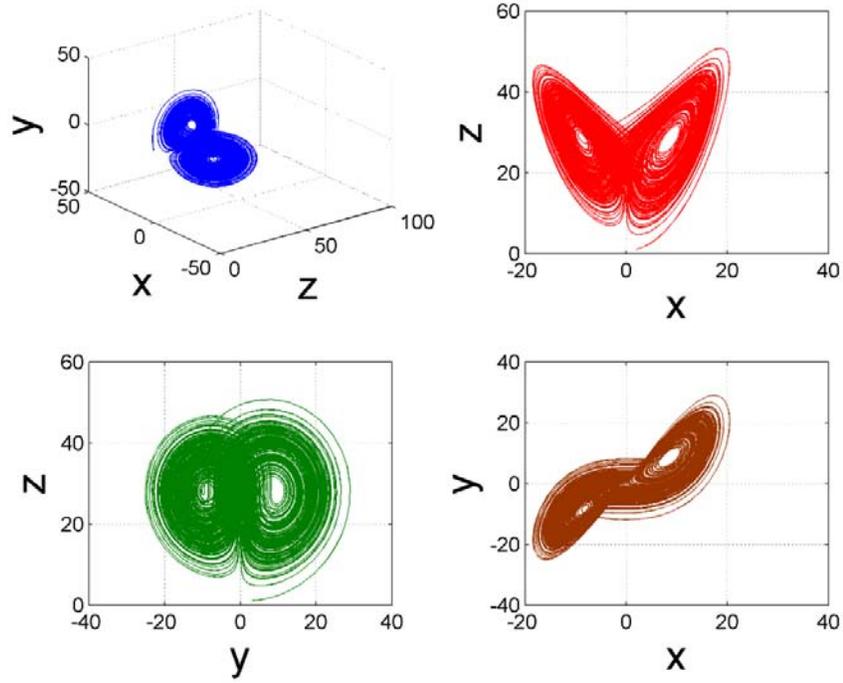

*Figure 21: Phase space dynamics of LorXY9*

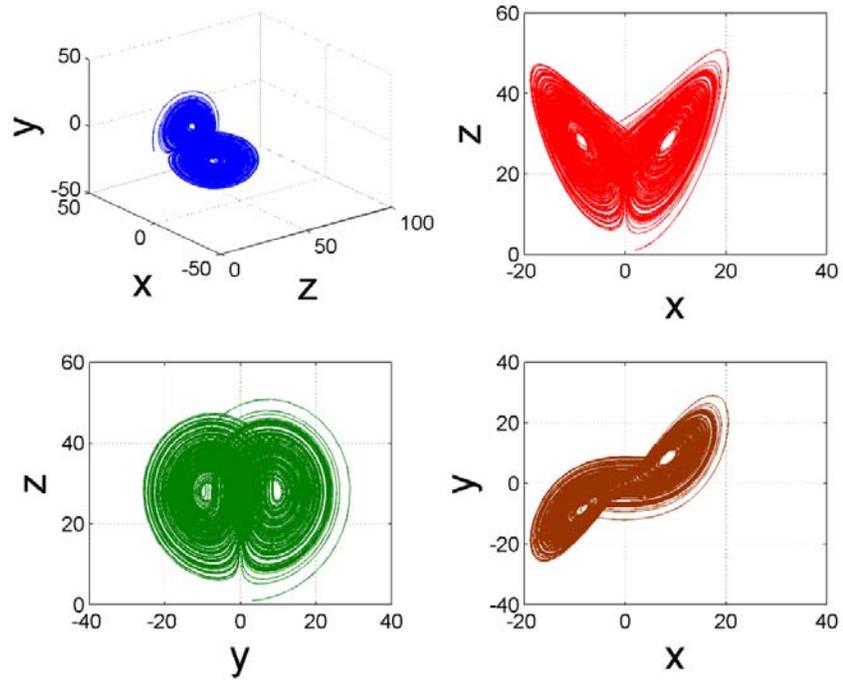

*Figure 22: Phase space dynamics of LorXY11*

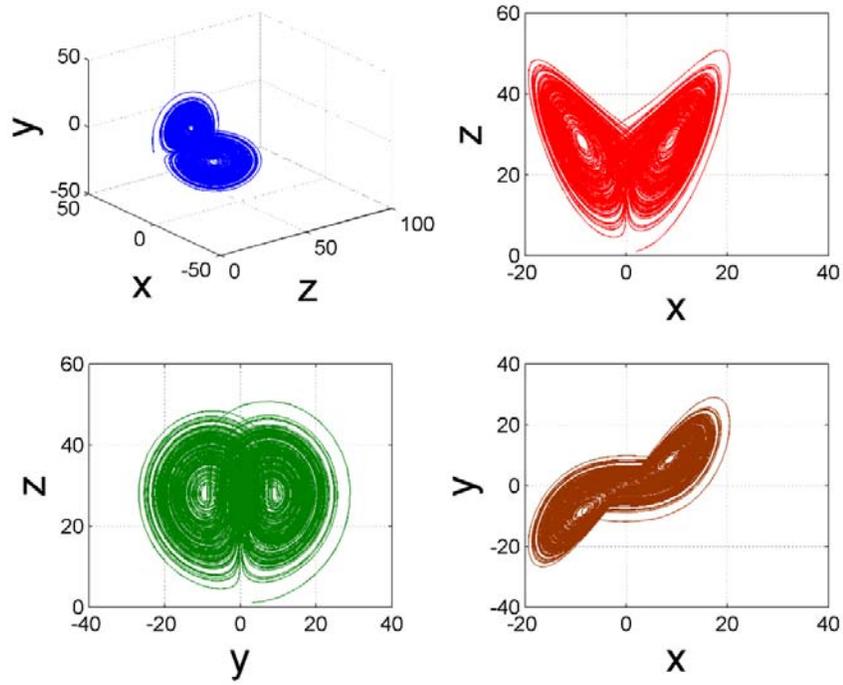

*Figure 23: Phase space dynamics of L₀rXY12*

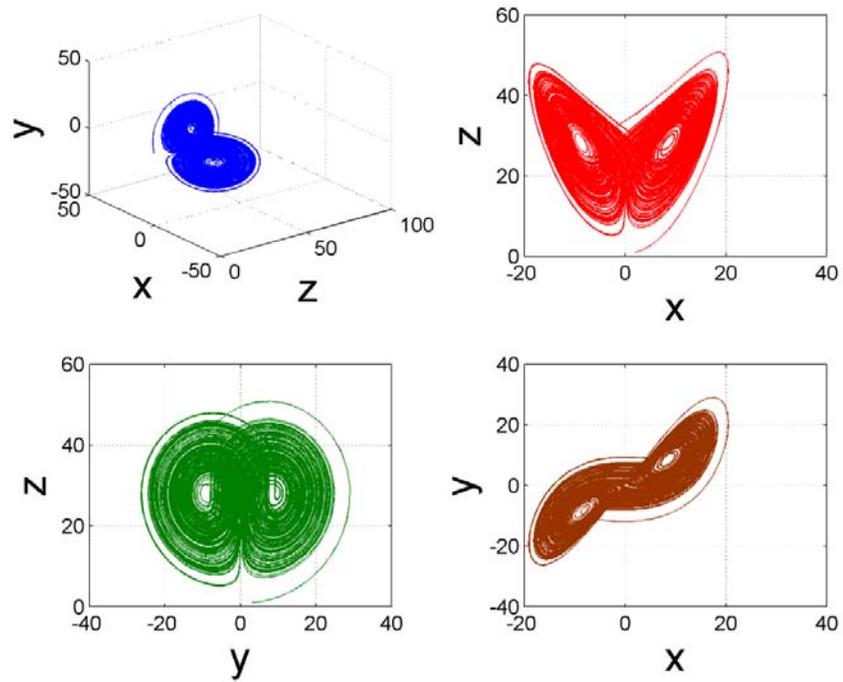

*Figure 24: Phase space dynamics of L₀rXY13*

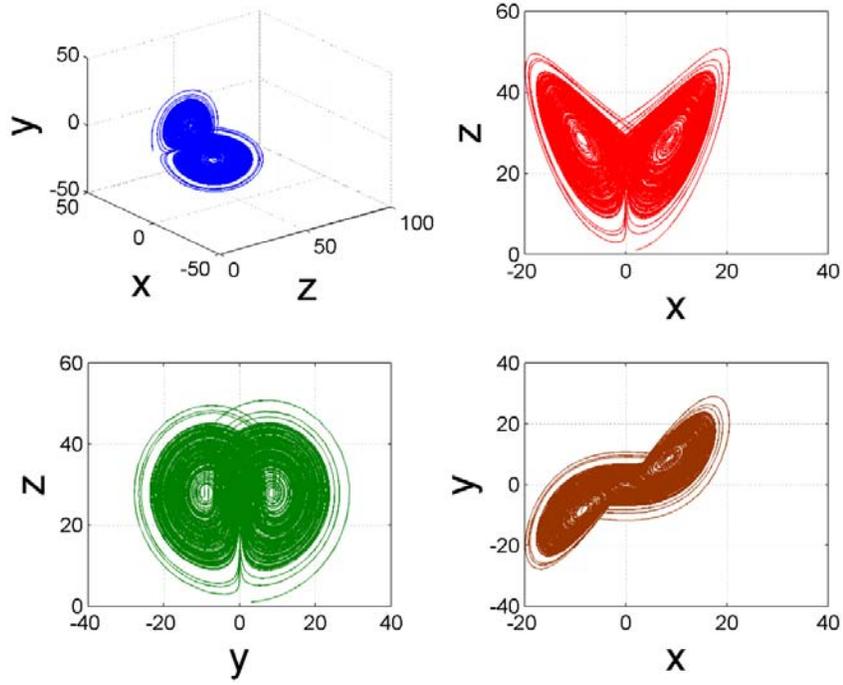

*Figure 25: Phase space dynamics of LorXY14*

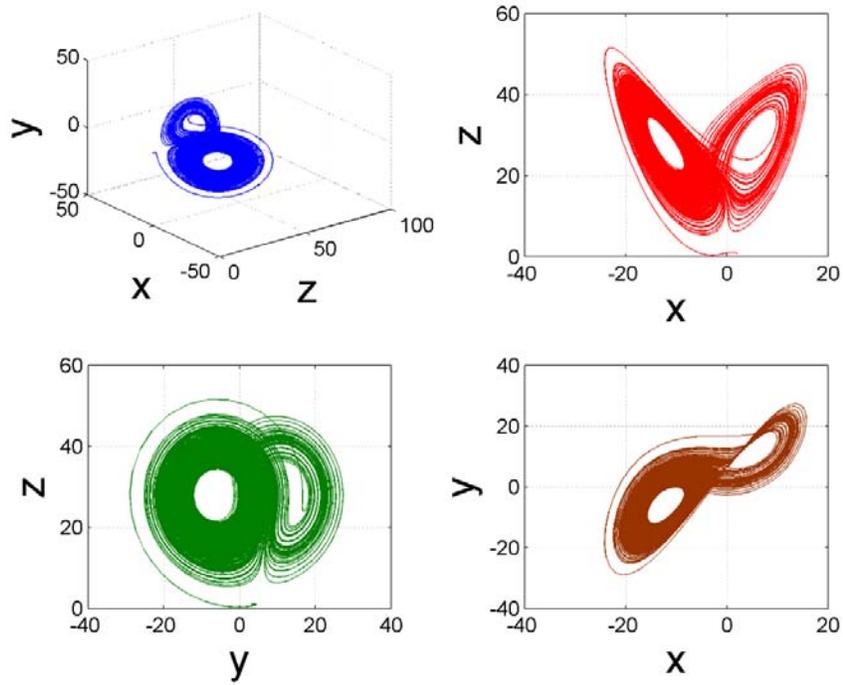

*Figure 26: Phase space dynamics of LorXY15*

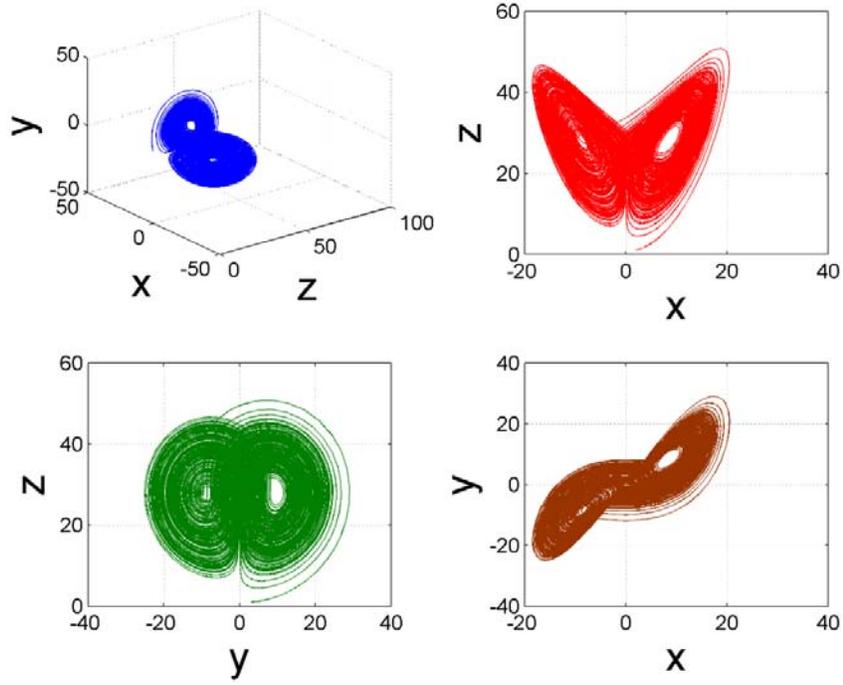

*Figure 27: Phase space dynamics of L$or$XY16*

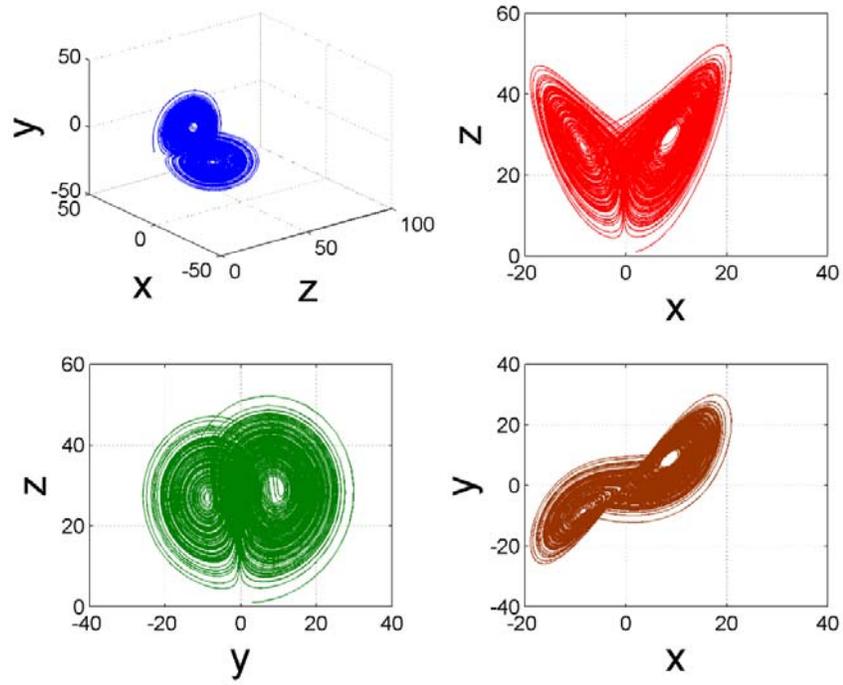

*Figure 28: Phase space dynamics of L$or$XY17*

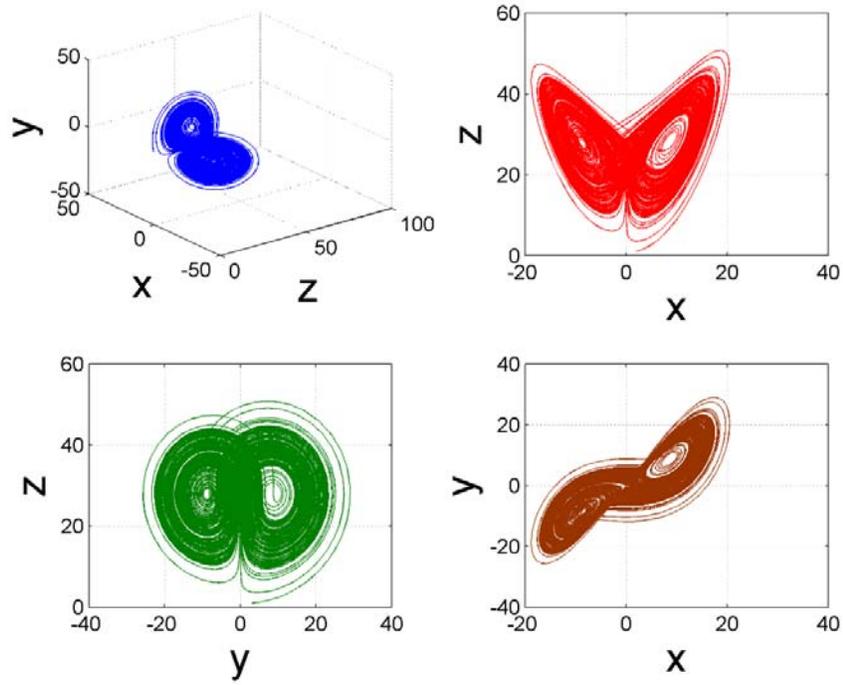

*Figure 29: Phase space dynamics of LorXY18*

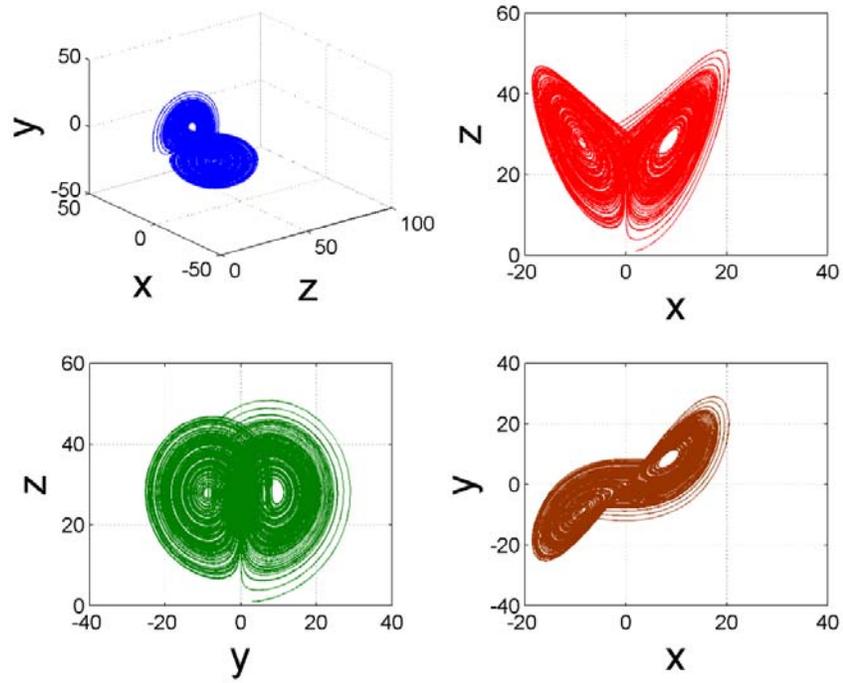

*Figure 30: Phase space dynamics of LorXY19*

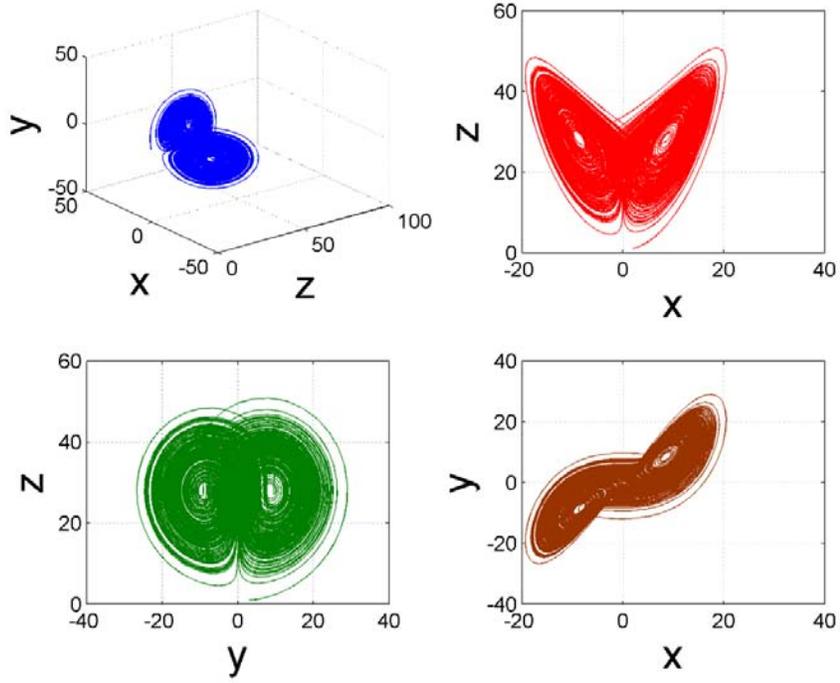

*Figure 31: Phase space dynamics of LorXY20*

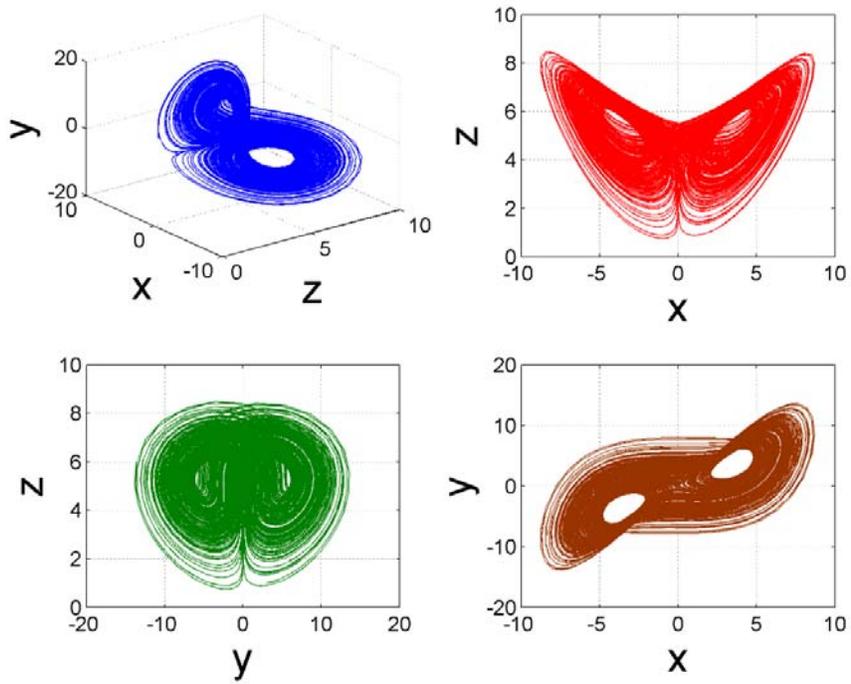

*Figure 32: Phase space dynamics of LorXY22*

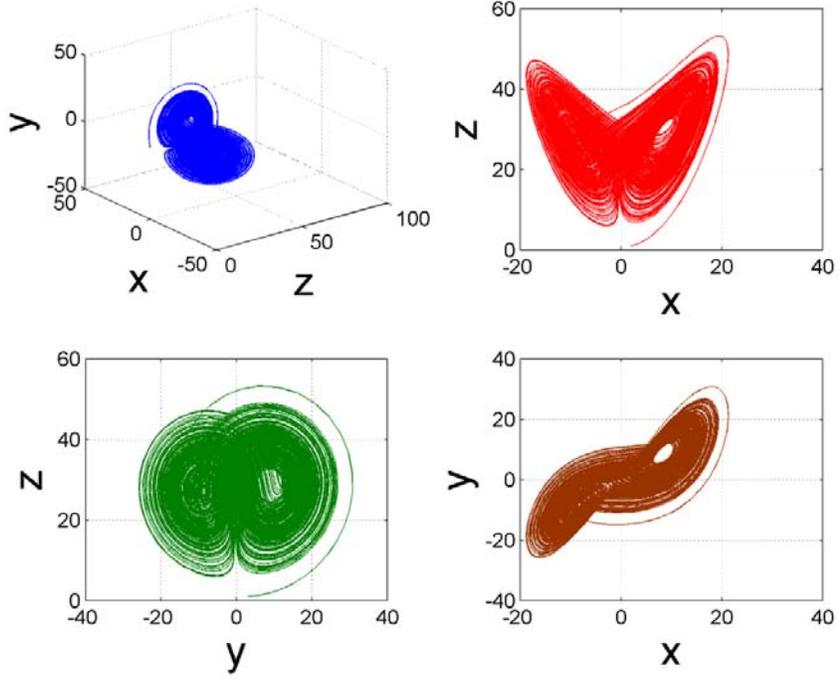

*Figure 33: Phase space dynamics of LorXY23*

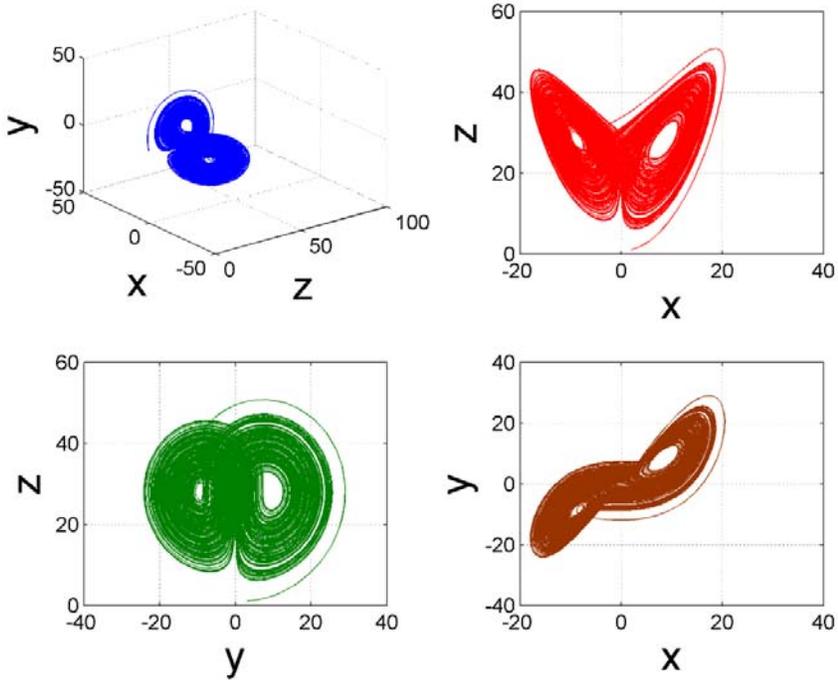

*Figure 34: Phase space dynamics of LorXY24*

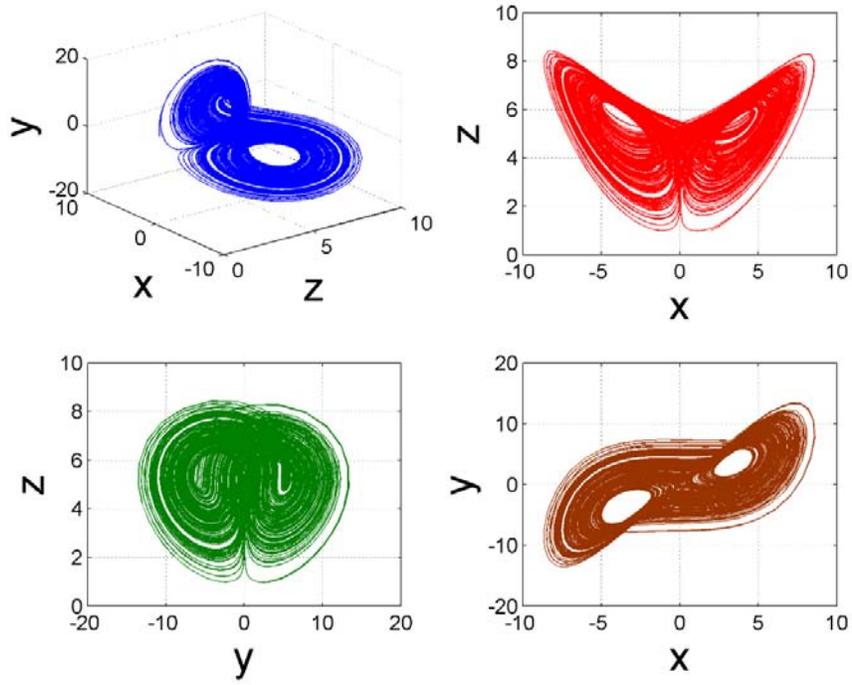

*Figure 35: Phase space dynamics of LorXY25*

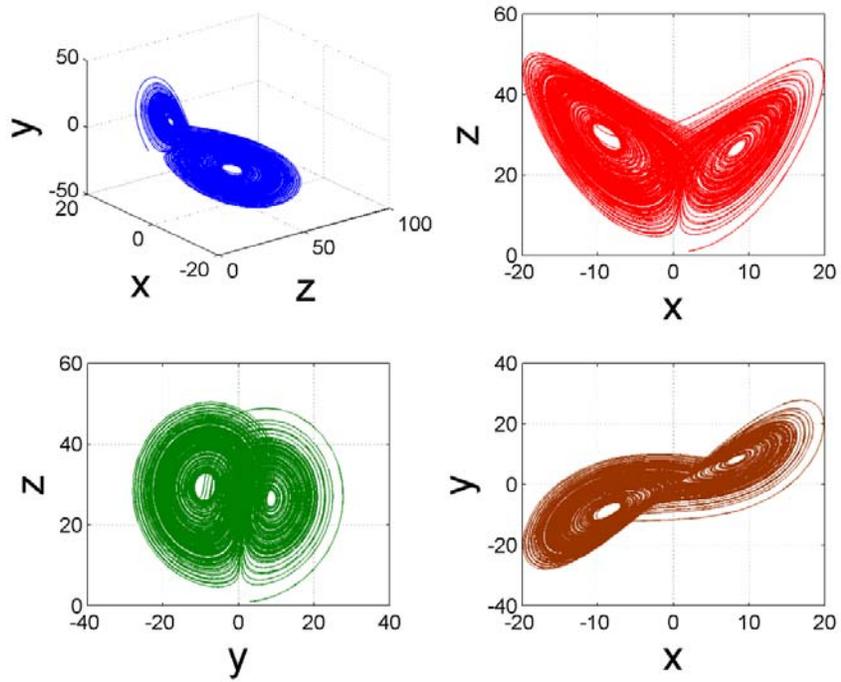

*Figure 36: Phase space dynamics of LorXY26*

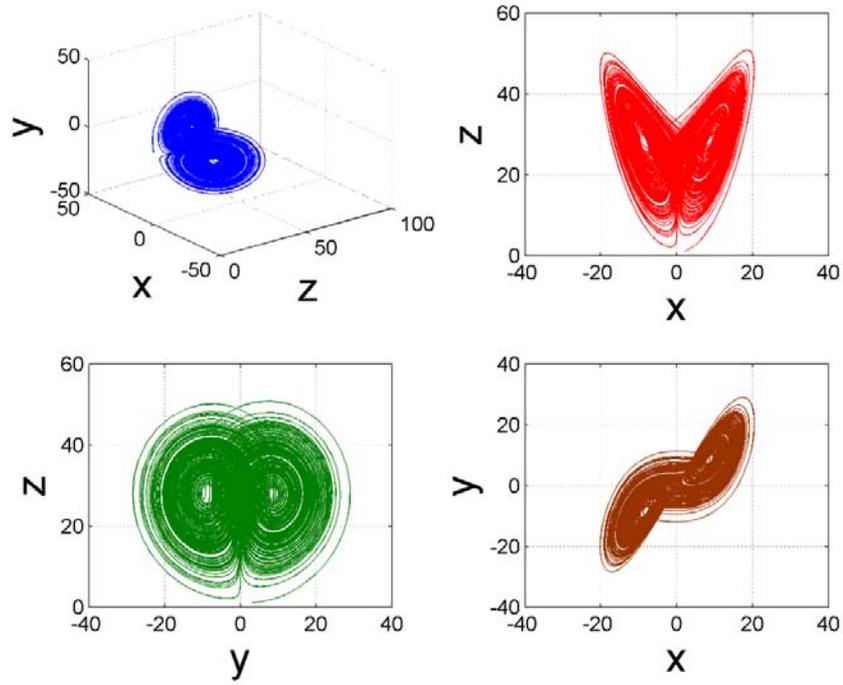

*Figure 37: Phase space dynamics of L*or*XY27*

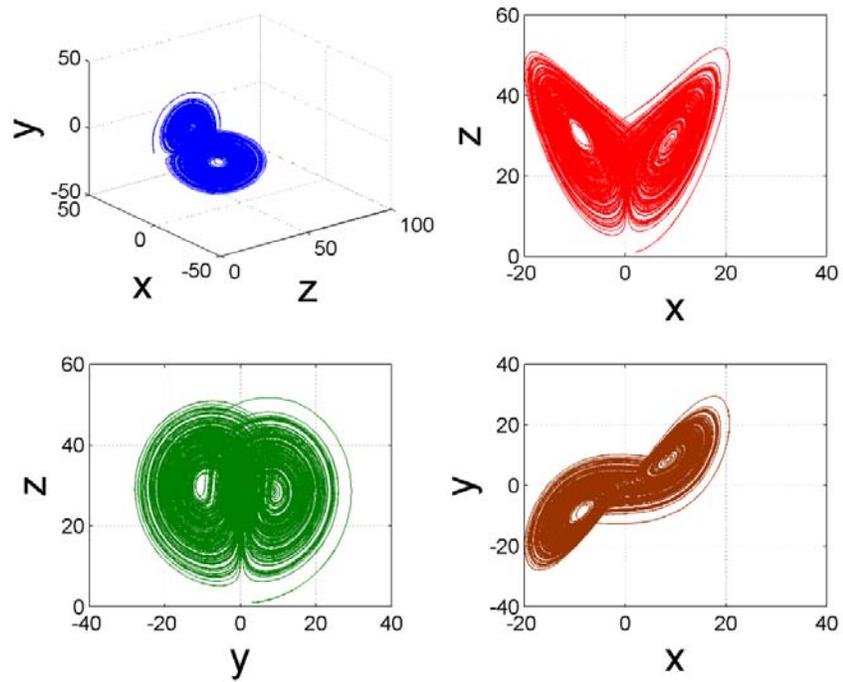

*Figure 38: Phase space dynamics of L*or*XY28*

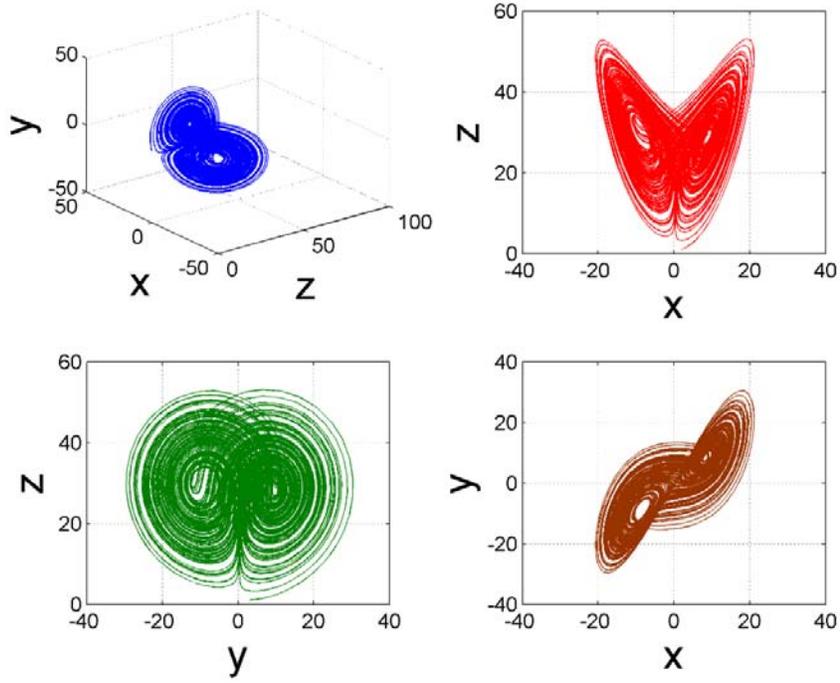

*Figure 39: Phase space dynamics of LorXY29*

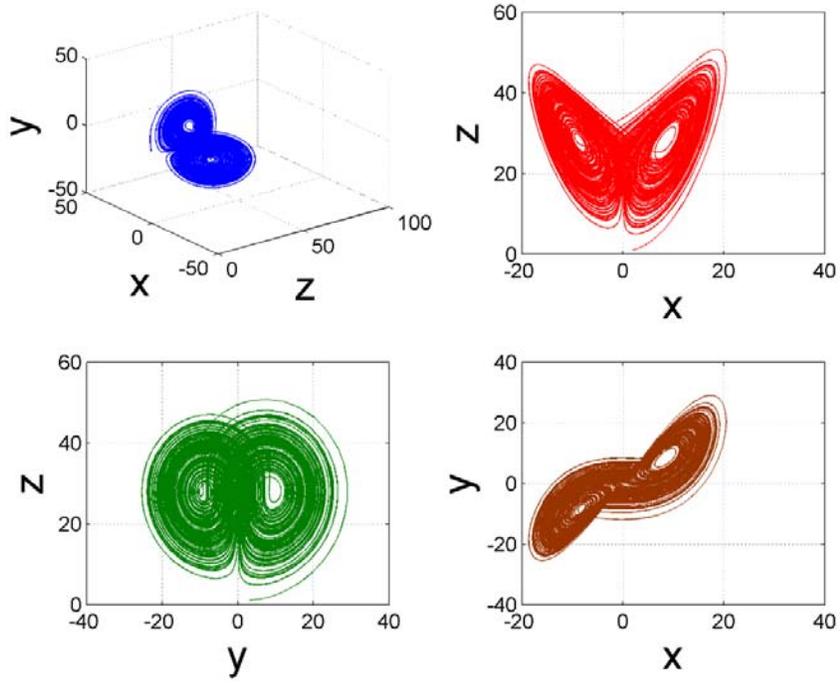

*Figure 40: Phase space dynamics of LorXY30*

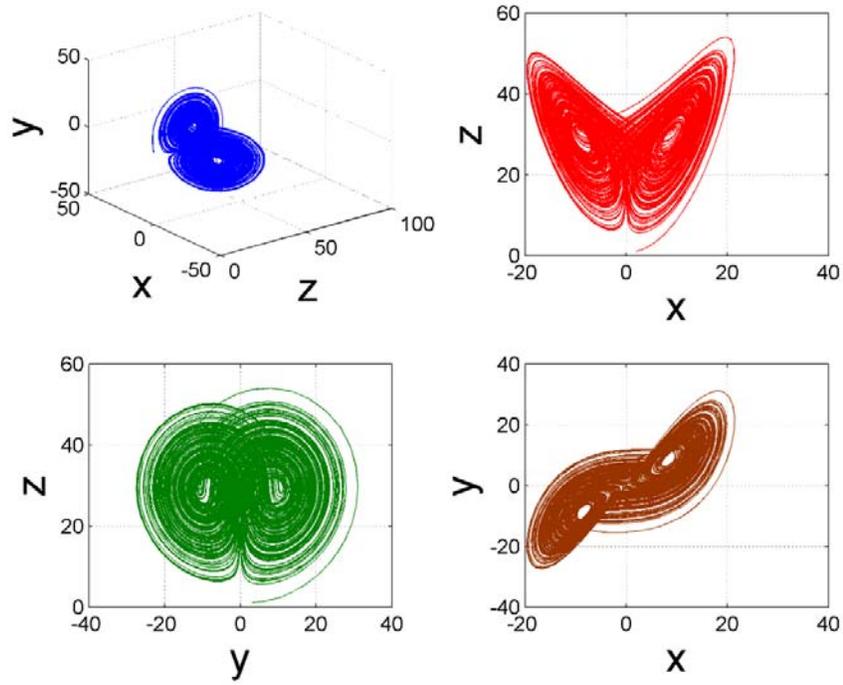

*Figure 41: Phase space dynamics of LorXY31*

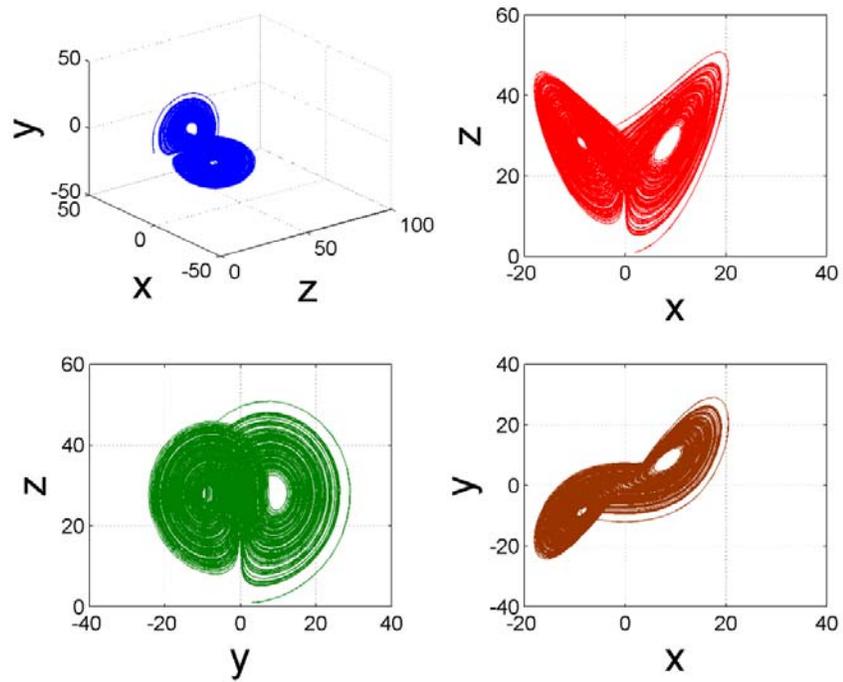

*Figure 42: Phase space dynamics of LorXY32*

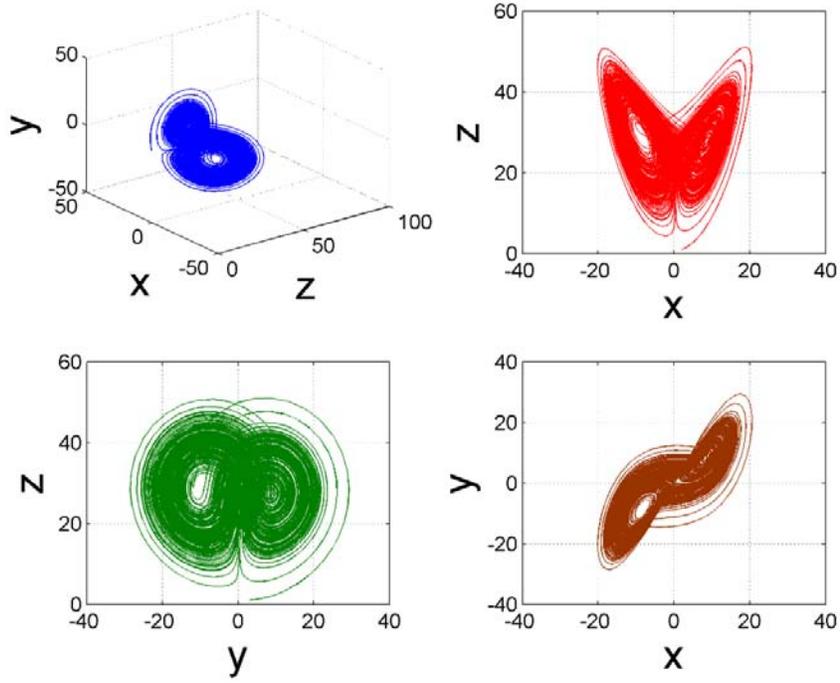

*Figure 43: Phase space dynamics of LorXY33*

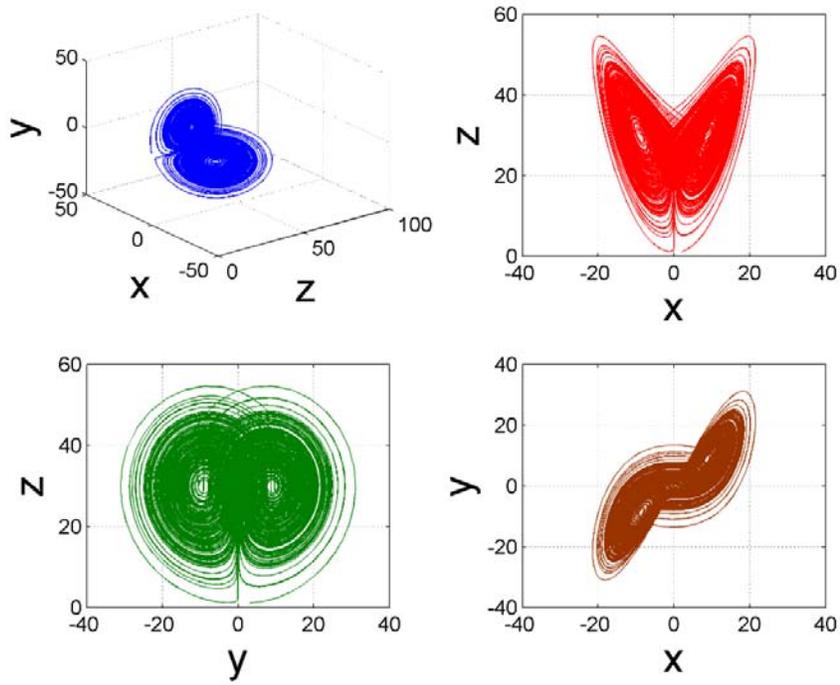

*Figure 44: Phase space dynamics of LorXY34*

## c) *Phase portraits of generalized Lorenz y-z family of attractors*

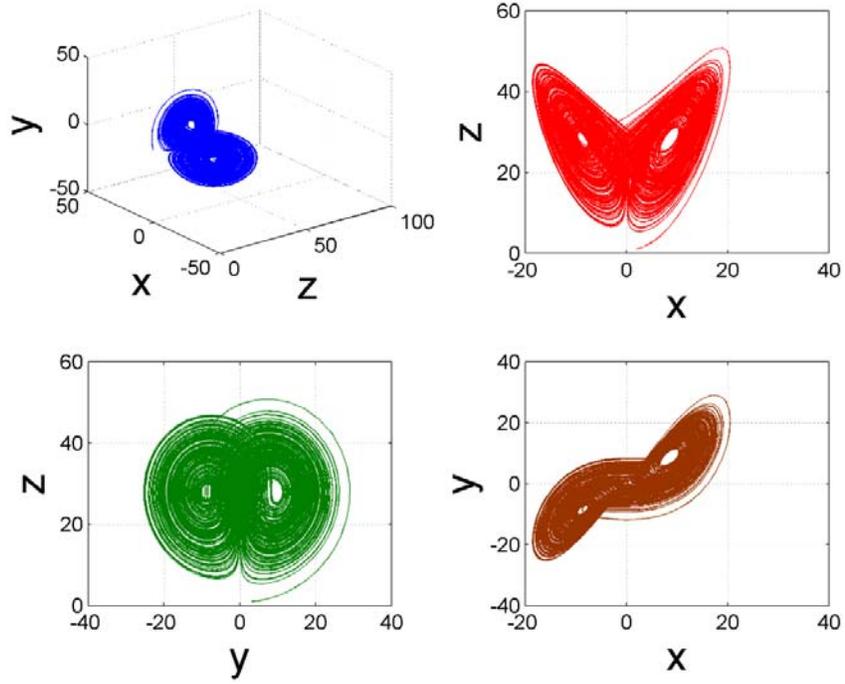

*Figure 45: Phase space dynamics of LorYZ1*

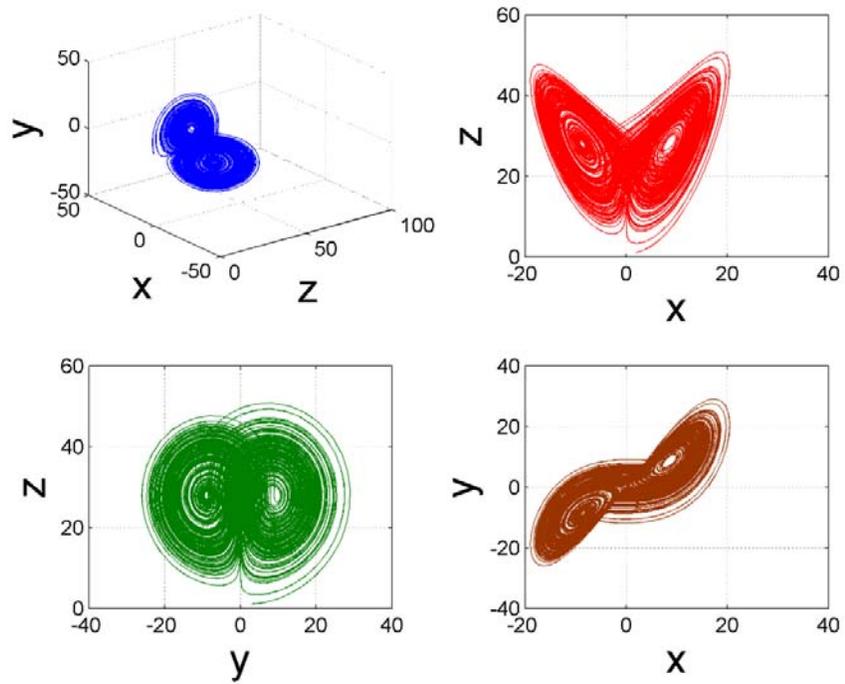

*Figure 46: Phase space dynamics of LorYZ2*

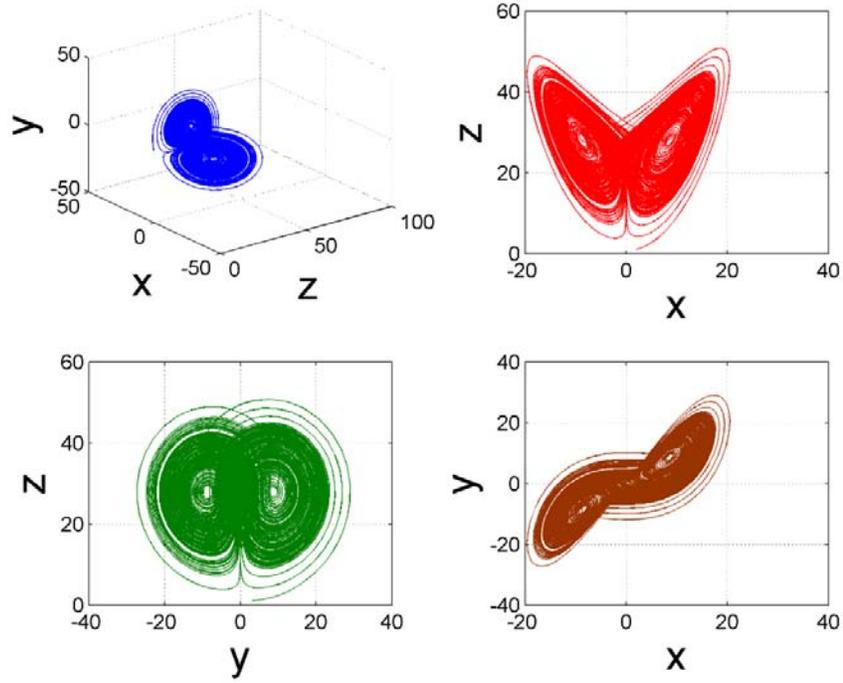

*Figure 47: Phase space dynamics of LorYZ3*

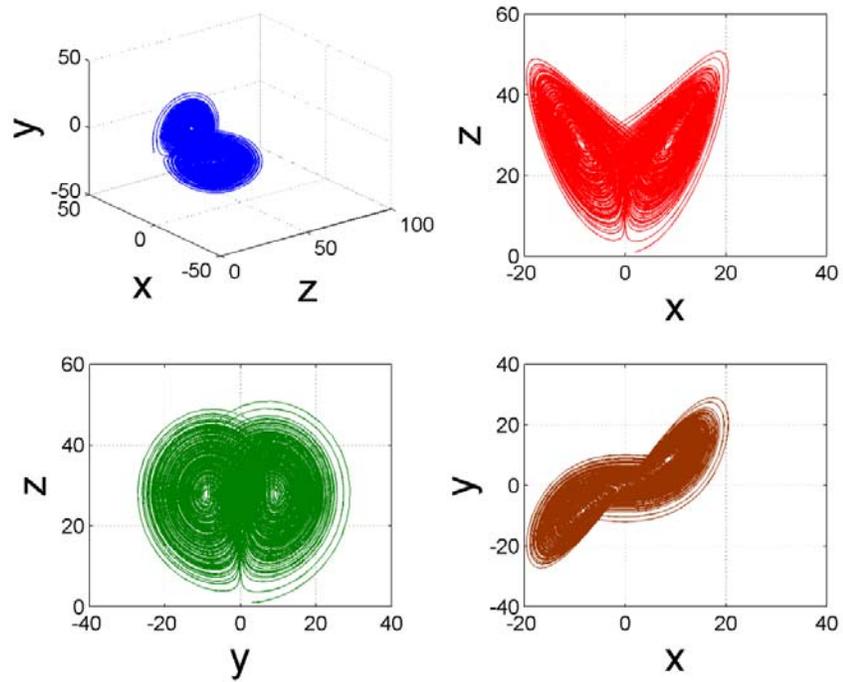

*Figure 48: Phase space dynamics of LorYZ4*

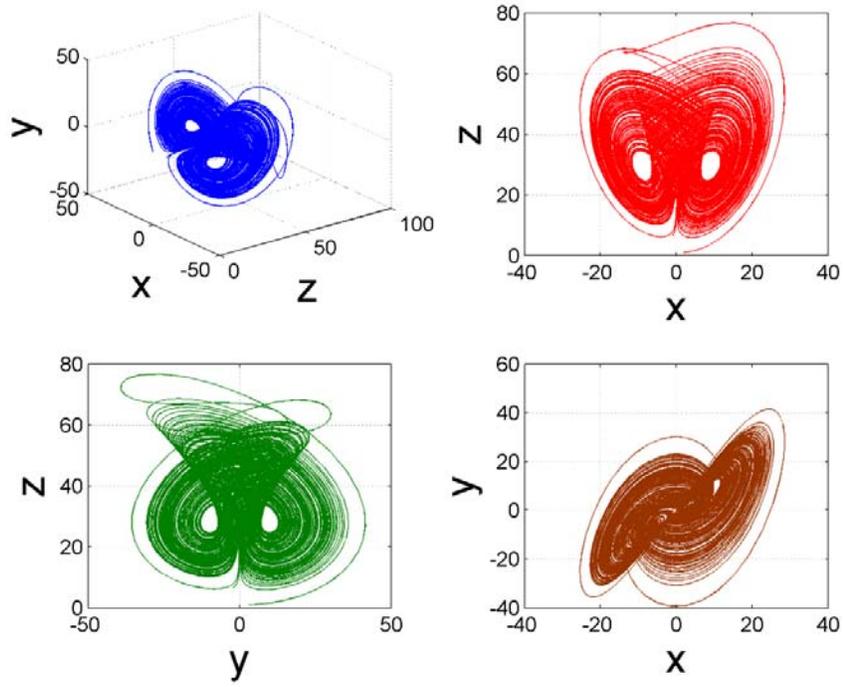

*Figure 49: Phase space dynamics of L*o*rYZ5*

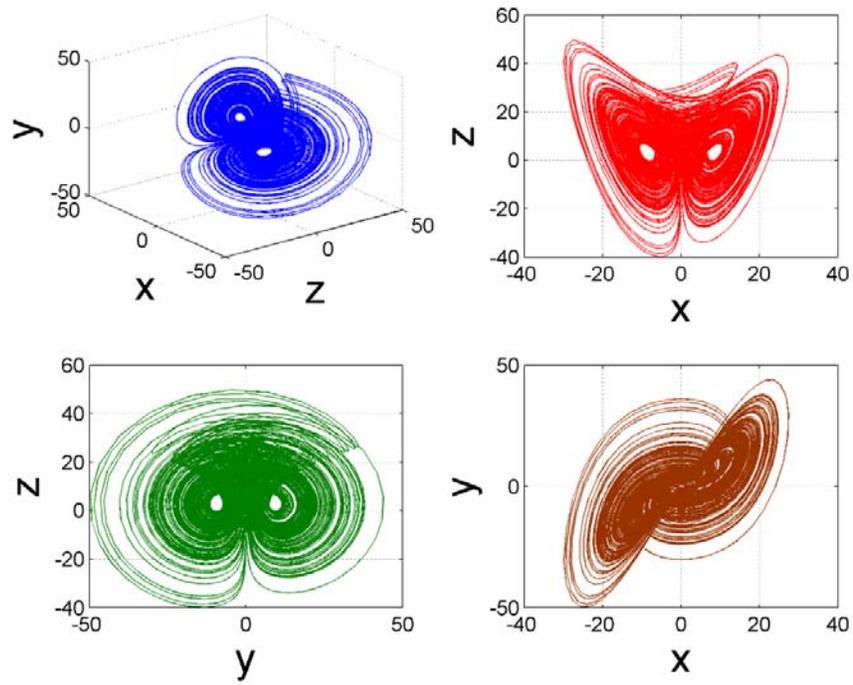

*Figure 50: Phase space dynamics of L*o*rYZ6*

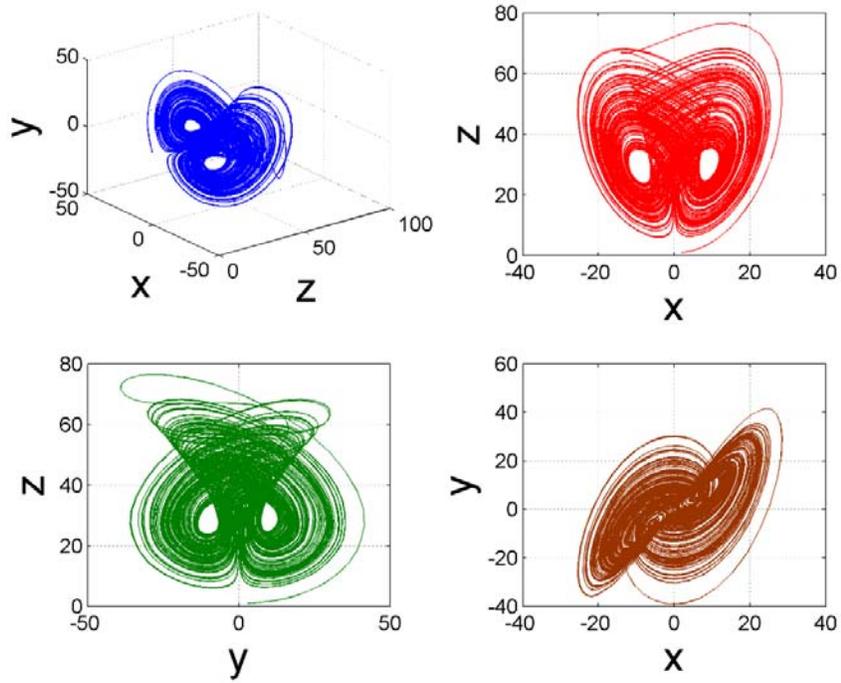

*Figure 51: Phase space dynamics of LorYZ7*

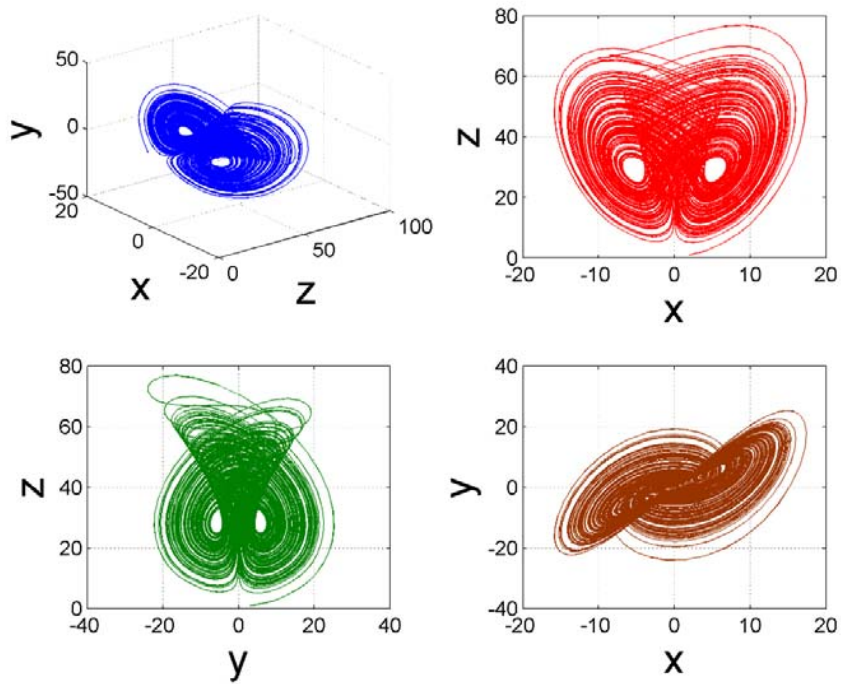

*Figure 52: Phase space dynamics of LorYZ8*

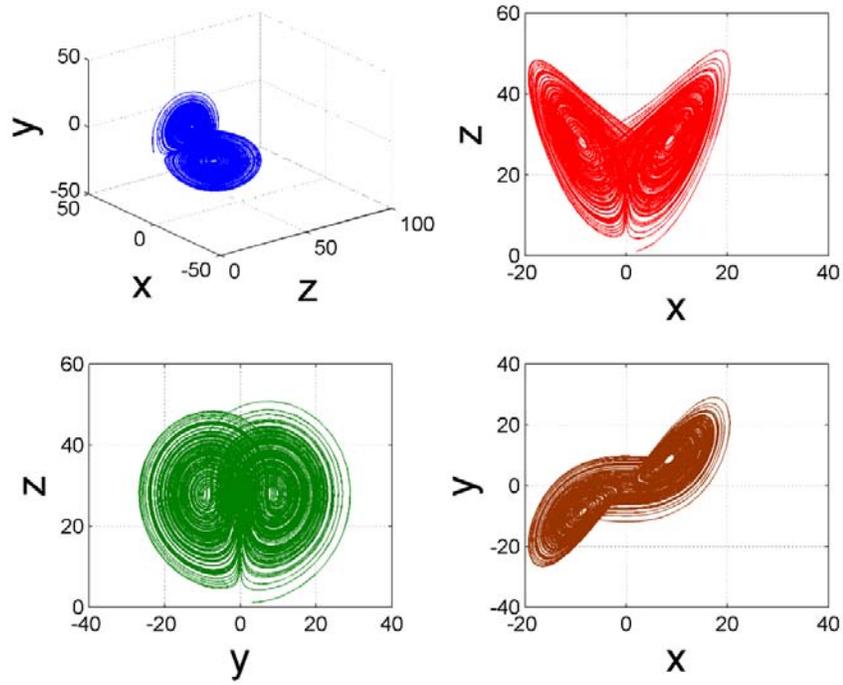

*Figure 53: Phase space dynamics of LorYZ9*

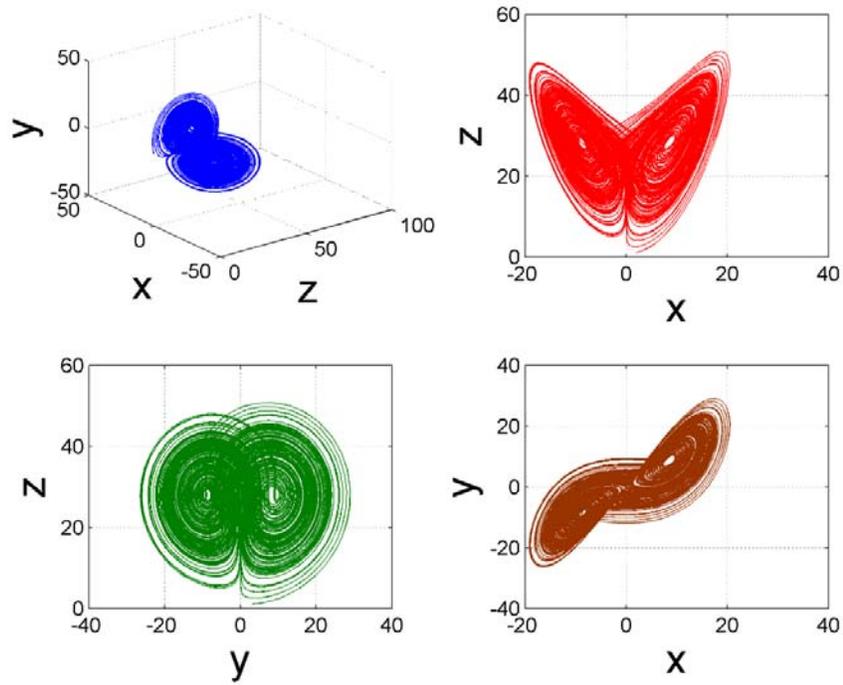

*Figure 54: Phase space dynamics of LorYZ10*

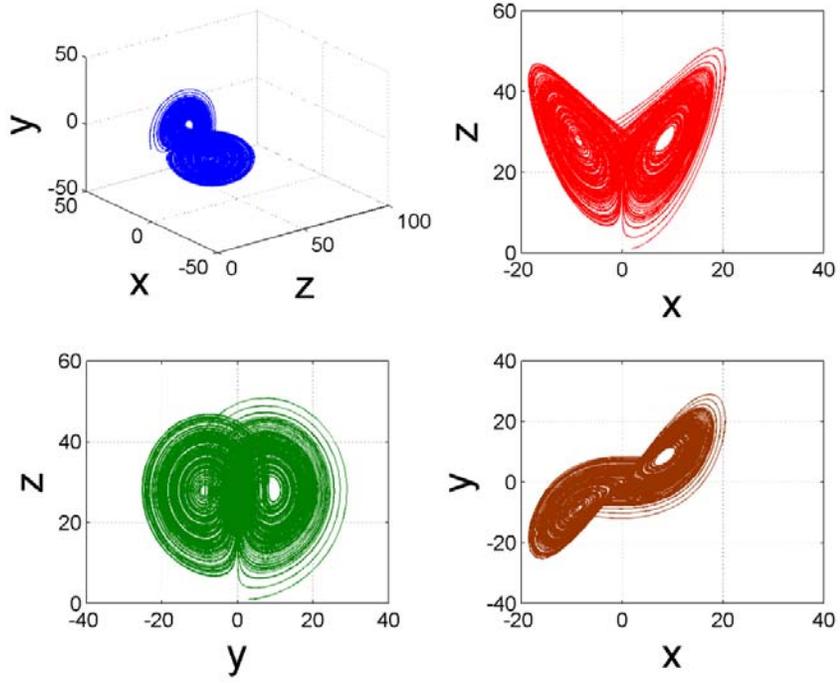

*Figure 55: Phase space dynamics of LorYZ11*

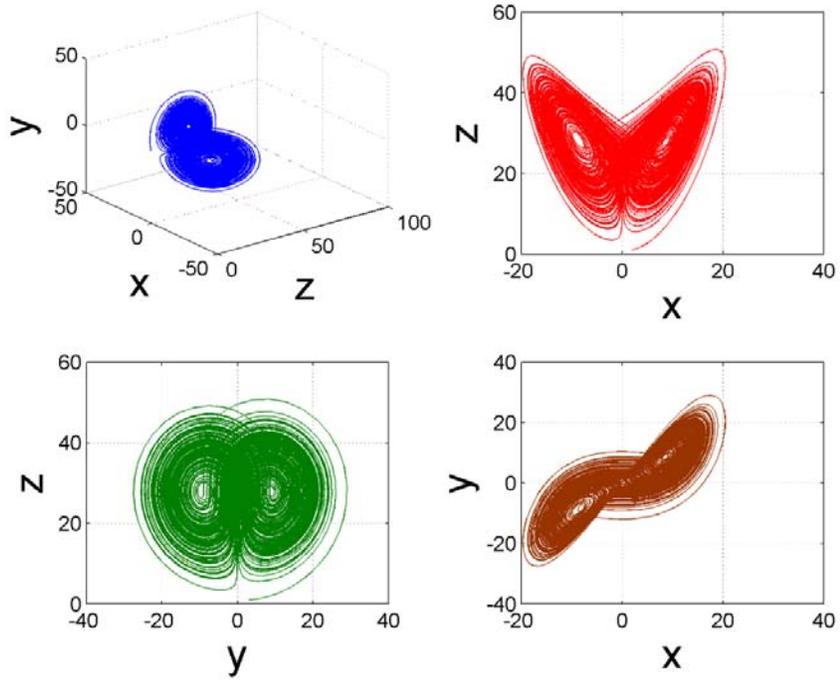

*Figure 56: Phase space dynamics of LorYZ12*

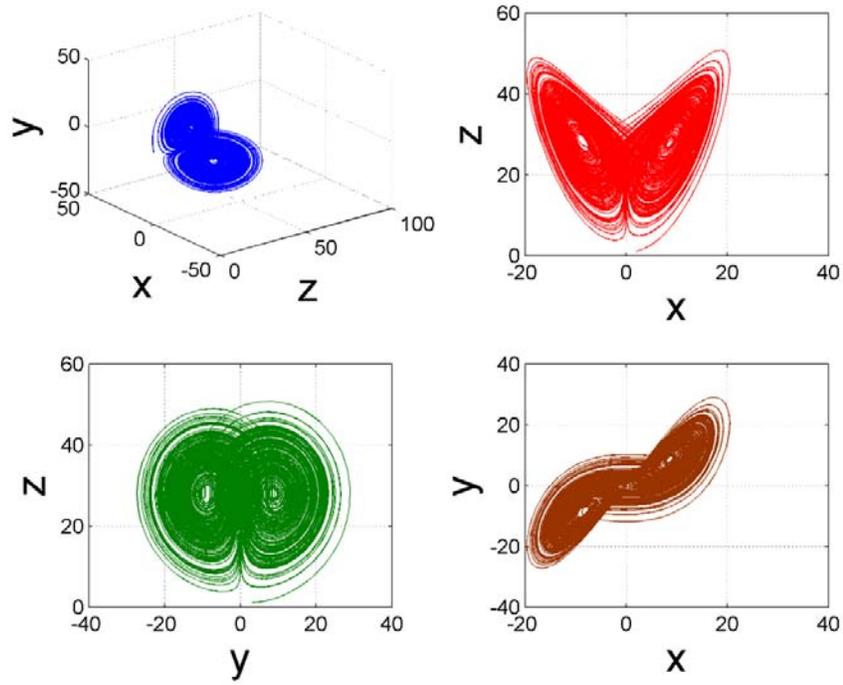

*Figure 57: Phase space dynamics of LorYZ13*

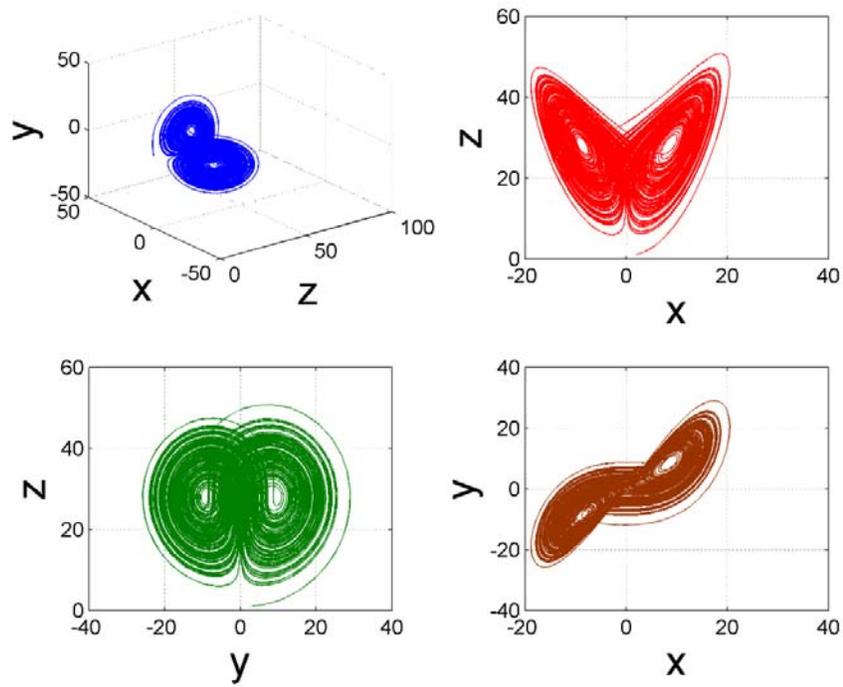

*Figure 58: Phase space dynamics of LorYZ14*

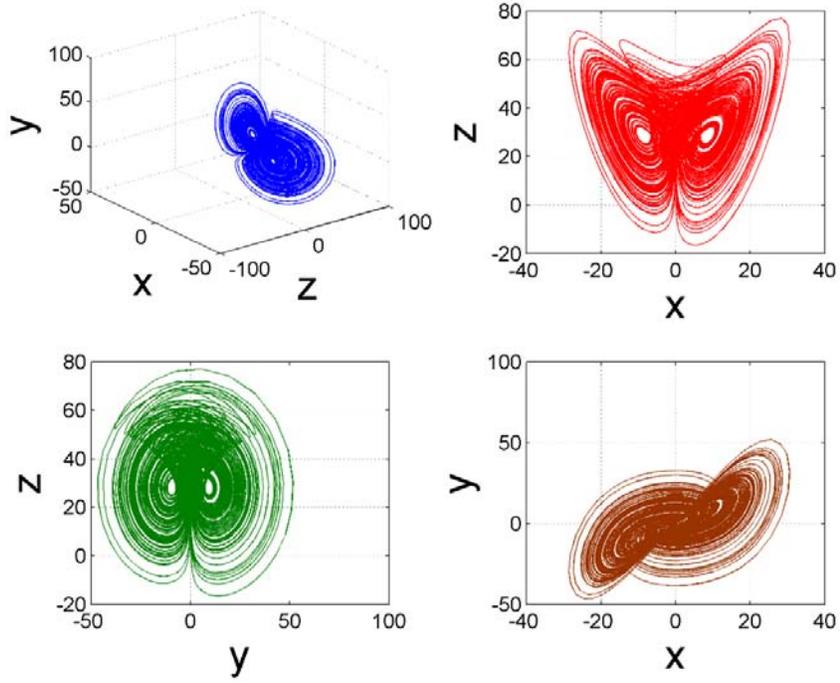

*Figure 59: Phase space dynamics of LorYZ15*

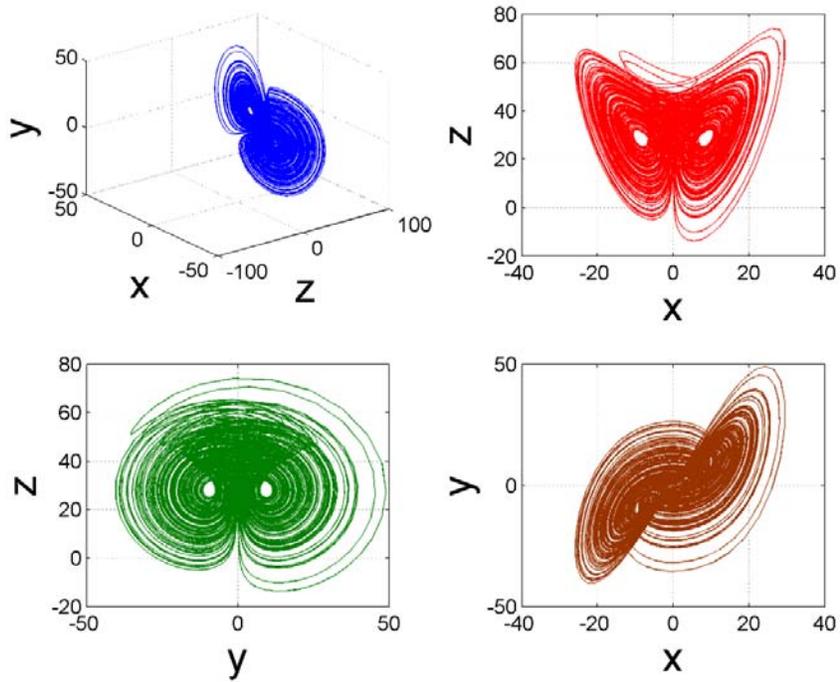

*Figure 60: Phase space dynamics of LorYZ16*

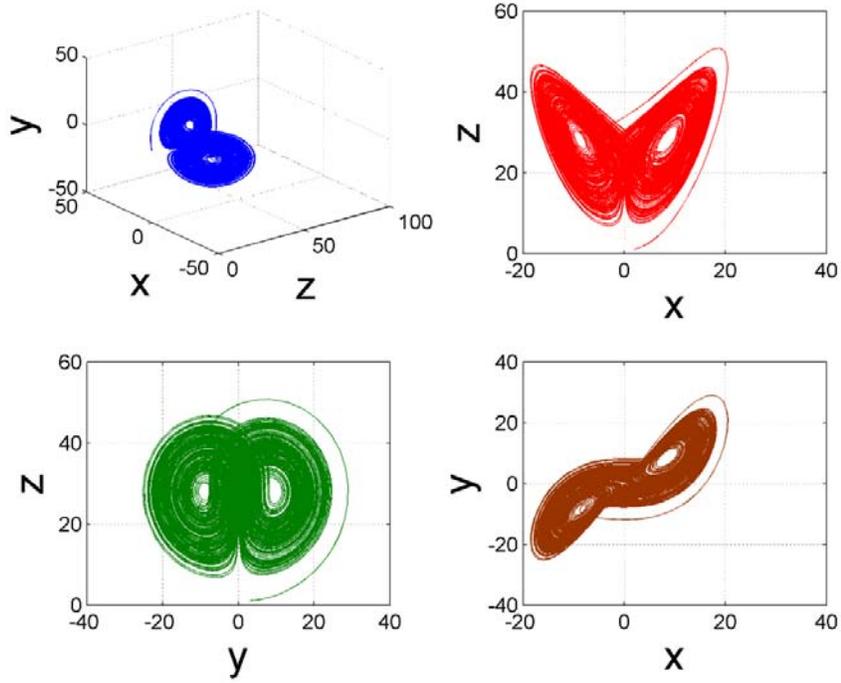

*Figure 61: Phase space dynamics of LorYZ17*

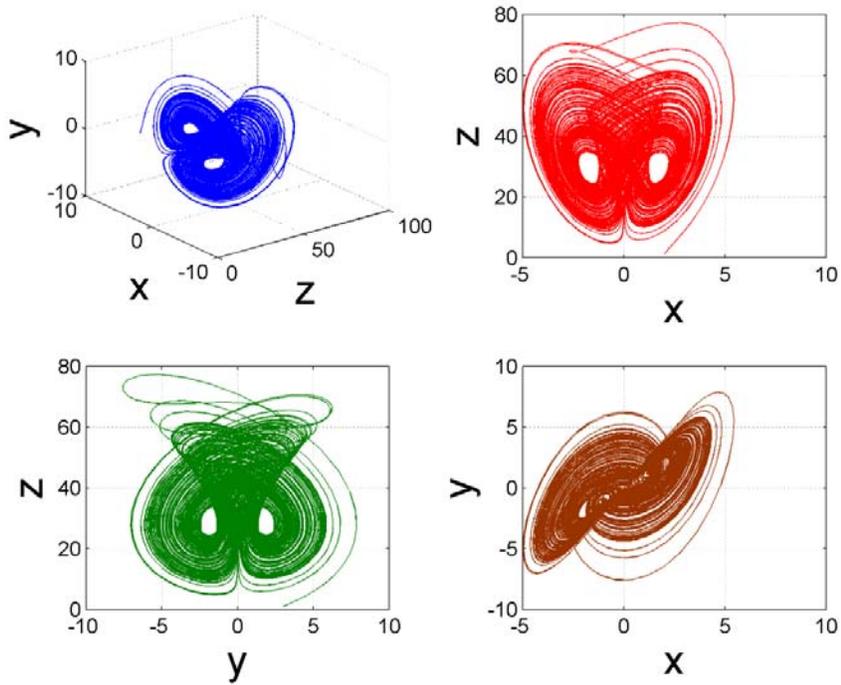

*Figure 62: Phase space dynamics of LorYZ18*

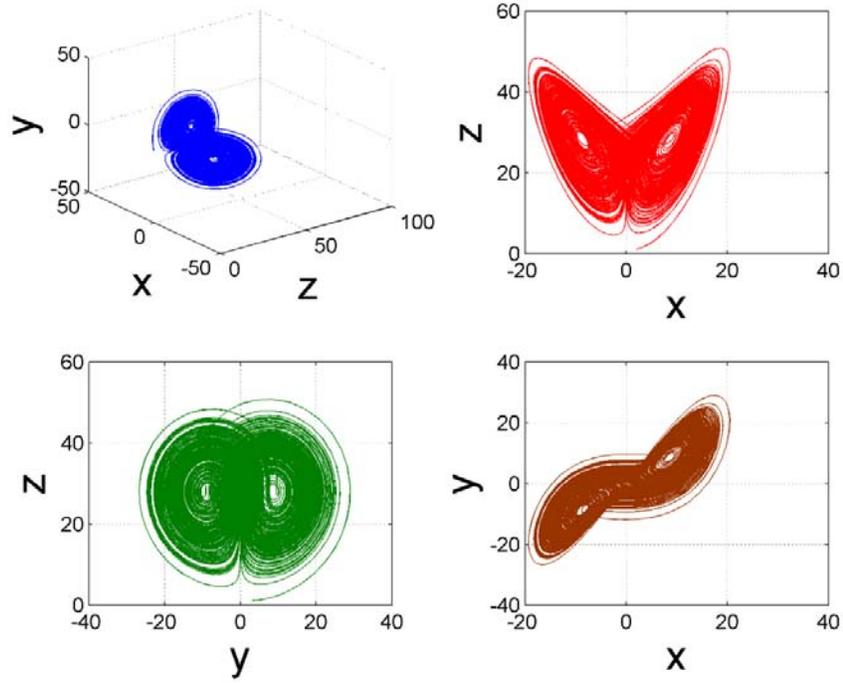

*Figure 63: Phase space dynamics of LorYZ20*

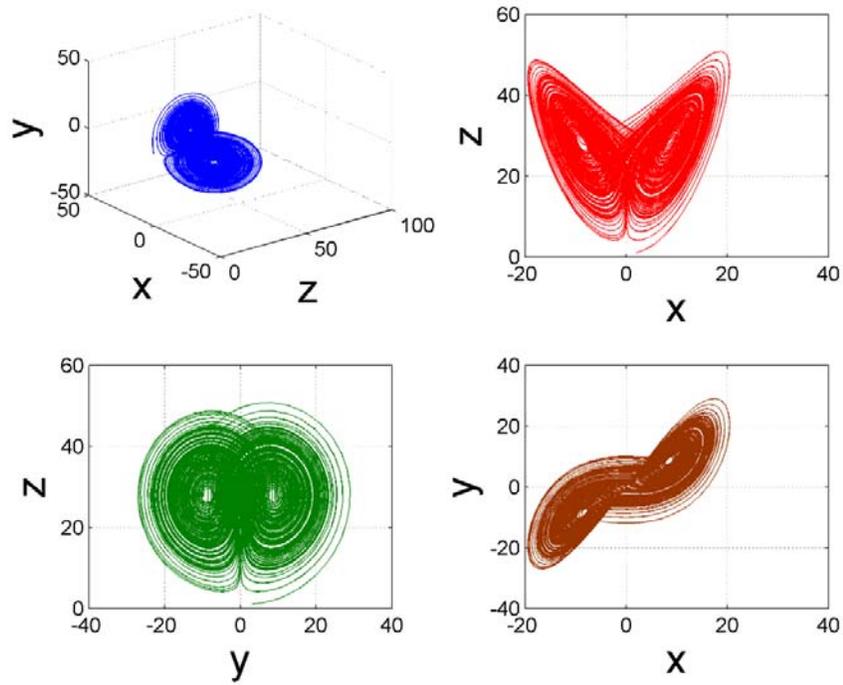

*Figure 64: Phase space dynamics of LorYZ21*

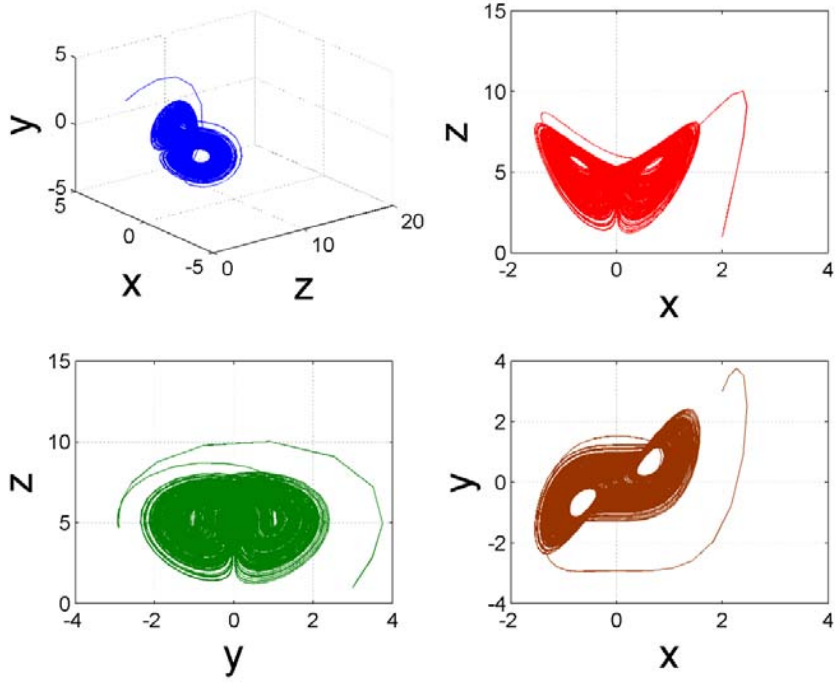

*Figure 65: Phase space dynamics of LorYZ23*

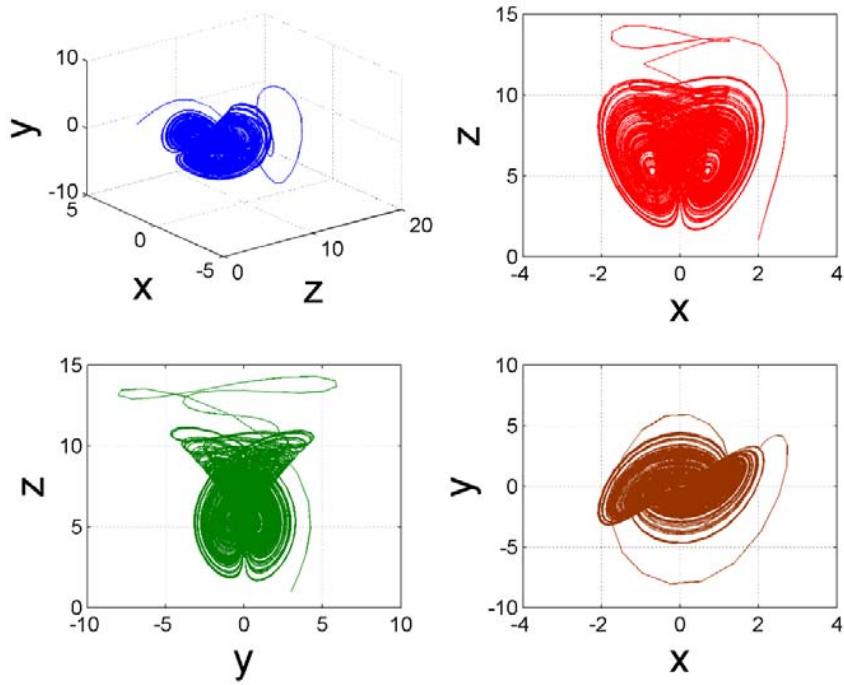

*Figure 66: Phase space dynamics of LorYZ24*

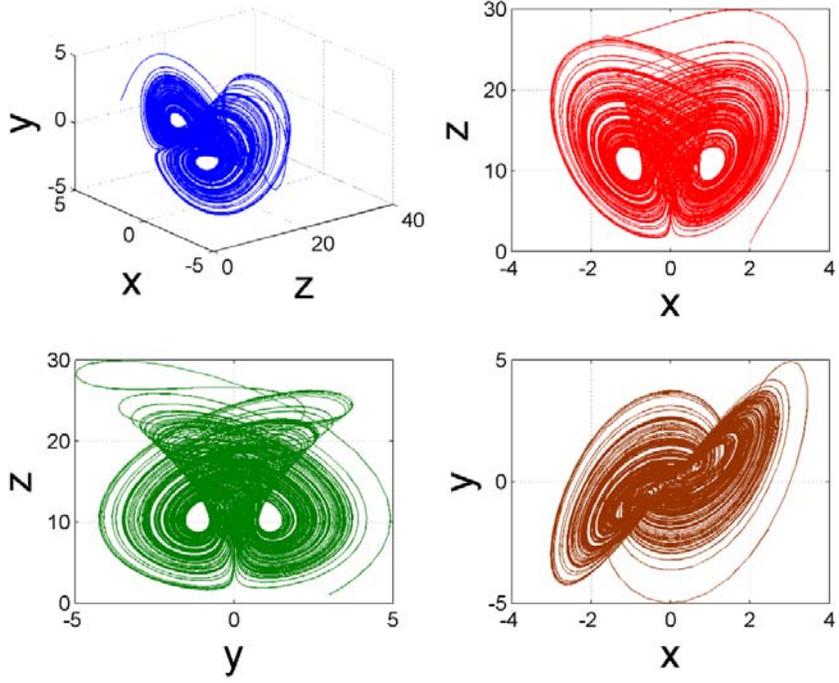

*Figure 67: Phase space dynamics of L₀rYZ25*

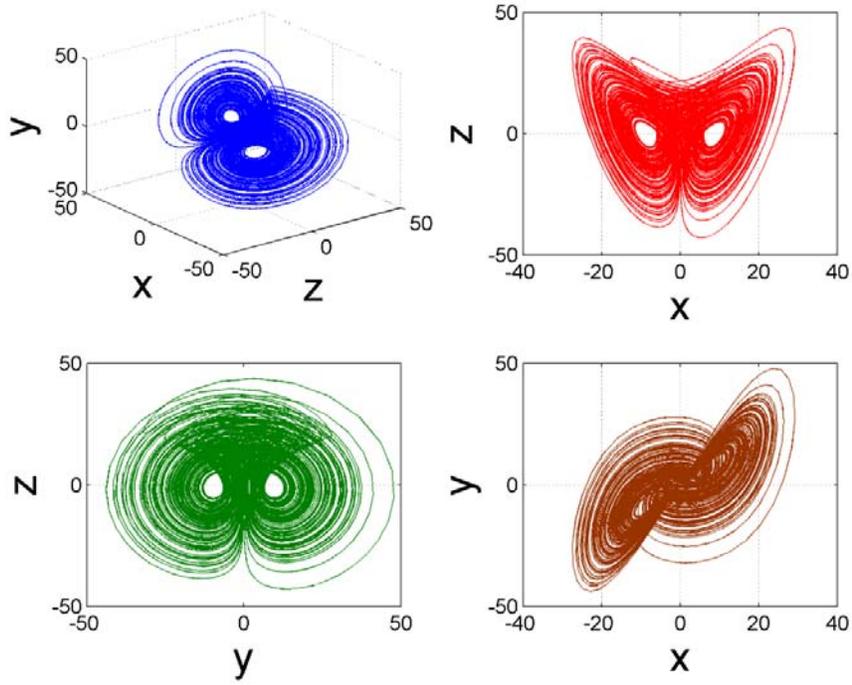

*Figure 68: Phase space dynamics of L₀rYZ26*

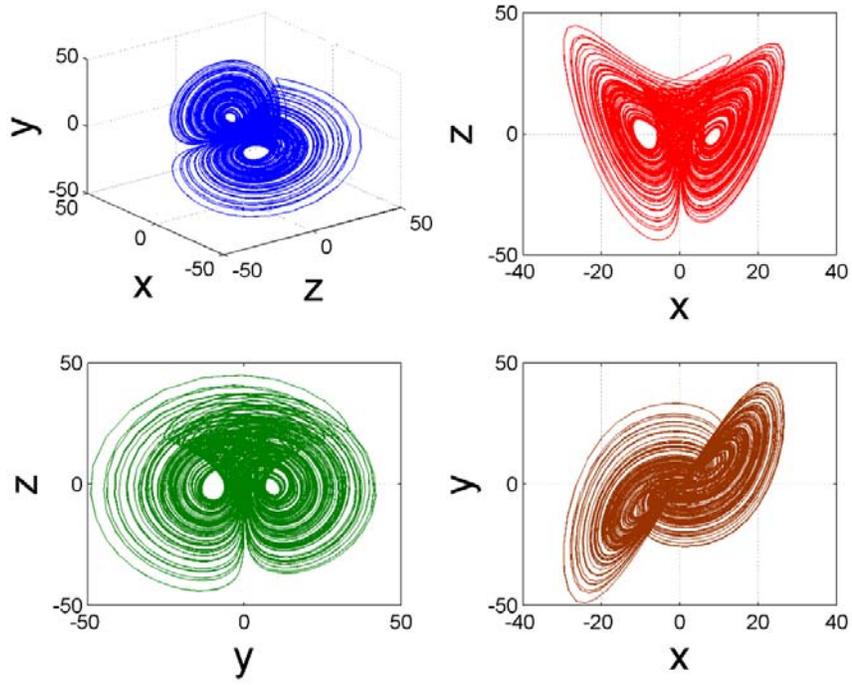

*Figure 69: Phase space dynamics of LorYZ28*

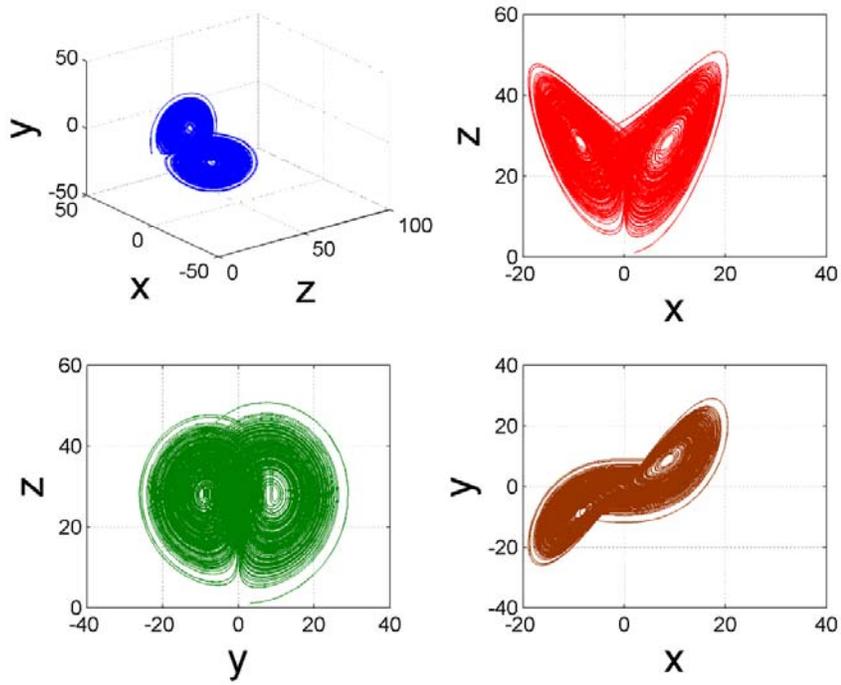

*Figure 70: Phase space dynamics of LorYZ29*

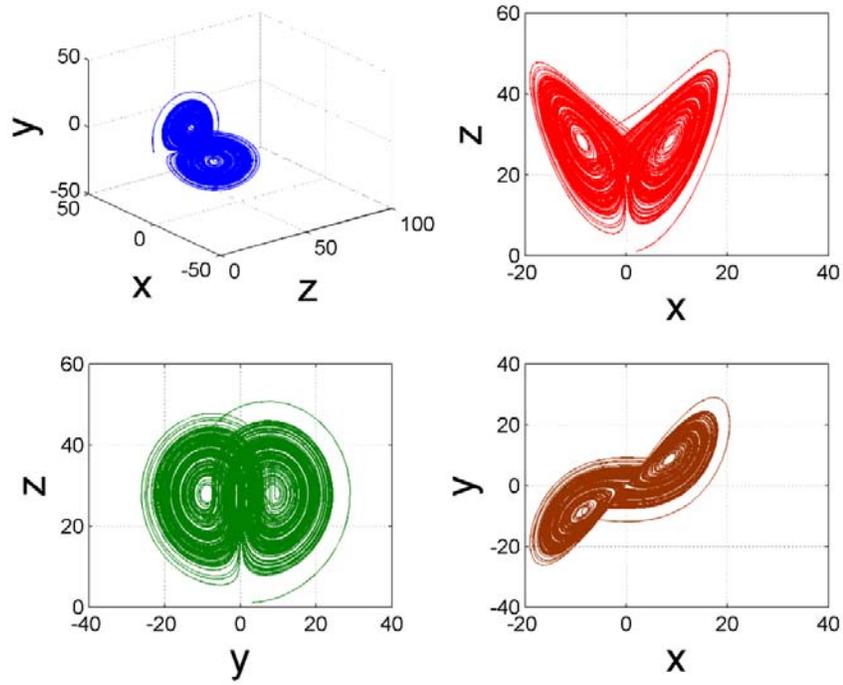

*Figure 71: Phase space dynamics of LorYZ30*

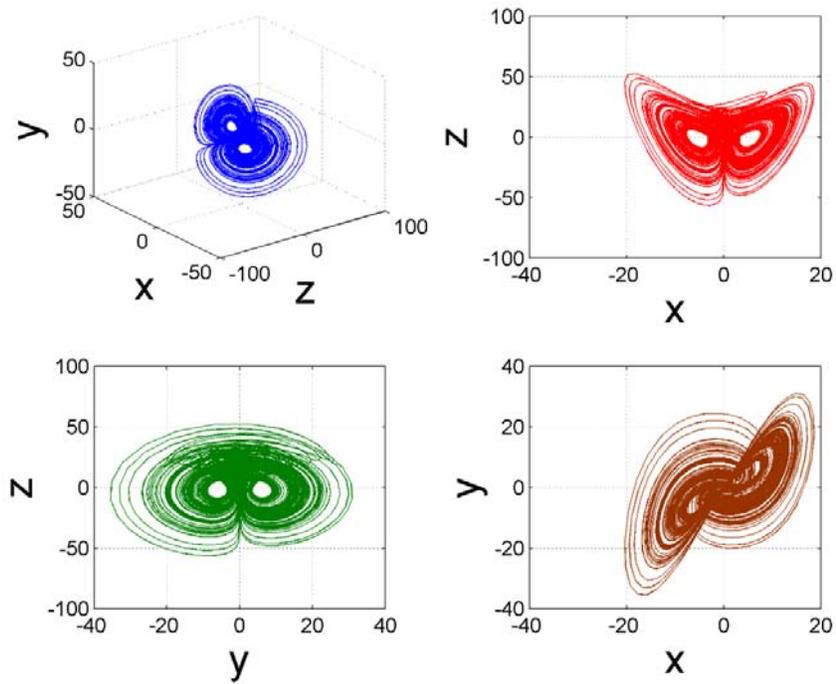

*Figure 72: Phase space dynamics of LorYZ31*

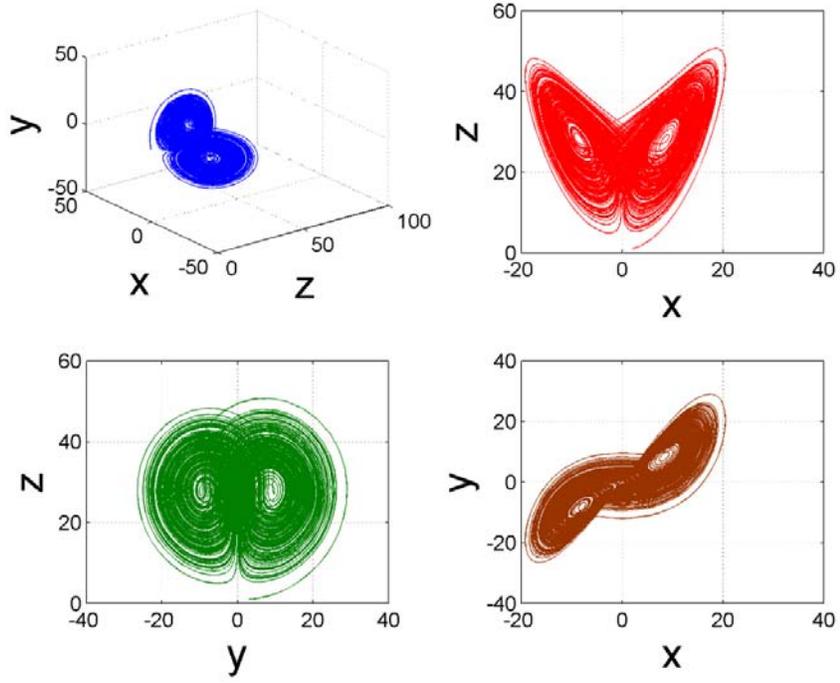

*Figure 73: Phase space dynamics of LorYZ32*

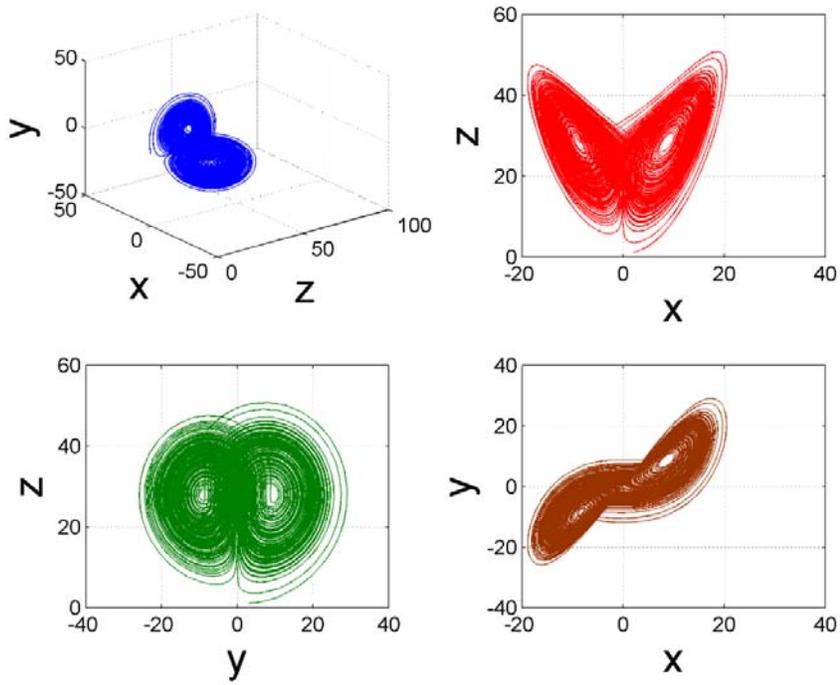

*Figure 74: Phase space dynamics of LorYZ33*

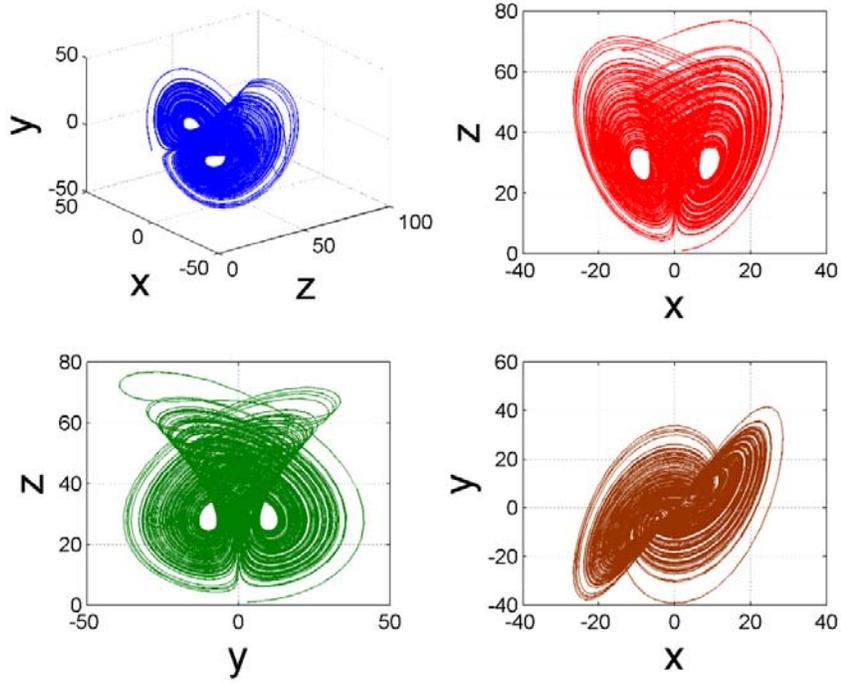

*Figure 75: Phase space dynamics of LorYZ34*

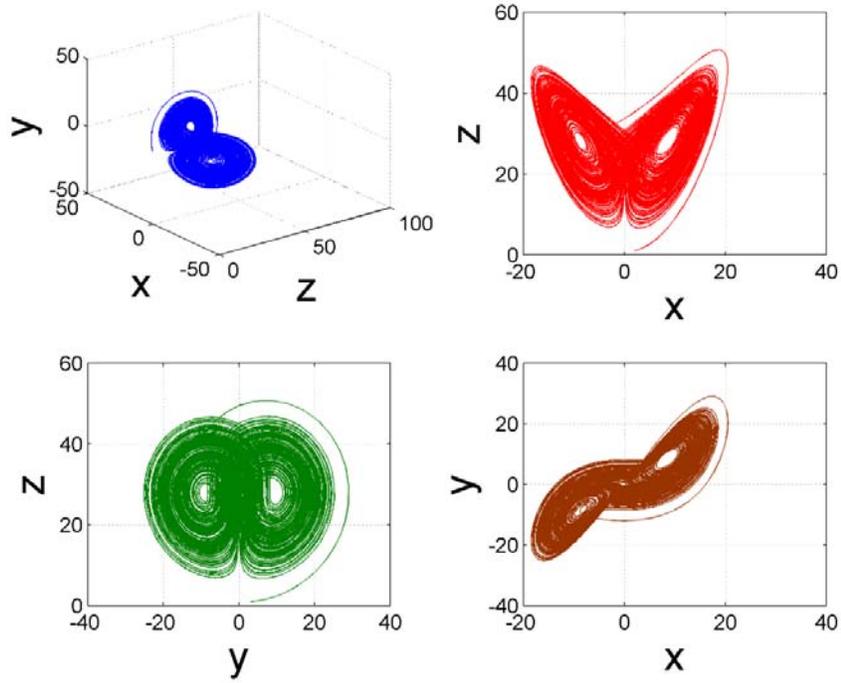

*Figure 76: Phase space dynamics of LorYZ35*

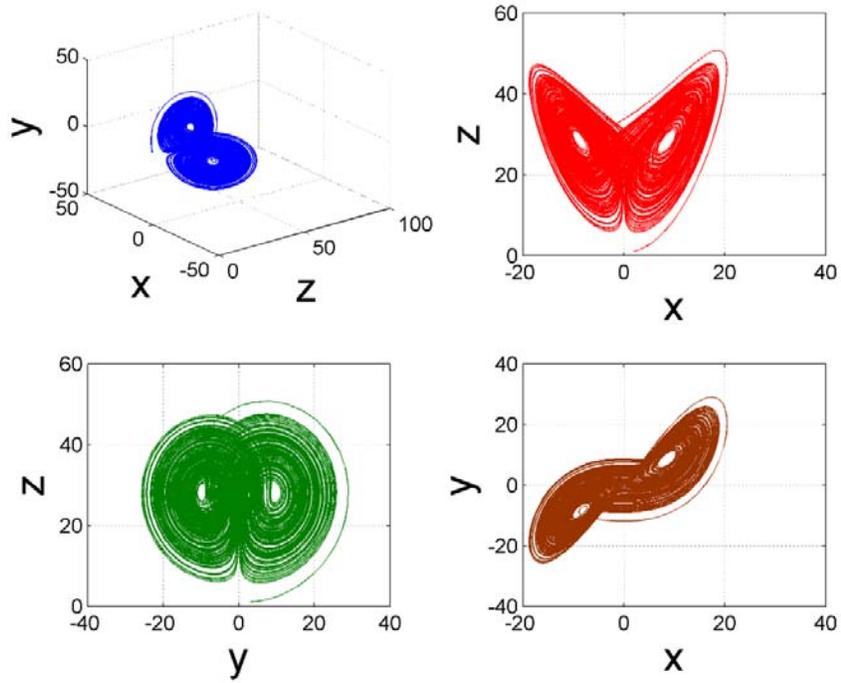

*Figure 77: Phase space dynamics of LorYZ36*

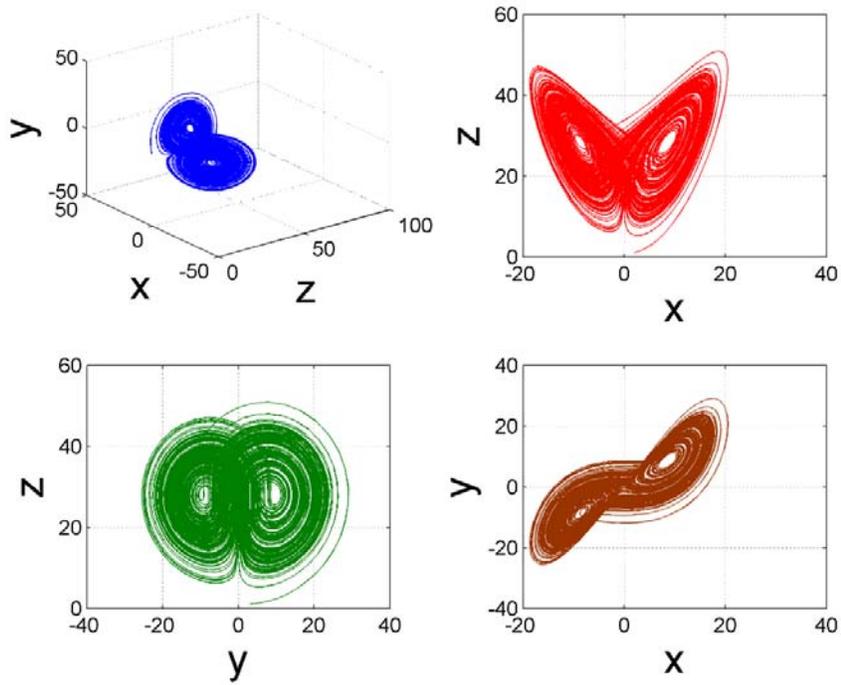

*Figure 78: Phase space dynamics of LorYZ37*

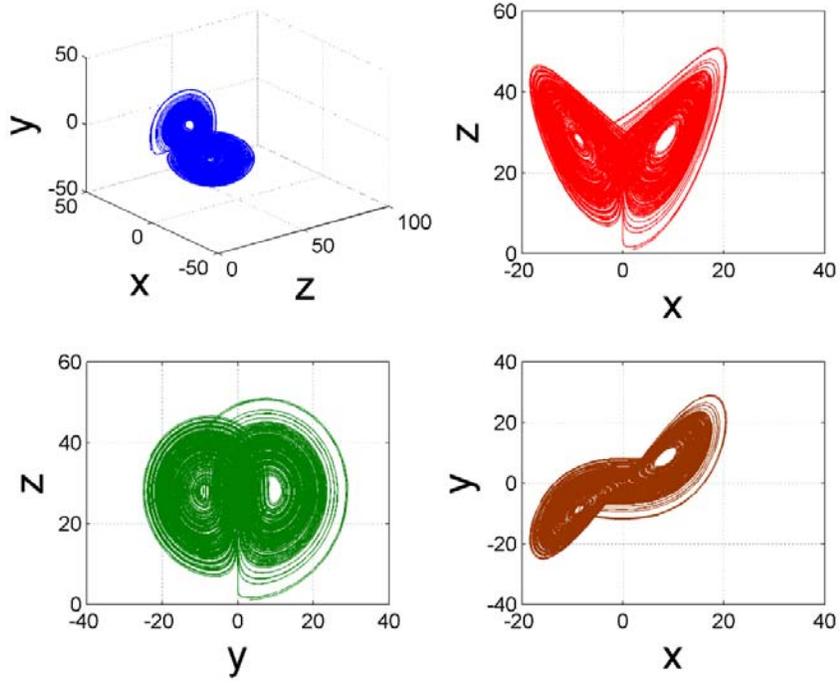

*Figure 79: Phase space dynamics of LorYZ38*

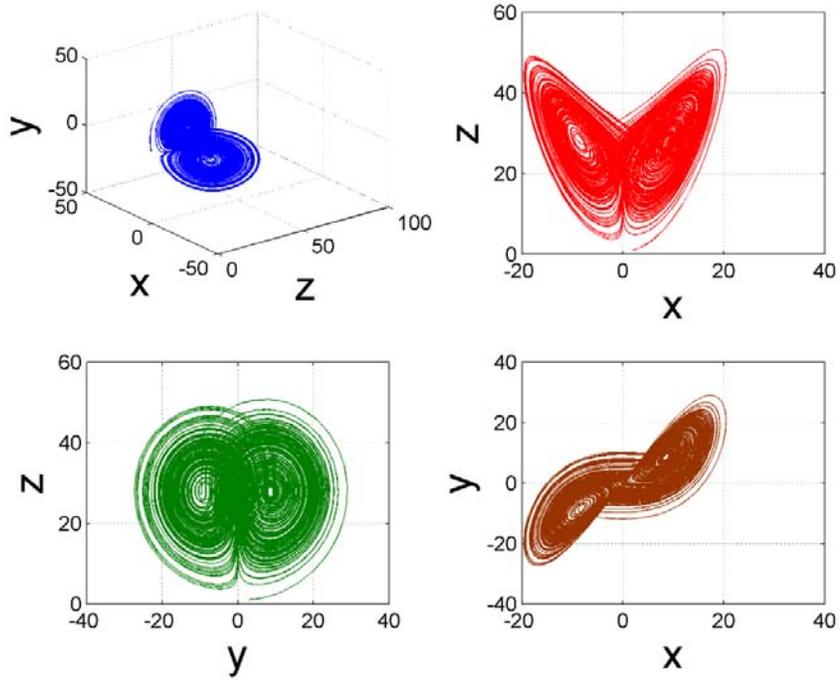

*Figure 80: Phase space dynamics of LorYZ39*

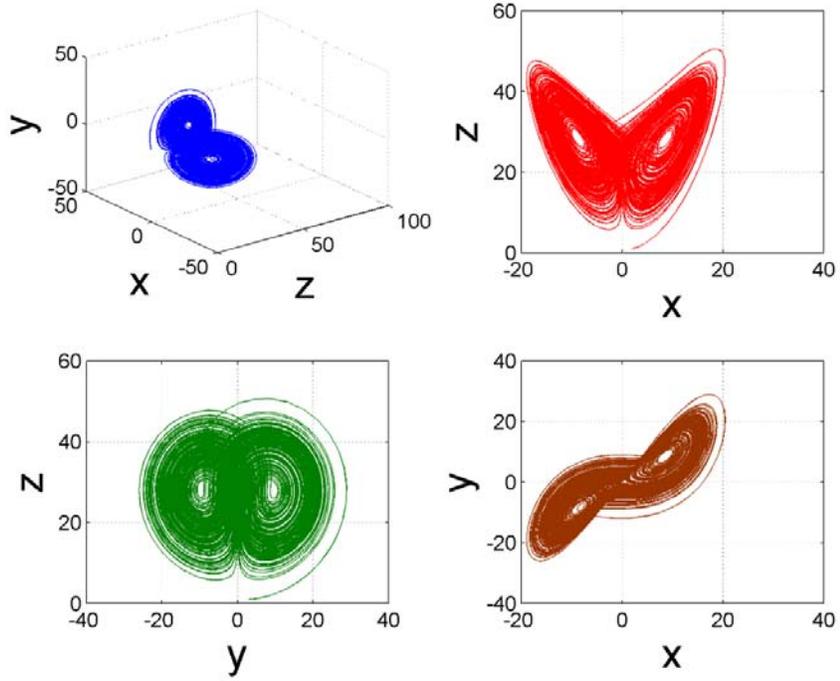

*Figure 81: Phase space dynamics of LorYZ40*

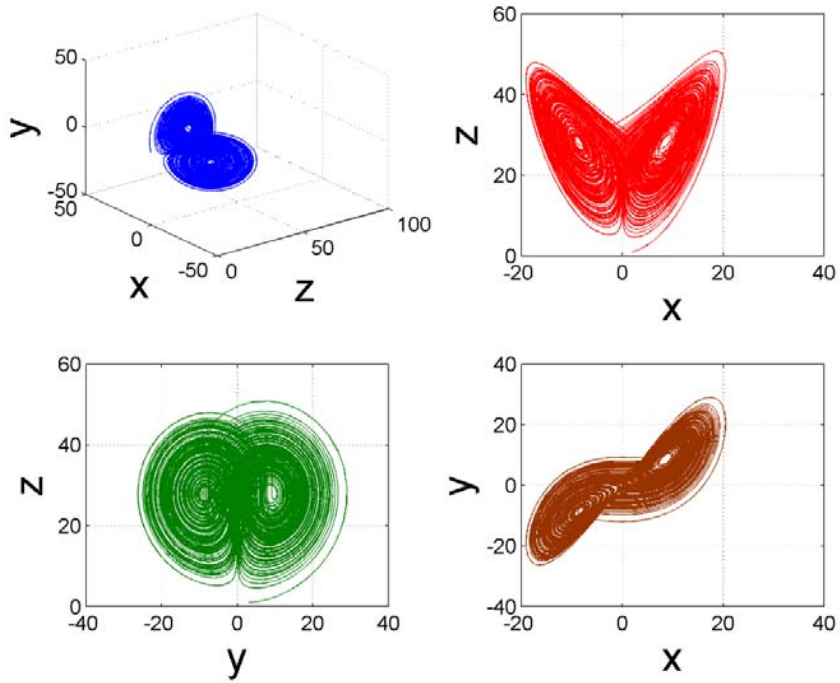

*Figure 82: Phase space dynamics of LorYZ41*

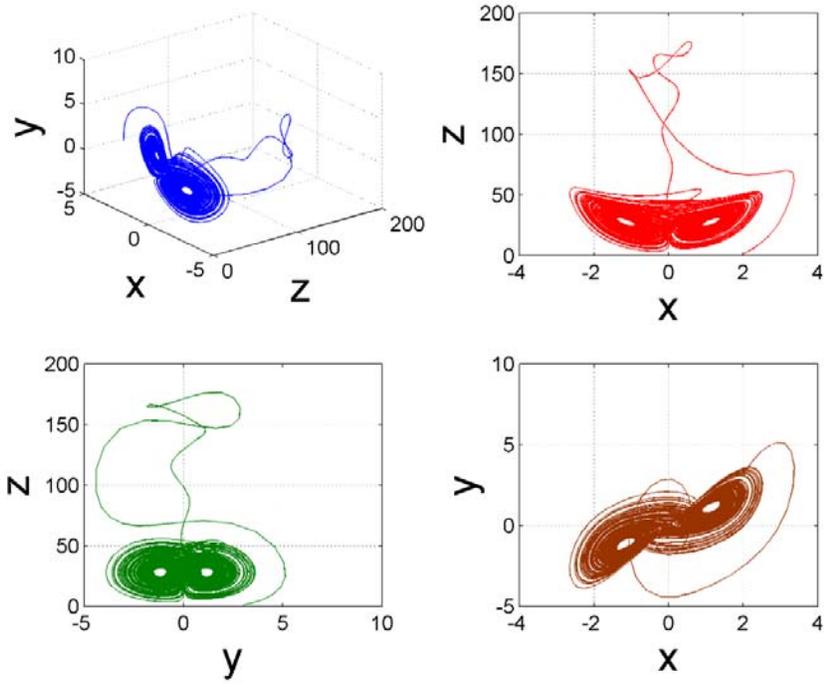

*Figure 83: Phase space dynamics of LorYZ42*

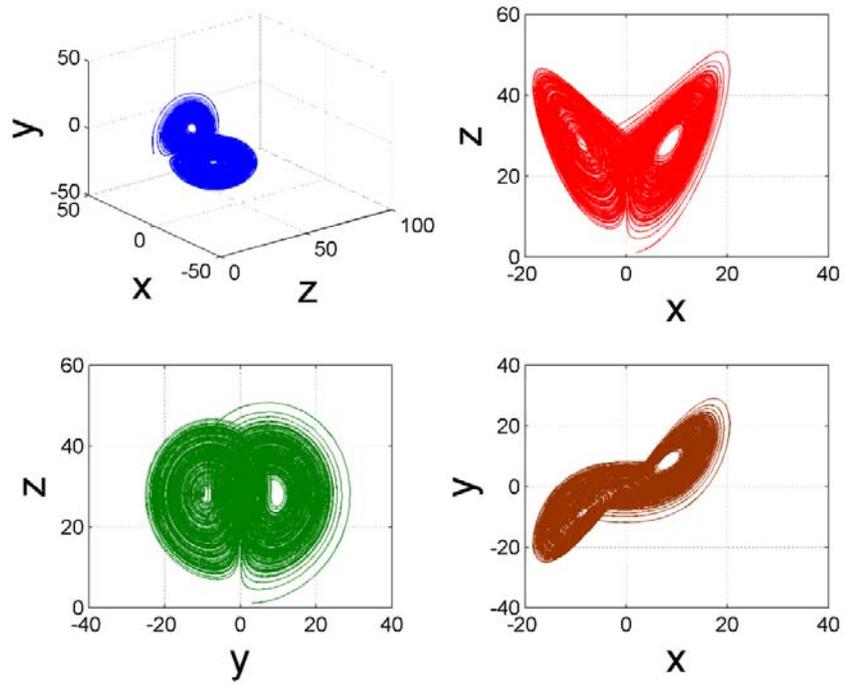

*Figure 84: Phase space dynamics of LorYZ43*

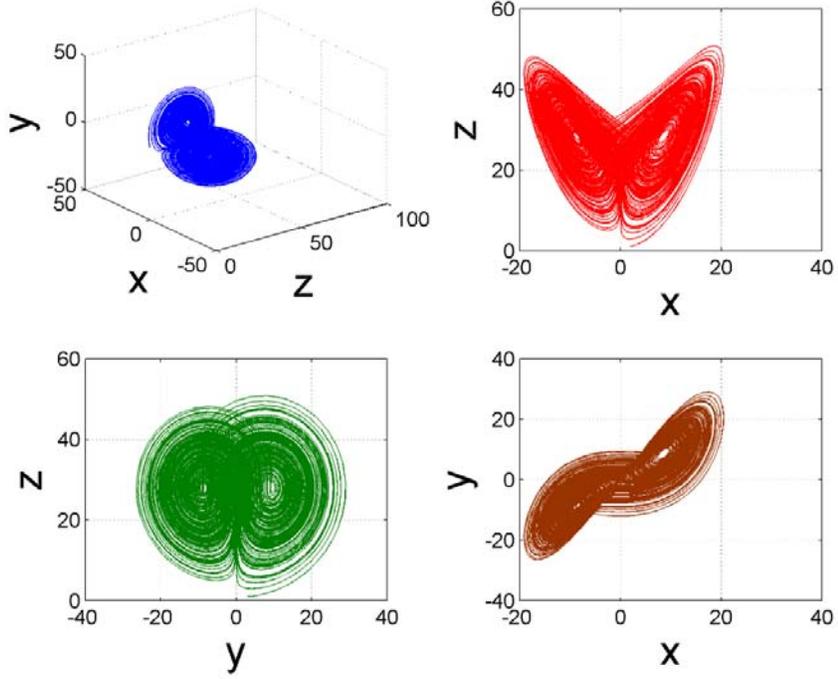

*Figure 85: Phase space dynamics of LorYZ44*

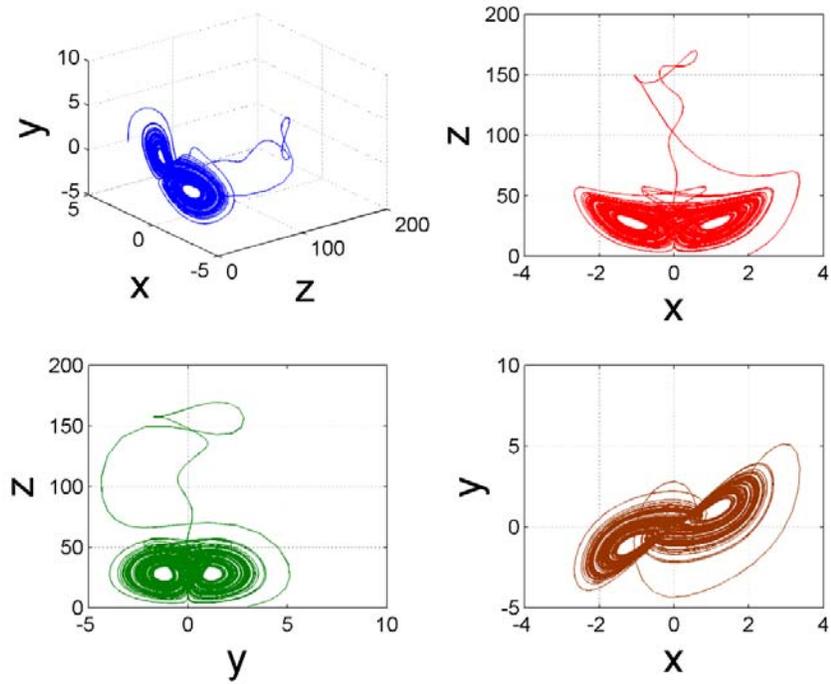

*Figure 86: Phase space dynamics of LorYZ45*

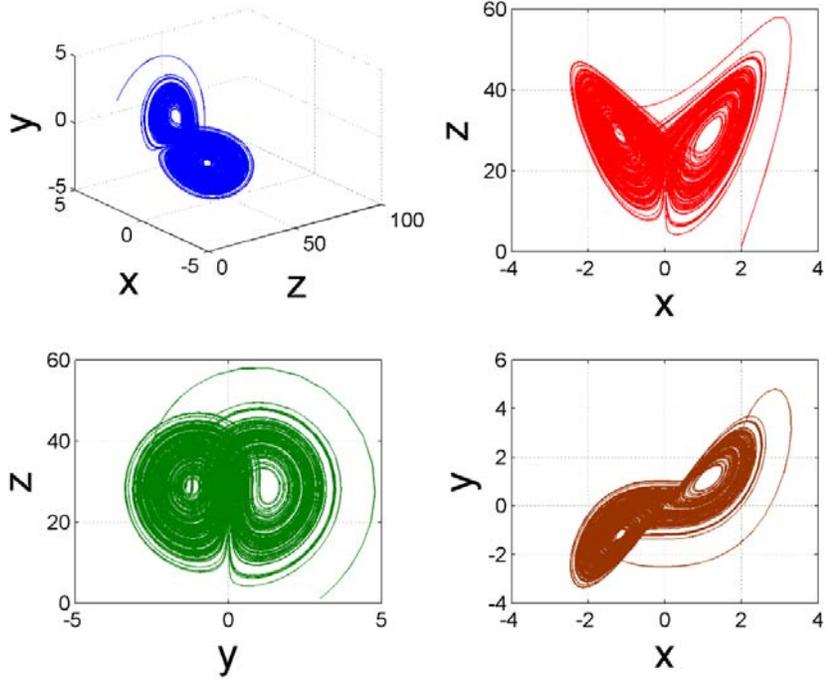

*Figure 87: Phase space dynamics of L₀rYZ46*

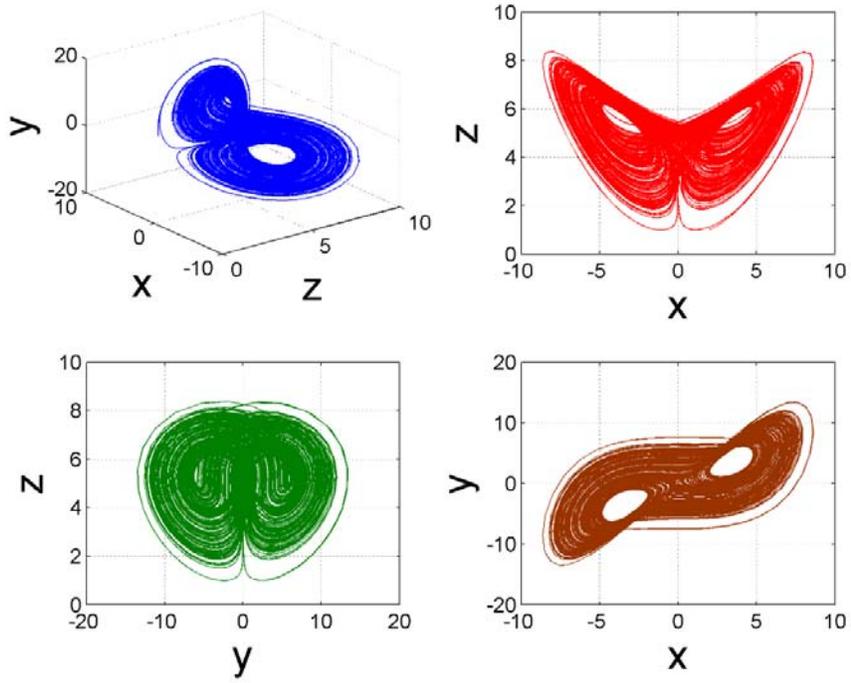

*Figure 88: Phase space dynamics of L₀rYZ48*

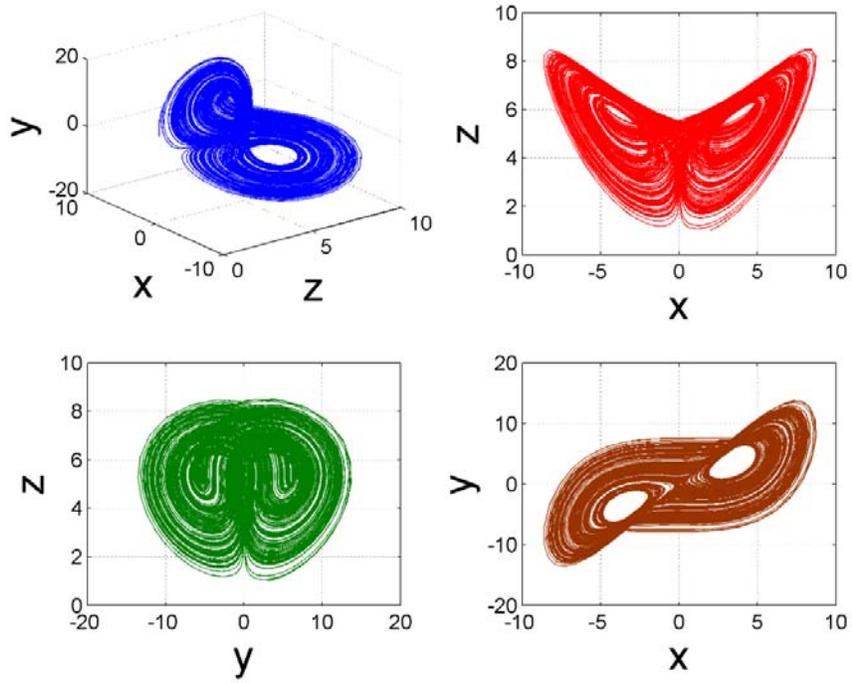

*Figure 89: Phase space dynamics of LorYZ49*

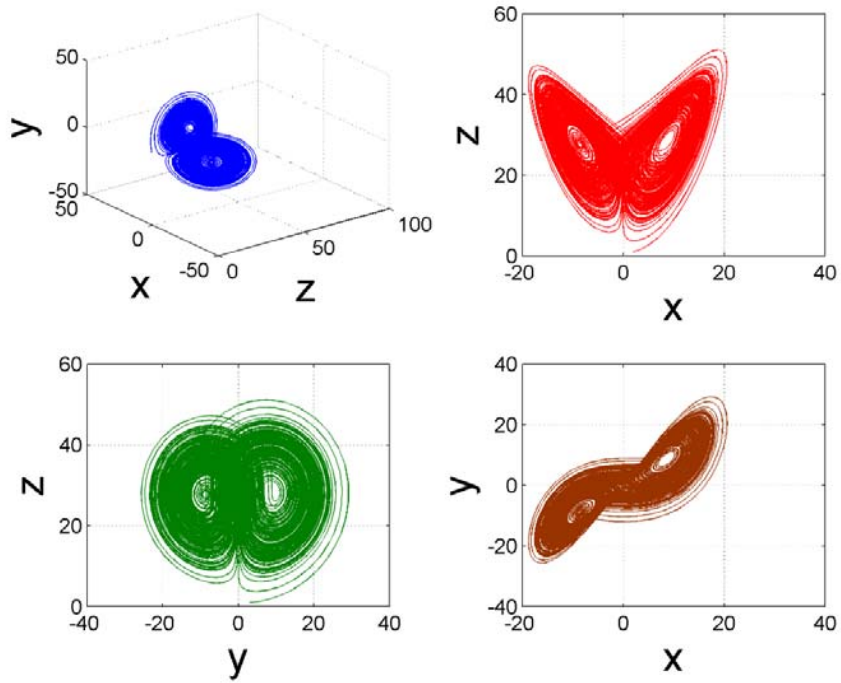

*Figure 90: Phase space dynamics of LorYZ50*

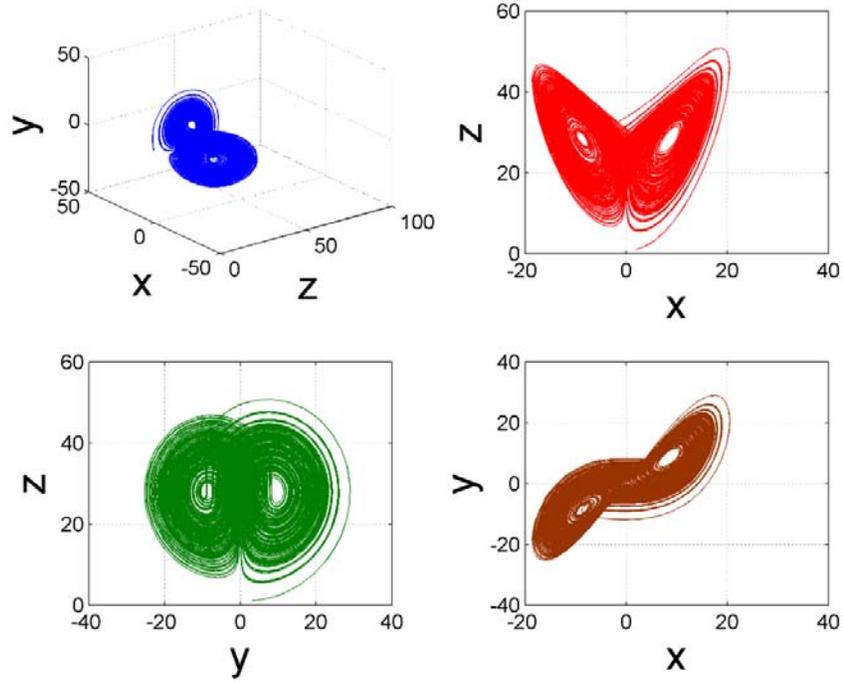

*Figure 91: Phase space dynamics of LorYZ51*

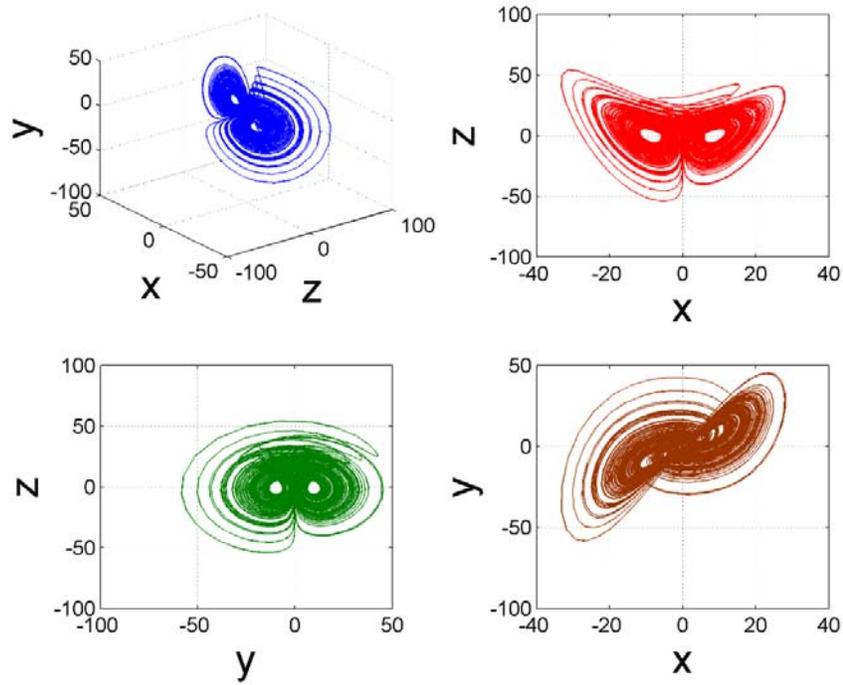

*Figure 92: Phase space dynamics of LorYZ52*

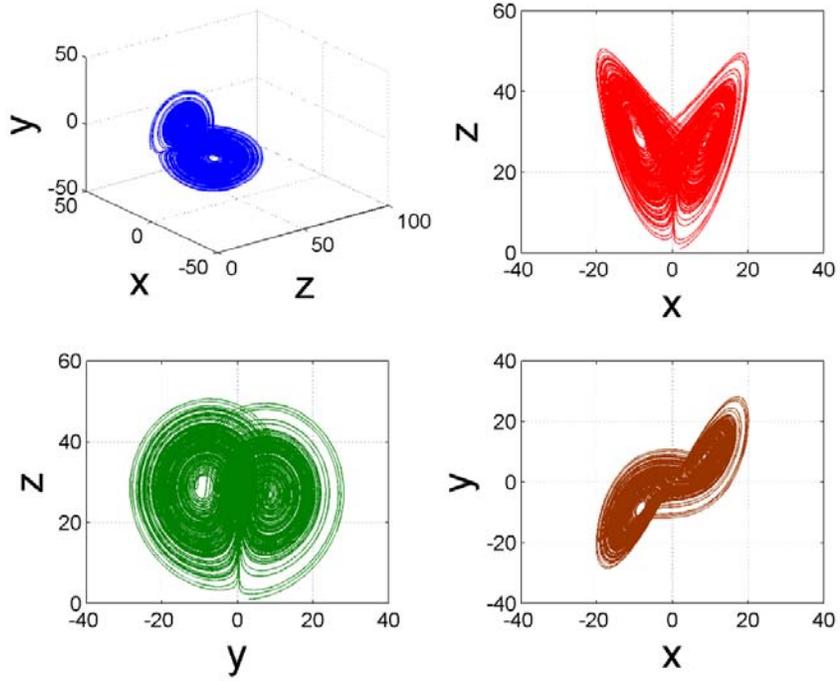

*Figure 93: Phase space dynamics of LorYZ53*

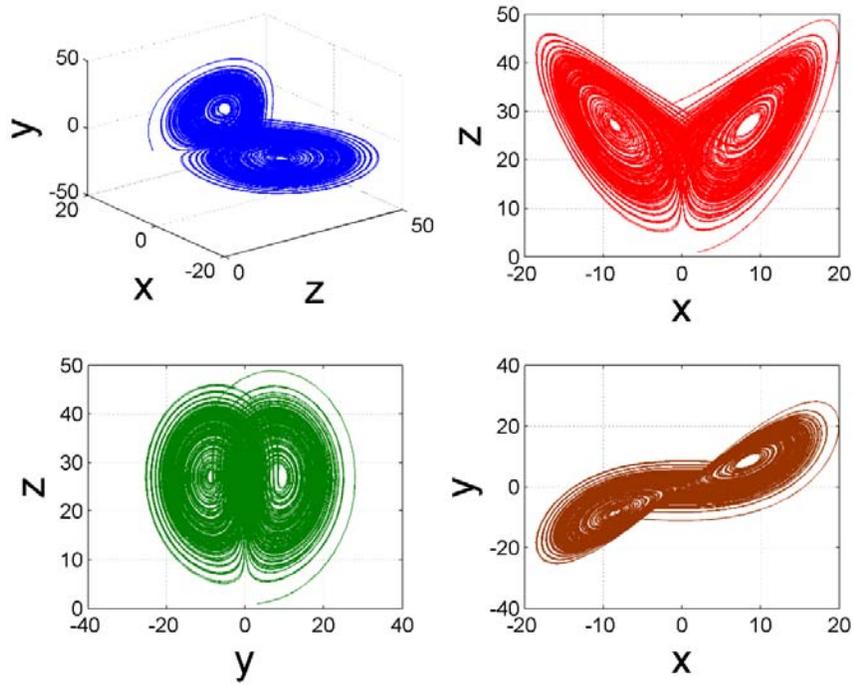

*Figure 94: Phase space dynamics of LorYZ54*

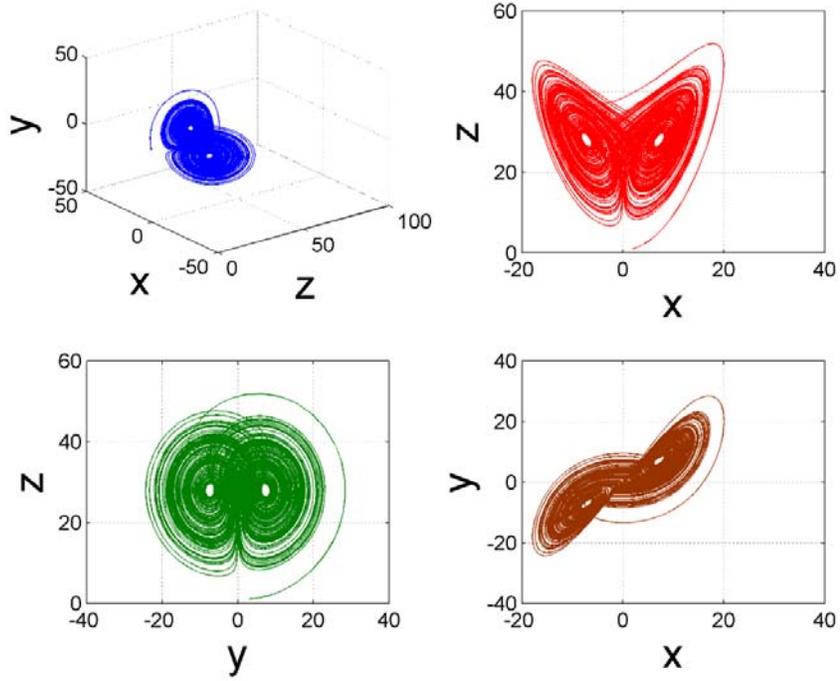

*Figure 95: Phase space dynamics of LorYZ55*

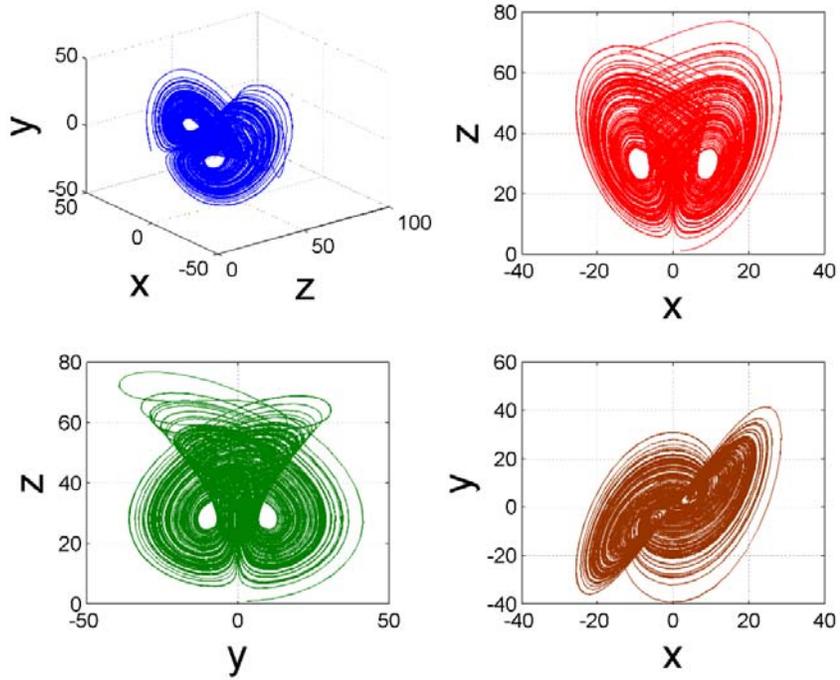

*Figure 96: Phase space dynamics of LorYZ56*

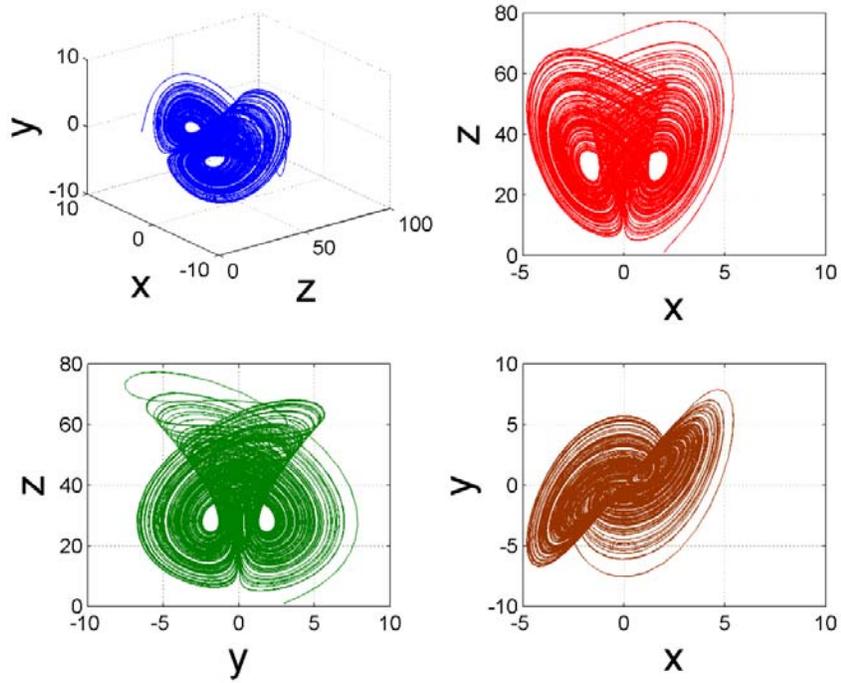

*Figure 97: Phase space dynamics of LorYZ57*

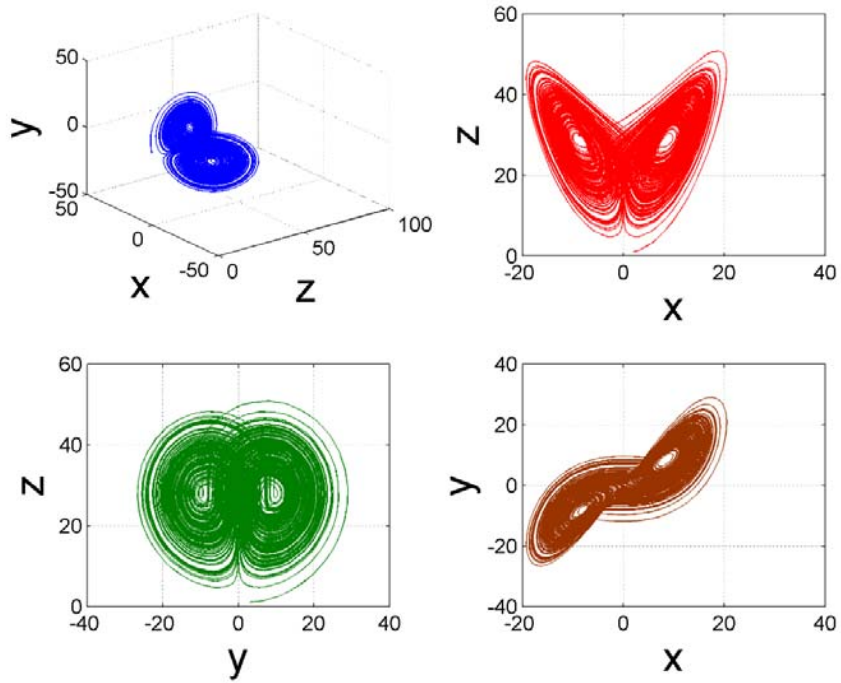

*Figure 98: Phase space dynamics of LorYZ58*

## d) *Phase portraits of generalized Lorenz x-y-z family of attractors*

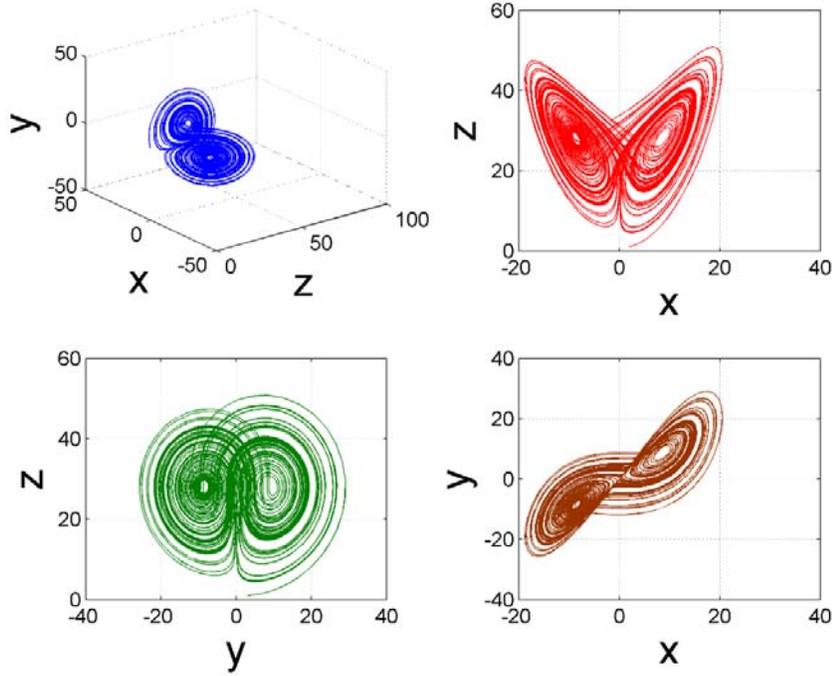

*Figure 99: Phase space dynamics of LorXYZ1*

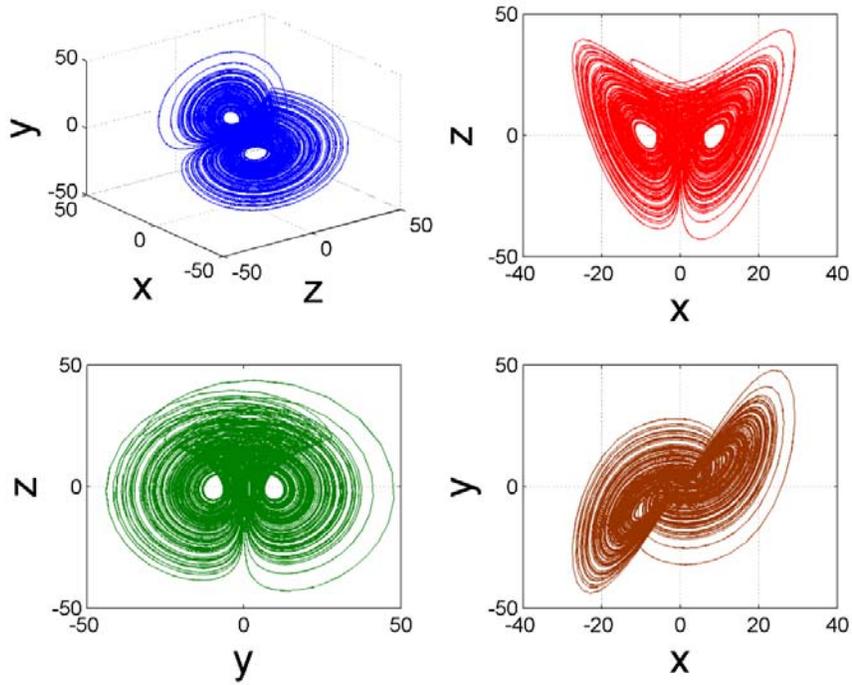

*Figure 100: Phase space dynamics of LorXYZ2*

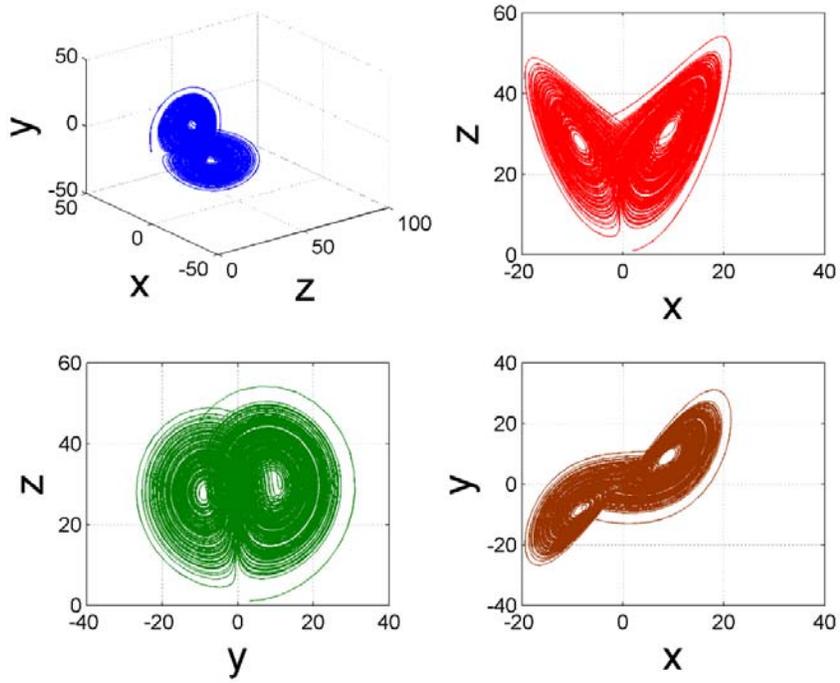

*Figure 101: Phase space dynamics of LorXYZ3*

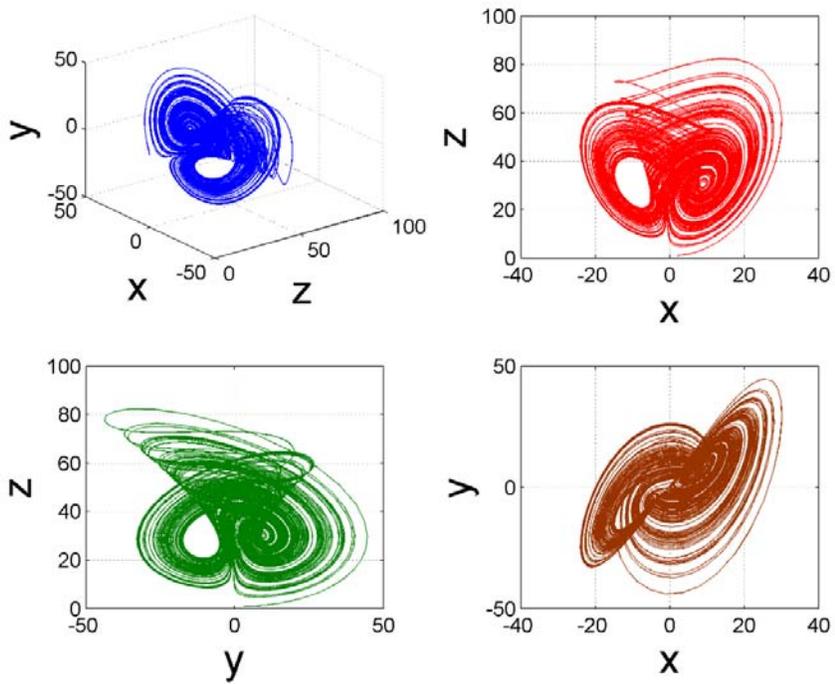

*Figure 102: Phase space dynamics of LorXYZ4*

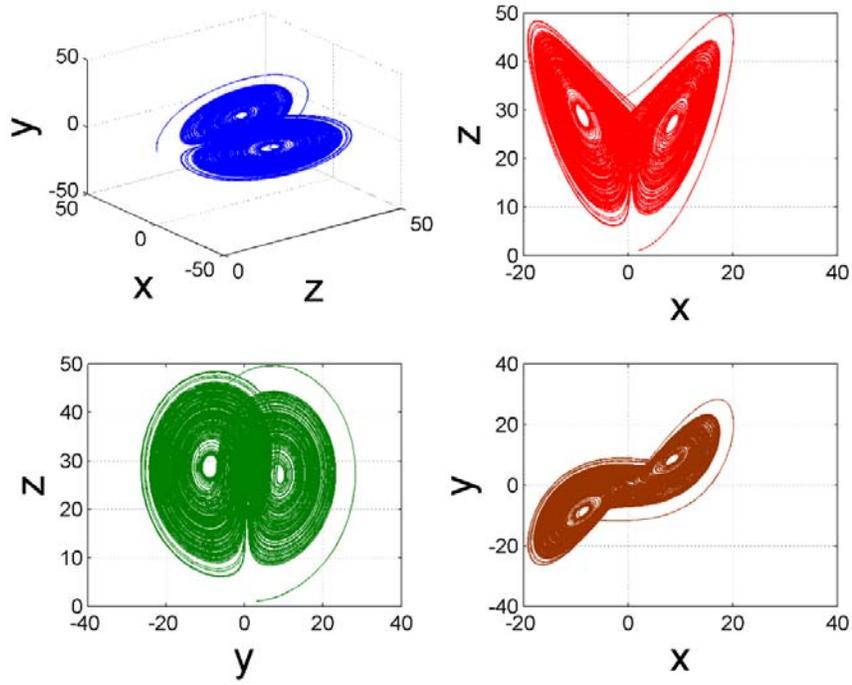

*Figure 103: Phase space dynamics of LorXYZ5*

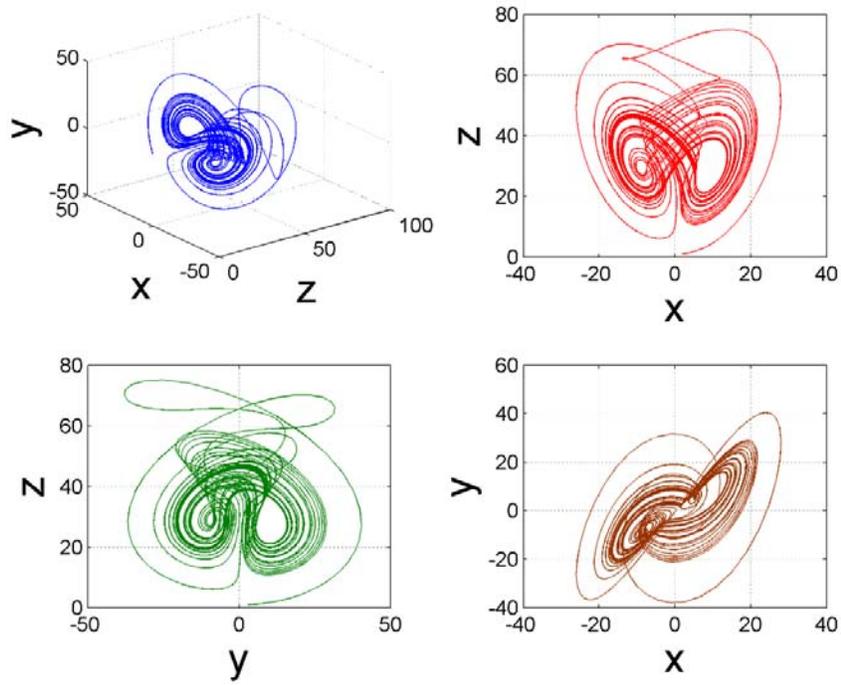

*Figure 104: Phase space dynamics of LorXYZ6*

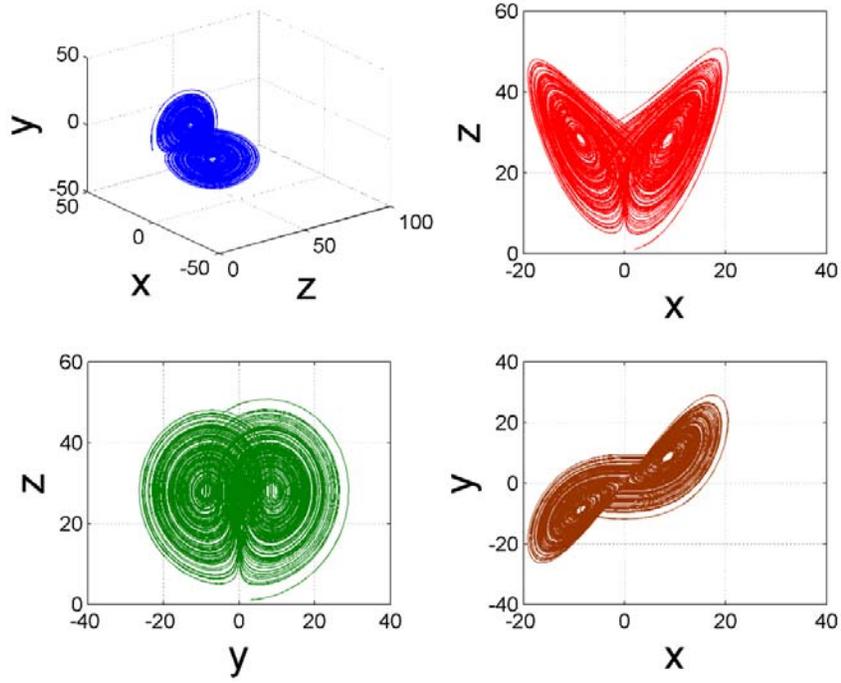

*Figure 105: Phase space dynamics of LorXYZ7*

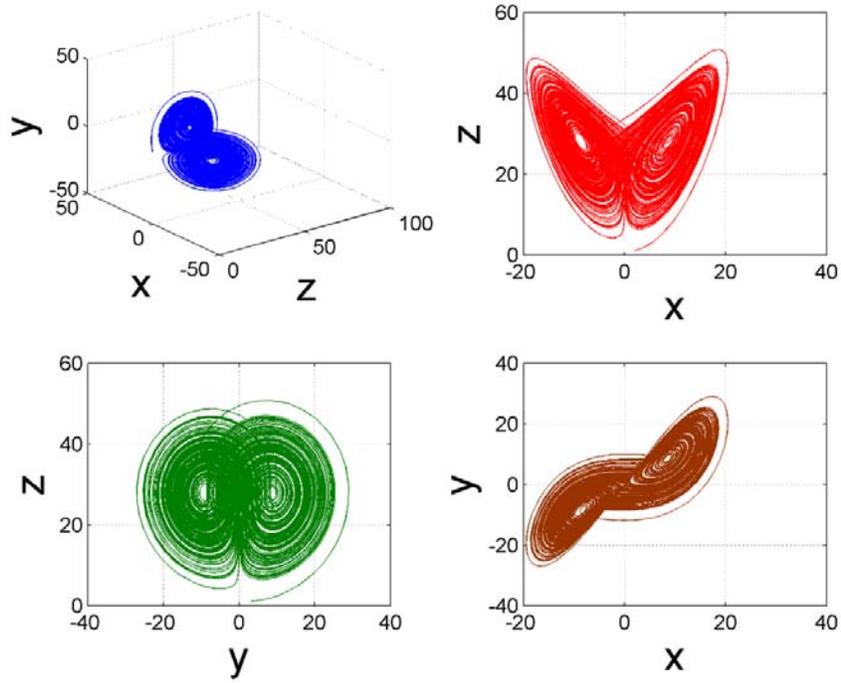

*Figure 106: Phase space dynamics of LorXYZ8*

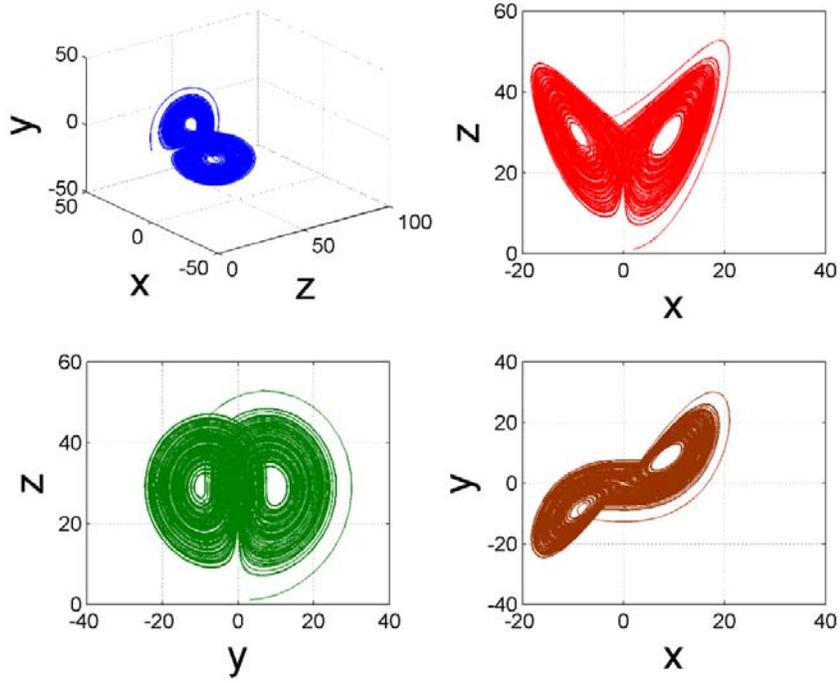

*Figure 107: Phase space dynamics of LorXYZ9*

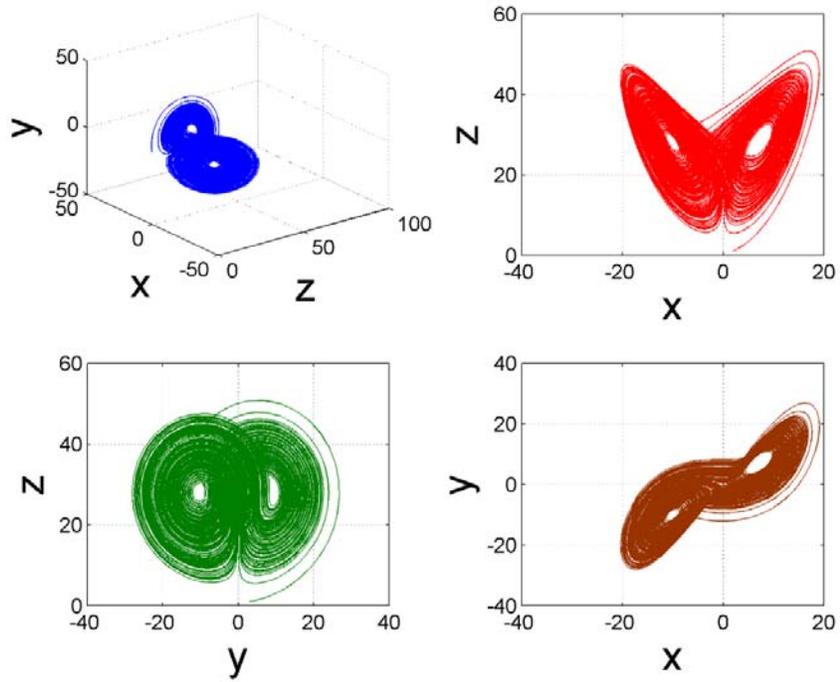

*Figure 108: Phase space dynamics of LorXYZ10*

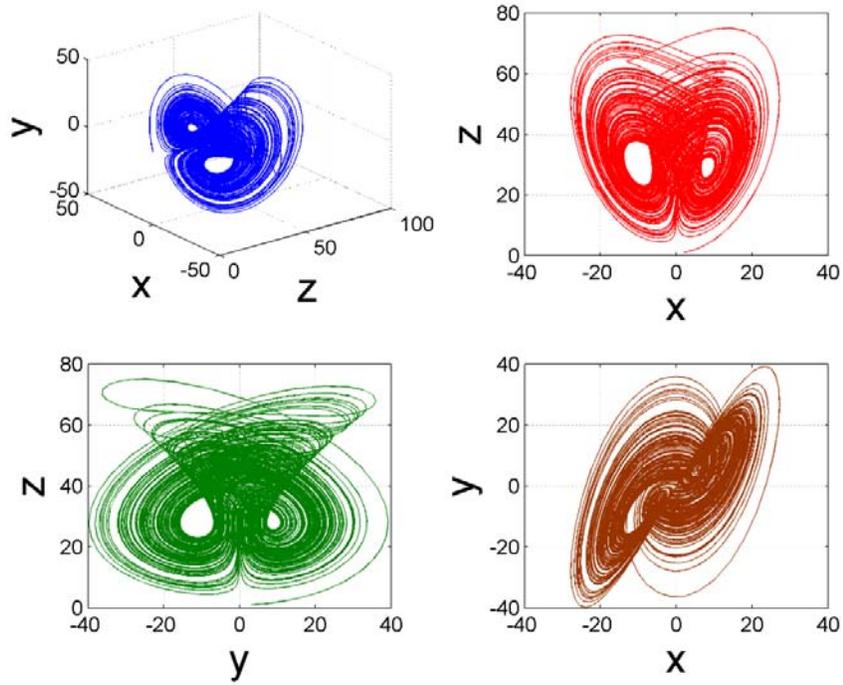

*Figure 109: Phase space dynamics of LorXYZ11*

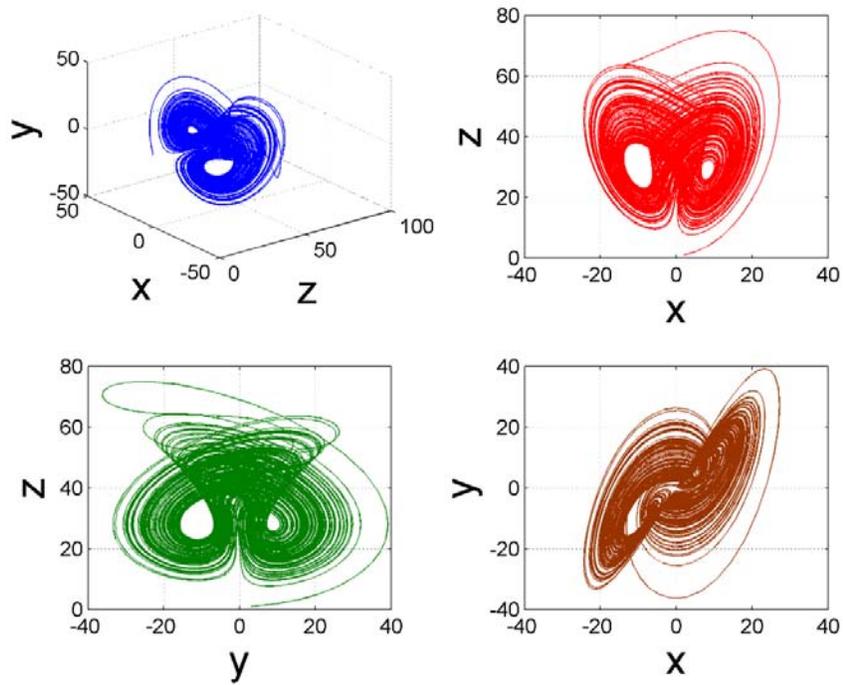

*Figure 110: Phase space dynamics of LorXYZ12*

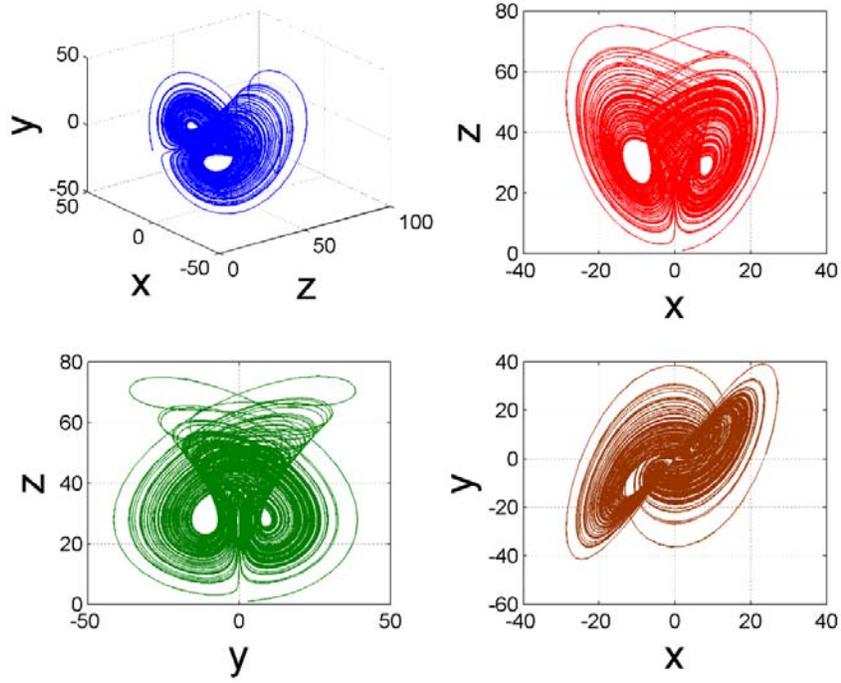

*Figure 111: Phase space dynamics of LorXYZ13*

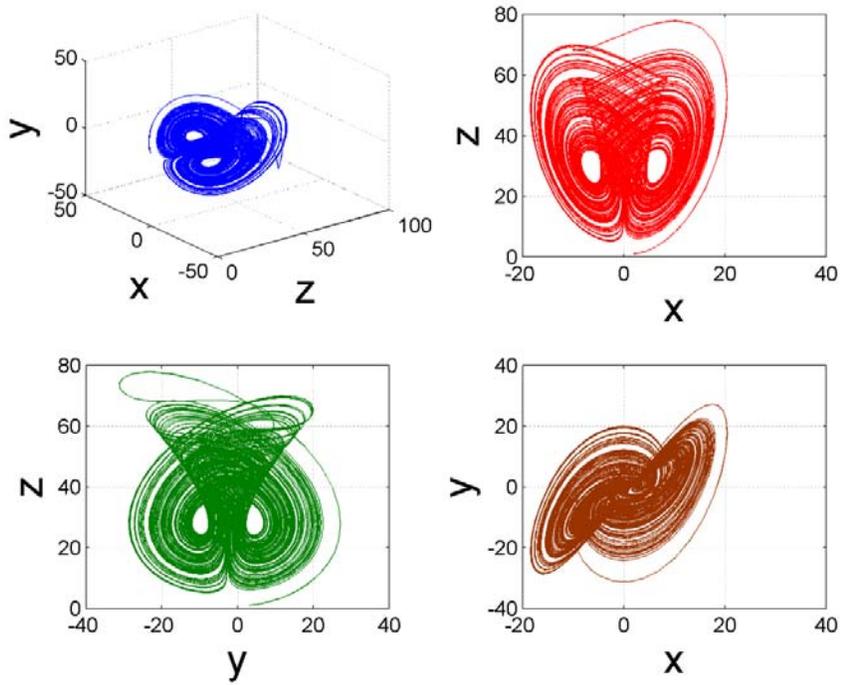

*Figure 112: Phase space dynamics of LorXYZ14*

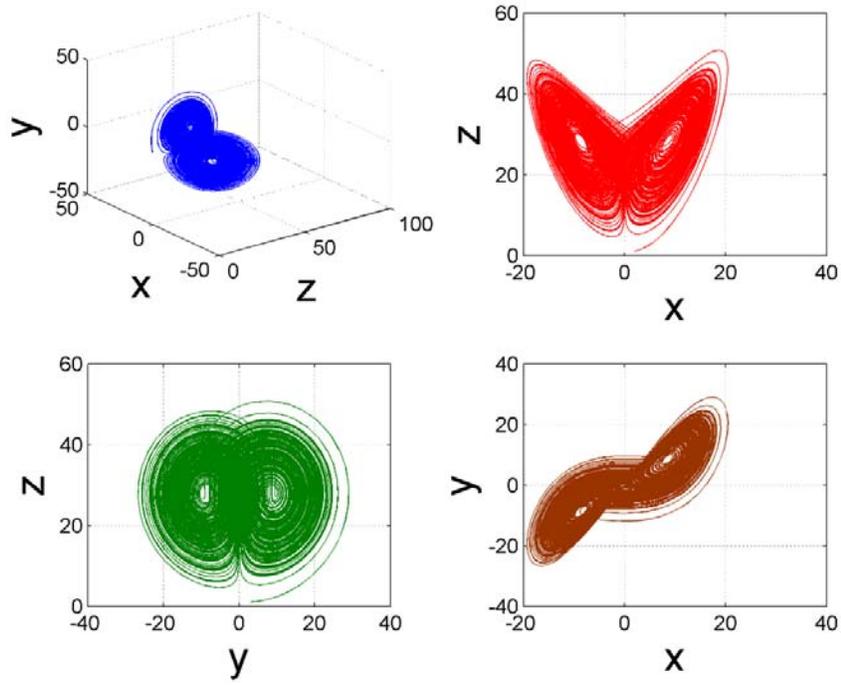

*Figure 113: Phase space dynamics of LorXYZ18*